\documentclass[a4paper, 11pt]{article}
\usepackage[left=2cm,right=2cm,top=2cm,bottom=2cm]{geometry}
\usepackage{lineno,hyperref,graphicx}
\usepackage{amsmath}
\usepackage{amssymb}
\usepackage{dsfont}
\usepackage{nicefrac}
\usepackage{here}
\usepackage{multirow}
\usepackage{diagbox}
\usepackage{amssymb}
\usepackage{caption}
\usepackage{subcaption}
\usepackage{color}
\usepackage{titlesec}
\usepackage{ulem}
\newcommand\LM[1]{{\color{black}{#1}}}

\newcommand\GP[1]{{\color{black}{#1}}}

\newcommand*\dif{\mathop{}\!\mathrm{d}}

\begin{document}

\title{A coupled implicit-explicit time integration method for compressible unsteady flows}
\author{Laurent Muscat, Guillaume Puigt, Marc Montagnac, Pierre Brenner}
\maketitle


\begin{abstract}
This paper addresses how two time integration schemes, the Heun's scheme for explicit time integration
and the second-order Crank-Nicolson scheme for implicit time integration, can be coupled spatially. This coupling is the prerequisite to
perform a coupled Large Eddy Simulation / Reynolds Averaged Navier-Stokes computation in an industrial context, using
the implicit time procedure for the boundary layer (RANS) and the explicit time integration procedure in the LES region.
The coupling procedure is designed in order to switch from explicit to implicit time integrations as fast as possible,
while maintaining stability. After introducing the different schemes, the paper presents the initial coupling procedure
adapted from a published reference and shows that it can amplify some numerical waves. An alternative procedure, studied in a coupled
time/space framework, is shown to be stable and with spectral properties in agreement with the requirements of industrial
applications. The coupling technique is validated with standard test cases, ranging from one-dimensional to three-dimensional flows.
\end{abstract}



\section{Introduction \label{sec:intro}}

Turbulence occurring in an unsteady compressible fluid flow is characterized by the cascade
of vortices distributed over a large range of temporal and spatial scales ~\cite{Chapman_1979_AIAA}.
When the Reynolds number $Re$ is large (typically for $Re\simeq 10^6 - 10^9$),
the resolution of the entire turbulence spectrum using Direct Numerical Simulation is still out of reach
today for industrial configurations, even on the most powerful supercomputers.
Currently, two major classes of approaches are mainly used to model turbulence effects in these situations.

The first technique consists in modeling all effects of the turbulence on the flow and leads to
Reynolds-Averaged Navier-Stokes (RANS) equations.
All turbulence effects are
represented by their averaged effects on the mean flow, and turbulence fluctuations are accounted for through
a turbulence model.
The RANS approach is today the preferred technique for industrial
applications, even if it is unable to represent unsteady effects of the turbulence.
In particular, its accuracy strongly depends on the turbulence
model used to represent the turbulence effects on the mean flow.
The main interest of this approach remains
its relatively low CPU cost for reasonable accuracy, especially for computing boundary layers.
Implicit time integrators are generally considered in that case
since they allow either fast convergence to steady-state solutions or large time steps
for unsteady flows. Indeed, implicit schemes are generally chosen so to be unconditionally stable.
These schemes generally need the linearization of the implicit residual, which makes the Jacobian of the
residual appear. Finally, this leads to look for the solution increment as the solution of a linear system.
Examples of implicit time integrators are the family of implicit
Runge-Kutta methods (IRK) proposed by Kunstmann~\cite{JKuntzmann_1961_ZAMM} and
Butcher~\cite{Butcher_1964_MC} or the Gear's scheme~\cite{Gear_Book_1971}. \LM{I}mplicit methods have the drawback to require extra
computational costs compared with explicit techniques, because a linear system must be solved at all time steps.

Another way to account for turbulence effects is Large Eddy Simulation (LES). Indeed,
the principle of LES is to compute the largest turbulent scales and to model the smallest scales that
cannot be captured by the mesh. Of course, the LES approach introduces additional terms to account for the
intractable turbulence spectrum but the universal nature of the smallest scales makes the models more general
than for the RANS approach.
In practice, LES requires a very fine mesh to capture turbulence scales, which partly explains why the overall CPU cost
is much higher than for RANS.
This ratio of CPU cost is today typically
about $10^3-10^5$, depending on the amount of modeled scales in LES, and therefore on the local mesh refinement.
However, the intrinsically unsteady nature of LES makes it favorable for flows for which
the hypothesis for averaging turbulence effects is not fully justified as for instance in wakes behind bodies
at high incidence, or for simulating transition effects.
Explicit time integration schemes are often used in LES to correctly capture the unsteady phenomenon.
Indeed, schemes with good spectral properties are the pillars for capturing turbulence spectrum accurately
and explicit schemes enable a simple and efficient control on spectral properties (dissipation and dispersion) 
and accuracy (order of accuracy)~\cite{Sengupta_2003_JCP,Sengupta_2011_JCP}.
However, explicit schemes suffer from small time steps that are limited by the Courant-Friedrichs-Lewy (CFL) number.
Hence, computing boundary layers with a fine mesh at the wall with LES becomes very demanding from the CPU time point of view.

Runge-Kutta (RK) methods~\cite{Runge_MathAnn_1895, Kutta_ZMathPhys_1901} are examples of explicit time integrators.
Some RK schemes were designed to respect good mathematical properties (low dissipation and low dispersion~\cite{Bogey_2004_JCP},
Total Variation Bounded~\cite{Cockburn_MC_52_1989,Cockburn_JCP_84_1989,Cockburn_MC_54_1990,Cockburn_JCP_141_1998},
Total Variation Diminishing~\cite{Gottlieb_MoC_67_1998}).
Standard Runge-Kutta methods are one-step and multi-stage schemes: they compute
the solution at the next discrete time from the solution at the current time using additional intermediate solutions.
An obvious way to extend the order of accuracy of the approach is to account for the solution history.
As the number of stages increases non-linearly with the order of accuracy for standard RK methods,
Williamson~\cite{Williamson_1980_JCP} designed low-storage RK schemes up to fourth order with
only two stages in order to decrease memory cost. In the same way, linear multi-step methods, such as
Adams-Bashforth methods~\cite{Norsett_1969_Springer}, enable the computation of the new solution using several past solutions.
For non-stiff problems, explicit Runge-Kutta or linear multi-step methods give accurate results.
However, in case of stiffness, such standard methods can suffer from instability~\cite{Higham_1993_BIT}.
Therefore, for many stiff problems, the use of explicit methods requires very small time steps for the calculation to be feasible or stable.

The formula of Navier-Stokes equations for RANS or LES are very similar, and composed of a term with a time derivative,
a convection term and a diffusion term. They both include
turbulence effects by means of turbulent viscosity and diffusivity (defined using either an averaging or a filtering procedure).
All the above comments on RANS and LES methods suggest that both methods may be coupled to take benefit from each of them
in its best domain of confidence. The idea is to compute the turbulent boundary layer with the RANS approach, in a region
where the LES can be less accurate because the mesh is generally not enough refined to capture turbulence effects. Far from
the wall, it could be better to switch to LES in order to capture the largest turbulence vortices.
Several ways to couple RANS and LES models can be found in the literature such as the Wall-Modeled LES~\cite{Catchirayer2018}
or the Detached Eddy Simulation~\cite{Spalart_2009,TerracolSagautDeck}.

In this context, the present study concerns the design of a time integration procedure adapted to the spatial coupling
of RANS and LES regions using implicit and explicit time integrators already included in the solver called FLUSEPA\footnotemark[1].
This solver, developed by ArianeGroup, is used in the certification process of launchers during take-off and reentry,
when complex physical phenomena occur. \LM{During the last years, the solver received attention and was the subject of many improvements 
\cite{Limare_2016_AIAA,Pont_JCP_2017,Charrier_2018_1,COUTEYENCARPAYE_2018_JCS,Amandine_2018_AIAA,Amandine_CS_XX_2019}
.}

The starting point of our study is the Heun's~\cite{Heun_ZMathPhys_1900} explicit time integration scheme available
in this code. It has produced results in agreement with given LES industrial requirements and it is in addition 
made compatible with local mesh refinement, as in \cite{Krivodonova_JCP_229_2010}.
The remaining question is how to couple this second-order explicit scheme with a second-order implicit scheme,
making a hybrid version of the underlying schemes that still remains second-order accurate.

Hybridization of explicit and
implicit schemes is an important subject of research. In this context, IMplicit/EXplicit (IMEX)
schemes consist in using a different time integration for the different terms in the original set of
equations. Stiff terms are solved implicitly and regular terms explicitly. But since only one time integration
term appears in the original set of equations, the IMEX approach is generally justified by the Strang's
splitting procedure~\cite{Strang_1968}. The pillar of IMEX schemes lies on the separation of time
integration schemes based on the different terms of the equations (see~\cite{Kim_1985_JCP} and
the analysis performed by Ascher {\it et al.}~\cite{Ascher_1995_SIAM}).
However, in our case, the coupling of RANS and LES
leads to {\textit{spatially}} couple explicit and implicit schemes as implicit schemes are
suitable for RANS equations and explicit time integration for LES. To our knowledge, much less
attention was paid on techniques to couple explicit and implicit schemes spatially. 
\GP{At this point, it must be highlighted that for practical applications, the monitoring 
of the convergence of the implicit formulation has a strong influence on the quality of the results ,
especially for LES. This is a reason why a full implicit formulation for both RANS and LES areas is not retained.}

In 1986, Fryxell {\it et al.}~\cite{Fryxell_1986_JCP} built an hybrid implicit-explicit method
to perform one-dimensional Lagrangian hydrodynamics simulation based on the extension of a Godunov-type scheme to implicit regime.
Dai and Woodward~\cite{Dai_1996_JCP} proposed an iterative approach of this framework for Euler equations.
Their scheme is second-order accurate in time and space but quite complicated to
implement. This is especially a consequence of the requirement of solving the solution by an iterative
process. In addition, the authors mention that the extension to multi-dimensional system is
still an open question.  Collins {\it et al.}~\cite{Collins_1995_JCP} proposed the hybrid
explicit-implicit scheme for Eulerian hydrodynamics, denoted CCG, that ensures the total variation diminishing (TVD) property at all CFL
numbers for a linear equation. This hybrid scheme allows a second order of accuracy in space and time on pure explicit mesh cells.
Unfortunately, the CCG scheme leads to incorrect numerical solutions and fails to maintain the TVD property in
case of nonlinear equations. Men'shov and Nakamura~\cite{Men_Shov_2004_AIAA} proposed an extension
of this formulation to non-linear hyperbolic equation with Maximal Norm Diminishing (MND) property.
There still remains a strong limitation: the switching procedure of CCG and Men'shov and Nakamura schemes
(from explicit to implicit and {\textit{vice versa}}) cannot maintain a second order of accuracy in space and time over the whole mesh. 
Timofeev and Norouzi~\cite{Timofeev_2016_inProcessing} overcame this
difficulty by blending second-order accurate schemes (in space and time) for explicit, implicit and
hybrid cells using a smooth parameter to couple the considered time integrators. Their method is one of the
most advanced \LM{ones} since it maintains TVD property. 
In 2006, second order accurate explicit and implicit time integrators were hybridized by T\'{o}th {\it et al.} 
\cite{Toth_2006_JCP} and coupled with an adaptive mesh refinement (AMR) strategy for three-dimensional applications in
magnetohydrodynamics using structured grids. The procedure is quite complex to implement and needs to blend 
the first and second order fluxes on the interfaces.

In 2017, May and Berger \cite{May_2016_JSC} proposed a technique to switch between explicit and implicit
approaches called \textit{flux bounding}.
The coupling of explicit and implicit schemes is performed by the
coupling of explicit and implicit flux contributions on the interface in the context of finite volume
discretization procedure. Unfortunately, it led to a result that is not second-order accurate in time
and space for "transient" cells for which residuals were computed using explicit and implicit flux
contributions.
Finally, it is important to notice that Persson~\cite{Persson_AIAA_2011} also provided a Runge-Kutta based IMEX scheme
that allows to couple spatially explicit and implicit time integration schemes. However, this technique does not respect
our requirement to focus on Heun's scheme {\it i.e.} the explicit time integrator in FLUSEPA\footnotemark[1].
\footnotetext[1]{\label{note:FLUSEPA}FR4009261, 27/07/2013}
In this paper, a new way to couple spatially second-order accurate explicit and implicit time integrators
for compressible unsteady flows is introduced. The goal is to propose a coupling technique that ends to a second-order accurate
scheme over the whole computational domain. The remainder of this paper unfolds as follows.
In Sec.~\ref{sec:1b_Equations}, the standard form of the Navier-Stokes
is recalled and the explicit and implicit schemes chosen for the present study are introduced. 
Our explicit and implicit basic schemes, Heun and implicit Runge Kutta schemes, are defined in Sec.~\ref{sec:BasicTI}.
In Sec.~\ref{sec:2_Norouzi}, the method presented
in~\cite{Timofeev_2016_inProcessing} is recalled as it will serve as the basis of the new scheme. 
In Sec.~\ref{sec:HCS1}, a first version of coupling Heun and implicit Runge Kutta (IRK2) time integrators is introduced with its
dedicated space-time stability analysis. The first hybridized scheme is not acceptable and a second version of the coupling scheme 
is presented in Sec.~\ref{sec:AION}. The space-time stability analysis is performed again. 
Before concluding, Sec.~\ref{sec:Validation} is dedicated to the validation of the new coupling scheme for 1D, 2D and 3D test cases.

\section{Discretization of the Navier-Stokes Equations}\label{sec:1b_Equations}

The Navier-Stokes conservation laws can be written in the following compact form,
\begin{equation}
\frac{\partial W}{\partial t}   +\nabla \cdot F(W) + \nabla \cdot G(W,\nabla W)  = 0,
\label{eq:euler}
\end{equation}
where $W$ is the vector of conservative variables and $\nabla W$ its gradient, $F$ and $G$ are respectively the flux density for convection and viscous effects.
The computational domain $\Omega$ is split into $N$ non-overlapping rigid immobile cells $\Omega_j$ and
Eq.~\eqref{eq:euler} are integrated over every mesh cell. The Gauss relation ties
the volume integrals of the divergence terms to the interface fluxes, leading to
\begin{equation}
\frac{\dif}{\dif t} \iiint_{\Omega_{j}}  {W}  \dif \Omega = -\iint_{A_{j}} F(W) \cdot \vec{n} \dif S   -\iint_{A_{j}}   {G(W, \nabla W)} \cdot \vec{n} \dif S,
\label{eq:weakform}
\end{equation}
where $\Omega_{j}$ is the $j$-th control volume with its border $A_{j}$ and $\vec{n}$ is the outgoing unit normal
vector.
From now on, it is possible to define the averaged quantity of the conservative variables,
\begin{equation}
\overline{W}_j= \frac{1}{|{\Omega_j}|} \iiint_{\Omega_j} W \dif \Omega.
\end{equation}
This formula allows to rewrite the conservation laws~\eqref{eq:weakform} discretized with a finite-volume formulation in the following compact form,
\begin{equation}
  \frac{\dif \overline{W}_j}{\dif t}=R(\overline{W}_j),\label{eq:compact_form}
\end{equation}
with $R(\overline{W}_j)$ is the residual computed using the averaged quantities in cell $j$.
Actually, the residual is defined as the sum of the flux over the whole
boundary of a cell. Here, convection and diffusion fluxes are discretized by means of a $k$-exact
formulation coupled with successive corrections~\cite{Pont_JCP_2017,Haider_2014_SpringerBook}.
Indeed, any high-order scheme needs a polynomial representation of
the unknowns locally around a cell of interest. Theoretically, the polynomial approximation can also be
defined as the Taylor expansion of the local solution. Standard $k$-exact schemes need the definition of a
large stencil in order to link Taylor expansion coefficients and averaged quantities (see~\cite{Gooch_2002_JCP}
as an example of standard $k$-exact scheme). This large stencil is a drawback to cope with parallel computations
since the number of data to exchange between domains depends on the partitioned (initial) mesh. The successive corrections were
designed in order to avoid geometrical reconstructions for parallel computations. The number of ghost cells
are limited to the minimum of one ghost cell per volume owing an interface on the boundary.
Starting from the solution, the polynomial reconstruction is built using the truncation error of the Taylor expansion
by increasing the order of the approximation.
The starting point is the availability of the averaged quantities in the mesh cells and the definition of a
discrete gradient scheme. The discrete gradient scheme applied to the averaged quantities leads to a given order of accuracy.
A local analysis enables to estimate the truncation error performed on the gradient and by removing
the last non-zero term in the truncation error, the gradient scheme accuracy
is made more accurate by one order. This gradient is then usable to define a more accurate representation of the unknown through
a local polynomial reconstruction. The process can then be repeated for the Hessian in order
to increase accuracy on the computation of the successive derivatives~\cite{Haider_2014_SpringerBook,Pont_JCP_2017}.
Of course, since the truncation error must be computed, it is
necessary to locally exchange the set of local derivatives (first, the gradients, then the second-order derivatives,
then the third-order derivatives and so on).
The TVD property is ensured with a MUSCL reconstruction.
For a second-order scheme, the flux on an interface is estimated using an affine extrapolation of the cell-centered unknowns
onto the interface. If $f$ denotes a mesh interface with a unit normal vector $\vec{n}$ directed from the left cell $j$ to
the right cell $i$, and if $C_i$, $C_j$ and $C_f$ respectively denote the cell centers and the interface center, the standard reconstruction on the vector of primitive variables $V$, is
simply,
\begin{equation}
\begin{aligned}
V^L=& V^n_{j}+(\nabla V)_j^{n} \cdot \overrightarrow{C_jC_f}, \\
V^R=& V^n_{i}+ (\nabla V)_{i}^{n} \cdot \overrightarrow{C_iC_f}, \\
\end{aligned}
\end{equation}
where the gradient $\nabla V$ needs to be computed with the $k$-exact reconstruction.
These gradients are also used to compute the viscous flux.

Finally, the finite-volume formulation of Eq.~\eqref{eq:compact_form} can be expressed as the following Cauchy problem,
\begin{equation}
\left\{
\begin{aligned}
& \frac{\dif \overline{W}_j(t)}{\dif t} = R({\overline{W}}_j), \hbox{~}\forall t \in R^+,\\
&\overline{W}_j(0)  =W_j^{0},  \quad \forall j \in\{1,...,N\},
\end{aligned}
\right .
\label{Cauchy1}
\end{equation}
where ${W_j^{0}}_{1\le j\le N}$ denotes the initial solution.
For the sake of clarity, the averaging symbol will be dropped,
and ${W}$ will represent the averaged quantities over the control volumes.

\section{Heun and Crank-Nicolson/IRK2 Time Integration Methods\label{sec:BasicTI}}

Heun and Crank-Nicolson/IRK2 are considered as the two basic time integrators of
the present study from which a new method will be developed.

The Heun's explicit scheme~\cite{Heun_ZMathPhys_1900} is a second-order accurate predictor-corrector method.
The state $W^{n+1}$ is first predicted with a forward Euler scheme, and then corrected by a standard trapezoidal rule,
\begin{equation}
  \left\{
  \begin{array}{ll}
  \hbox{Predictor step:} & \displaystyle \widehat{W}= W^n + {\Delta t} R(W^n), \\
  \vspace*{-3mm} &  \\
  \hbox{Corrector step:} & \displaystyle W^{n+1}=W^n+\frac{\Delta t}{2 } \Big( R(W^n) + R(\widehat{W}) \Big). \\
  \end{array}
  \right.
  \label{eq:HEUN}
\end{equation}

The Crank-Nicolson/IRK2 implicit scheme~\cite{Crank_1947_PCP} is second-order accurate, and it can be expressed as,
\begin{equation}
\begin{aligned}
W^{n+1}=W^{n}+\frac{\Delta t}{2 }\Big( R(W^n) + R(W^{n+1}) \Big).
\label{IRK2_f}
\end{aligned}
\end{equation}
The residual being non linear in $W$, Eq.~\eqref{IRK2_f} is solved with an iterative Newton method.
Starting from the iteration $n$ in Eq.~\eqref{IRK2_f}, the next solution is defined as the root $X$ of $f(X)=0$ with
\begin{equation}
f(X) = X-W^{n}-\frac{\Delta t}{2 } \Big( R(X)+R(W^{n}) \Big).
\label{eq:Newton1}
\end{equation}
At the $p$-th step of the Newton algorithm, the next iterative $X^{p+1}$ is such that
\begin{equation}
{f'(X)}_{{|}_{X^p}} (X^{p+1} - X^p) = -f(X^p).
\label{eq:Newton2}
\end{equation}
The final algorithm is simply
\begin{equation}
\left\{
\begin{aligned}
X^0&=W^n \\
\displaystyle (I- \frac{\Delta t}{2 } \, J) \, \delta X^p &= -f(X^p)\\
X^{p+1}&=\delta X^p +X^p
\end{aligned}
\right.
\end{equation}
with $J=\frac{\partial R}{\partial W}$ the Jacobian matrix and $I$ the identity matrix.
The linear system in the iterative loop is solved with a Krylov iterative method.
The proposed Newton procedure is motivated by
the strong non-linearities encountered during reentry and take-off. For small non-linear effects,
the residual $R(W^{n+1})$ can be linearized around $W^n$.

The open question concerns how to spatially couple both second-order time integrator schemes, while maintaining
the second order of accuracy and ensuring the overall stability, and taking into account the fact
that both time integrators need two residuals computed at two different times. To this aim,
the hybrid time integrator proposed by Timofeev and Norouzi~\cite{Timofeev_2016_inProcessing} is first considered.
In the following, all numerical schemes will be expressed in a one-dimensional
domain without loss of generality, and for convective terms. Diffusive terms will be discussed if required.

\section{Principle of the Hybrid Time Integration}\label{sec:2_Norouzi}

The spatially-coupled explicit and implicit scheme proposed
by Men'shov and Nakamura~\cite{Men_Shov_2004_AIAA} and extended by
Timofeev and Norouzi~\cite{Timofeev_2016_inProcessing} is recalled hereafter
to make an introduction to the hybrid time integration.
The hybrid scheme~\cite{Timofeev_2016_inProcessing} applied to the Cauchy problem of Eq.~\eqref{eq:compact_form}
has a predictor-corrector formulation:
\begin{equation}
    \left\{
    \begin{array}{ll}
    \hbox{Predictor step:} &
    \displaystyle \widehat{W}_j = W^{n}_j + \omega_j {\Delta t} R(W^{n}_j) \\
    \vspace*{-3mm} &  \\
    \hbox{Corrector step:} & \displaystyle W^{n+1}_j = W^n_j+\frac{\Delta t}{|\Omega_j|} (F_{j+\frac{1}{2}}^{Hybrid} - F_{j-\frac{1}{2}}^{Hybrid})  \\
\end{array}
    \right. \text{for } 0 \leq \omega_j \leq 1.
    \label{eq:TNScheme}
\end{equation}

The main components of the scheme are presented in the following using
a face $f$, with the face center $C_f$, separating two cells, $i$ and $j$, with
respectively cell centers $C_i$ and $C_j$.

During the predictor step, the residual computation does not use any Riemann solver, and
the predicted state $\widehat{W}_j$ is computed using a fully upwind scheme.
So, the flux $F(\LM{V}^f_j)$\LM{, computed from the vector of the primitive variables ${V}$ and} involved in the computation of the residual $R({W^{n}_j})$ of the cell $j$
uses the \LM{reconstruction on the primitive} state $\LM{V}^f_j = \LM{V}^{n}_j + (\widetilde{\nabla \LM{V}})^{n}_j \cdot \overrightarrow{C_jC_f}$ \LM{\cite{Norouzi_AIAAP_3046_2011,Timofeev_ISSW_2011,Norouzi_WCCM_XI-ECCM_2014}}.
This prediction step is not conservative since the flux $F(\LM{V}^f_j) \neq F(\LM{V}^f_i)$.
Moreover, the gradient $\widetilde{\nabla \LM{V}}$ is computed from the standard gradient ${\nabla \LM{V}}$ and
a slope limiter.

For any cell $j$, the hybrid flux $F^{Hybrid}$ is computed using a Riemann solver
with the following reconstructed left and right \LM{primitive} states on a face:
\begin{equation}
    \begin{array}{lrl}
    \LM{V}^L =& \omega_j & \displaystyle \bigg[ \frac{\LM{V}^n_{j}+\widehat{\LM{V}}_{j}}{2}+(\widetilde{\nabla \LM{V}})_j^{n}\cdot \overrightarrow{C_jC_f} \bigg]+ \\
    & \vspace*{-3mm}\\
    & (1-\omega_j) & \displaystyle \bigg[ \LM{V}^{n+1}_{j}+(\widetilde{\nabla \LM{V}})_j^{n+1}\cdot \overrightarrow{C_jC_f} - \frac{1-\omega_j}{2}(\widetilde{\Delta_t \LM{V}})^{n+1}_{j} \bigg],\\
    \LM{V}^R = & \omega_{i} & \displaystyle \bigg[ \frac{\LM{V}^n_{i}+\widehat{\LM{V}}_{i}}{2}+ (\widetilde{\nabla \LM{V}})_{i}^{n} \cdot \overrightarrow{C_iC_f} \bigg]+ \\
    & \vspace*{-3mm}\\
    & (1-\omega_{i}) & \displaystyle \bigg[ \LM{V}^{n+1}_{i}+(\widetilde{\nabla \LM{V}})_{i}^{n+1}\cdot \overrightarrow{C_iC_f} - \frac{1-\omega_{i}}{2}
    (\widetilde{\Delta_t \LM{V}})^{n+1}_{i} \bigg].
    \end{array}
    \label{eq:Timofeev_MUSCL}
\end{equation}

The parameter $\omega_j$, subsequently named cell status, is defined for each cell $j$
such that $0 \leq \omega_j \leq 1$. It is used to switch smoothly from a time-explicit ($\omega_j = 1$)
to a time-implicit ($\omega_j = 0$) scheme.
The cell status $\omega_j$ is locally adapted to the flow and the stability constraint. First,
the global time step is chosen to be the maximum allowable time step on a part $\cal{D}$
of the whole computational
domain without source of stiff terms:
\begin{equation}
\Delta t = \min_{j \in \cal{D}} \hat{\nu} \frac{h_j}{\| \vec{v_j}\| + c_j},
\end{equation}
where $\hat{\nu} < 1$, $h_j$ is a reference length scale, $\vec{v_j}$ the velocity vector,
 and $c_j$ the speed of sound in
cell $j$.
Then, the parameter $\omega_j$ is defined as:
\begin{equation}
\begin{aligned}
\omega_j = \min\bigg( 1,\frac{1}{\nu_j} \bigg),\hbox{ with }\nu_j=\frac{(\| \vec{v_j}\| + c_j)\,\Delta t}{h_j}.
\end{aligned}
\end{equation}

In Eq.~\eqref{eq:Timofeev_MUSCL}, the term ${(\widetilde{\Delta_t \LM{V}})}^{n+1}_{j}$ comes
from a generalization of the TVD property to the coupled space/time reconstruction.
It can be seen as a time slope limiter.
This term is given by a Newton's algorithm which at any step $s$ reads:
\begin{equation}
(\widetilde{\Delta_t \LM{V}})^{n+1,s}_{j} = \left\{
\begin{array}{ll}
\displaystyle \max\left[0, \min\left(\min_j [\beta (\LM{V}_j^{n+1,s} - \LM{V}_j^{n}) +
(\widetilde{\Delta_t \LM{V}})^{n+1,s-1}_{j}], \LM{V}_j^{n+1,s} - \LM{V}_j^{n}\right)\right] & \\
& \hspace*{-3cm}\hbox{ if } \LM{V}_j^{n+1,s} - \LM{V}_j^{n} \ge 0,\\
& \vspace*{-0mm} \\
\displaystyle \min\left[0, \max\left(\max_j [\beta (\LM{V}_j^{n+1,s} - \LM{V}_j^{n}) +
(\widetilde{\Delta_t \LM{V}})^{n+1,s-1}_{j}], \LM{V}_j^{n+1,s} - \LM{V}_j^{n}\right)\right] & \\
& \hspace*{-3cm}\hbox{ if } \LM{V}_j^{n+1,s} - \LM{V}_j^{n} < 0.\\
\end{array}
\right.
\end{equation}
The parameter $\beta$ is computed from the local CFL number $\nu_j$ and $\omega_j$ at each step $s$:
\begin{equation}
\beta = \frac{2 \big(1-\omega_j^s\nu_j^s(2-\omega_j^s \nu_j^s)\big)}{\nu_j^2(1-\omega_i^s)^2}
\label{eq:def_beta}
\end{equation}

This method is second-order accurate in both space and time in fully explicit,
fully implicit and explicit-implicit domains.
Moreover, the scheme was proved to be TVD for 1D linear/non linear scalar
conservation laws. It was applied successfully to problems including shocks \cite{Timofeev_2016_inProcessing}. This scheme will
be referred as TN in the following.

It is clear that the underlying explicit (where $\omega_j = 1$) and implicit (where $\omega_j = 0$)
schemes are not Heun's and Crank-Nicolson/IRK2 time integrators.
In section \ref{sec:HCS1}, a first straightforward attempt to couple these latter schemes is explained.

\section{First Hybrid Coupled Scheme \label{sec:HCS1}}


This section presents the coupling of three schemes: Heun for the explicit part, IRK2 for the implicit
part, and the hybrid formulation TN~\cite{Timofeev_2016_inProcessing} for a smooth transition between the latter two schemes.
The resulting numerical time integration scheme is defined to ensure a unique flux on each interface, leading to a conservative formulation.

\begin{figure}[!htbp]
\begin{center}
\includegraphics[width=7cm]{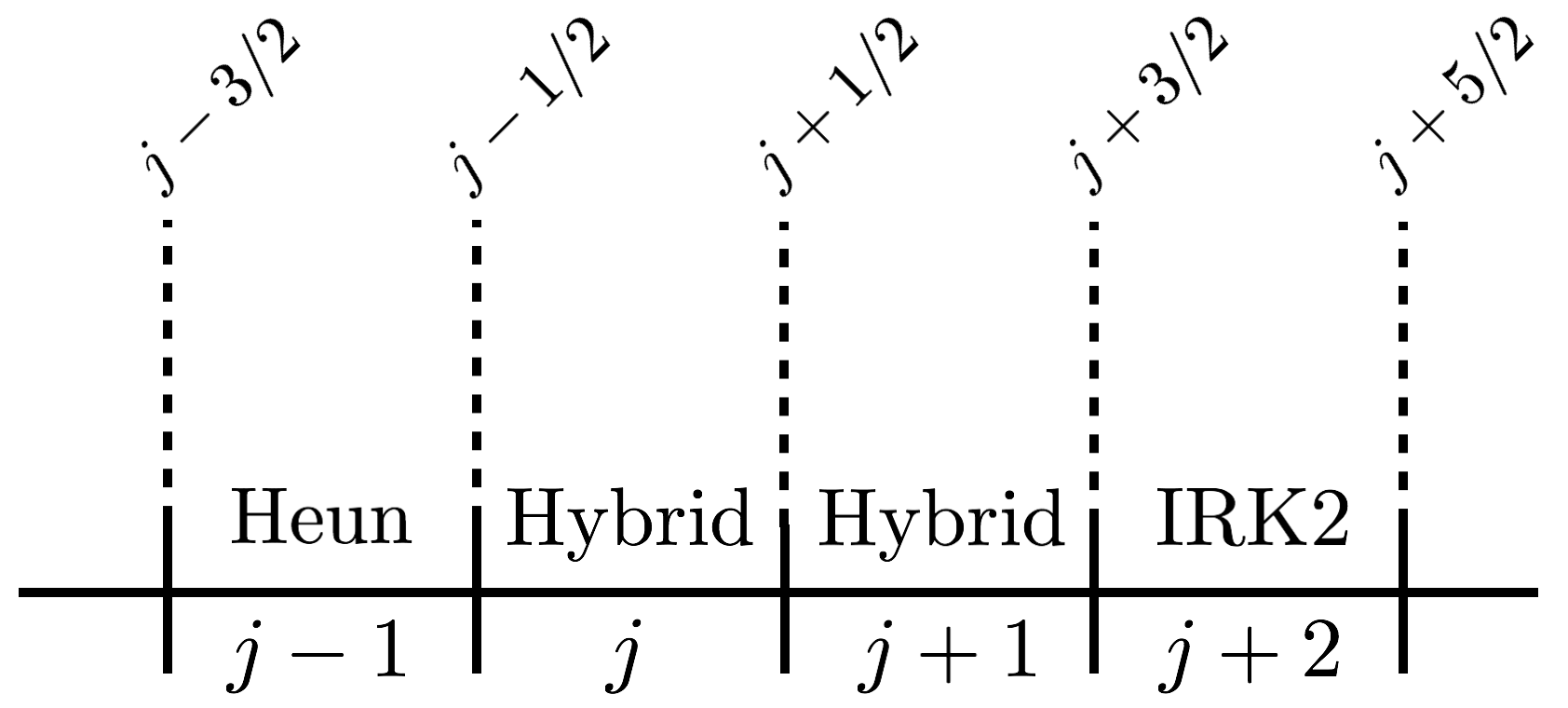}
\caption{1D example to explain cell status and the flux conservation property of the coupling scheme HCS1}
\label{fig:flux_conserv}
\end{center}
\end{figure}

An interface separates two cells that may have a different status given by the parameter $\omega_j$. The cell status is named
Heun for $\omega_j=1$, IRK2 for $\omega_j=0$, and Hybrid otherwise, as shown by the 1D example on Fig.~\ref{fig:flux_conserv}.

The flux formula used on an interface separating
two cells depends on the status of these cells as shown in Tab.~\ref{tab:flux-faces}.


\begin{table}[!htbp]
\caption{Flux formula depending on neighbour cells. The table is "symmetrical": 
the flux between two cells is independent of the direction of information propagation.\label{tab:flux-faces}}
\begin{center}
\begin{tabular}[b]{|l|c|c|c|c|}
\hline
\backslashbox{Left cell status}{Right cell status} & Heun & Hybrid & IRK2 \\
\hline
Heun & $F^{Heun}$ & $F^{Heun}$ & $\times$ \\
\hline
Hybrid & $F^{Heun}$& $F^{Hybrid}$ & $F^{IRK2}$ \\
\hline
IRK2 & $\times$ & $F^{IRK2}$ & $F^{IRK2}$ \\
\hline
\end{tabular}
\end{center}
\end{table}
According to Tab.~\ref{tab:flux-faces}, for the 1D example on Fig.~\ref{fig:flux_conserv},
the first version of the coupling scheme, referred as HCS1, is designed as,
\begin{equation}
\left\{
\begin{array}{ll}
\hbox{Predictor step:} & \displaystyle \widehat{W_j} = W_j^{n} + \omega_j {\Delta t} R({W}_j^{n})\\
\vspace*{-3mm} &  \\
\hbox{Corrector step:} & \displaystyle \left \{
    \begin{array}{ll}
    \displaystyle W^{n+1}_{j-1}=W^n_{j-1}+\frac{\Delta t}{2|\Omega_{j-1}|}( F_{j-\frac{1}{2}}^{n}+ \widehat{F}_{j-\frac{1}{2}} - F_{j-\frac{3}{2}}^{n}- \widehat{F}_{j-\frac{3}{2}} )  \\
    \vspace*{-3mm} &  \\
        \displaystyle W^{n+1}_{j}=W^n_{j}+\frac{\Delta t}{|\Omega_{j}|}( F_{j+\frac{1}{2}}^{Hybrid} - \frac{F_{j-\frac{1}{2}}^{n}+\widehat{F}_{j-\frac{1}{2}}}{2} )  \\
    \vspace*{-3mm} &  \\
    \displaystyle W^{n+1}_{j+1}=W^n_{j+1}+\frac{\Delta t}{|\Omega_{j+1}|}(\frac{F_{j+\frac{3}{2}}^{n}+ F_{j+\frac{3}{2}}^{n+1}}{2}-F_{j+\frac{1}{2}}^{Hybrid}     )  \\
    \vspace*{-3mm} &  \\
    \displaystyle W^{n+1}_{j+2}=W^n_{j+2}+\frac{\Delta t}{2|\Omega_{j+2}|}(F_{j+\frac{5}{2}}^{n}+ F_{j+\frac{5}{2}}^{n+1}- F_{j+\frac{3}{2}}^{n}- F_{j+\frac{3}{2}}^{n+1})  .
    \end{array}
\right.
\end{array}
\right.\label{eq:HCS1}
\end{equation}
The hybrid flux $F^{Hybrid}$ is computed with the state reconstruction method introduced in Eq.~\eqref{eq:Timofeev_MUSCL}.
The scheme Eq.~\eqref{eq:HCS1} ensures the unique definition of the interface flux, and then
flux conservation.

The differences with the TN scheme are:
\begin{itemize}
  \item In the predictor step, the residual is computed from the flux balance over all cell faces. The flux is computed
  using a linear extrapolation of the unknowns between cell centers and cell interfaces and accounting for a slope limiter.
  Since two different extrapolations are obtained on a face shared by two cells, the flux is computed by
  means of a (approximated) Riemann solver. As a consequence, $\widehat{W}_j$ has a physical meaning, something which was lost
  with TN scheme.
  \item The scheme recovers the Heun's scheme in the explicit regime ($\omega_j=1$).
  \item The scheme recovers the IRK2 scheme in the implicit regime ($\omega_j=0$).
\end{itemize}

{\bf Remark 1:} It is very important to make the transition between explicit and implicit time integration as fast
as possible. This is the reason why in practice, the IRK2 formulation is applied when $\omega_j\le 0.5$.

{\bf Remark 2:} For the Navier-Stokes equations, the standard
diffusion scheme available in the code was used for the implicit and explicit regions. The interface gradient is defined
from the limited cell-centered gradients by means of an average taking into account local metrics.
In the hybrid regions, the definition of the cell-centered gradient is modified as
\begin{equation}
{(\nabla \LM{V})}^{hyb}_j= \omega_j \bigg( \frac{(\widetilde{\nabla \LM{V}})^{n}_j+ (\widetilde{\widehat{\nabla \LM{V}}})_j }{2}  \bigg) +
(1-\omega_j) (\widetilde{\nabla \LM{V}})^{n+1}_j \quad \text{for }0.5< \omega_j < 1.
\label{eq:gradient}
\end{equation}
Every gradient in Eq.~\eqref{eq:gradient} is computed from the standard gradient and corrected using a slope limiter.

\subsection{Space-Time von Neumann Analysis}
\label{sec:VonNeumannHCS1}

The von Neumann analysis is a powerful tool to analyse the spectral properties
of any scheme. This analysis is generally performed separately either for spatial or temporal schemes
but it was shown in many recent papers (see~\cite{Sengupta_2012_AMC, Sengupta_2011_JCP, Sengupta_2003_JCP,Vanharen_JCP_2017}
for instance) that the true scheme behaviour is only obtained by a coupled analysis involving both space and time
discretizations.

Starting from the one-dimensional linear advection equation coupled with a harmonic initial condition and with periodic boundary conditions,
\begin{equation}
\left\{
\begin{array}{ll}
\displaystyle \frac{\partial y}{\partial t} +c \frac{\partial y}{\partial x} =0 &\\
\vspace*{-3mm}\\
y(x, 0) = \exp(ikx) & x \in [0, L] \\
\vspace*{-3mm}\\
y(0, t) = y(L, t) & t \in R^+ \\
\end{array}
\right.
\label{conv}
\end{equation}
with $i^2=-1$ and $c>0$, the comparison between the exact theoretical solution and the numerical approximation gives
information on both dissipation and dispersion of the fully discrete scheme. The fully discrete relation obtained from
Eq.~\eqref{conv} using a second-order finite volume scheme reads:
\begin{equation}
\begin{aligned}
y_j^{n+1}{}=G_j \, y_j^{n},\label{eq:defGj}
\end{aligned}
\end{equation}
where $y_j^n$ represents the harmonic solution at discrete time $n$ and in the center of cell $j$.
The complex coefficient $G_j$ represents the transfer function between two consecutive time solutions.
The coefficient $G_j$ can be expressed as $G_j = |G_j| \, \exp(i \arg(G_j))$, which highlights the dissipation
of the scheme related to $|G_j|$, and its dispersion related to $\exp(i \arg(G_j))$.
In the following, let $\mu_j = |G_j|$ be the dissipation coefficient, and $\phi_j = arg(G_j)/\hbox{CFL}$ the dispersion coefficient.

For standard finite element, difference or volume formulations, and focusing on a single scheme for space and
time, the transfer coefficient does not depend on the space position. However, in our case, several time discretization
schemes are blended, and the spectral properties of the global scheme will depend on the cell status $\omega_j$.
Accounting for the cell status variations can be done through the change in the mesh spacing. But
taking into account for a local change in space discretization
in the spectral analysis is cumbersome (see for instance~\cite{Vichnevetsky_MCS_23_1981}).
Once the mesh size $\Delta x$ is fixed, a manufactured choice of $\omega_j$ keeps the analysis quite simple to perform.
The number of 1D segments is fixed to $N=300$ and the function $\omega_j$ is defined as
\begin{equation}
\left\{
\begin{array}{ll}
  \displaystyle \omega_j=\alpha \, \omega_{j-1} &  \hbox{ for } \displaystyle\frac{N}{2}-50 \leq j \leq \frac{N}{2}  \\
  \vspace*{-3mm} &\\
  \displaystyle \omega_j=\frac{1}{\alpha} \, \omega_{j-1} & \hbox{ for } \displaystyle\frac{N}{2}+1 \leq j \leq \frac{N}{2}+50  \\
  \vspace*{-3mm} &\\
  \displaystyle \omega_j = 1 & \hbox{elsewhere.}
\end{array}
\right.
\end{equation}
with $\alpha=0.90$. The maximum CFL value for the Heun explicit scheme without amplification at any wavenumber is 0.6. Then, the spectral analysis will be performed for
CFL=0.1 and CFL=0.6.

\begin{figure}[!htbp]
\begin{center}
\includegraphics [width=8cm]{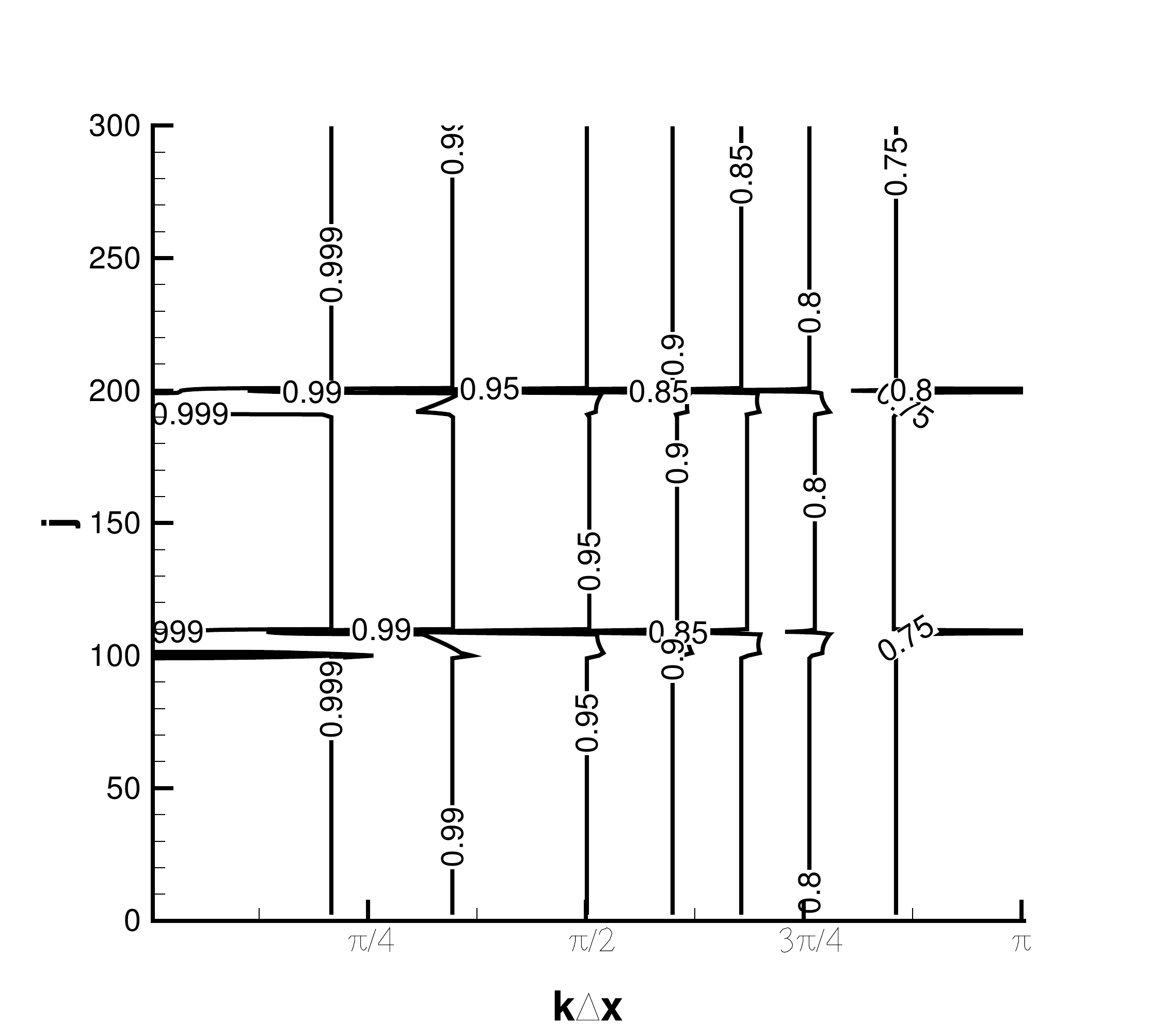}
\caption{Isocontours of the dissipation coefficient $\mu_j$ for the HCS1 scheme at CFL=0.1
\label{fig:HCS1_abs01}}
\end{center}
\end{figure}
\begin{figure}[!htbp]
\begin{center}
\includegraphics [width=8cm]{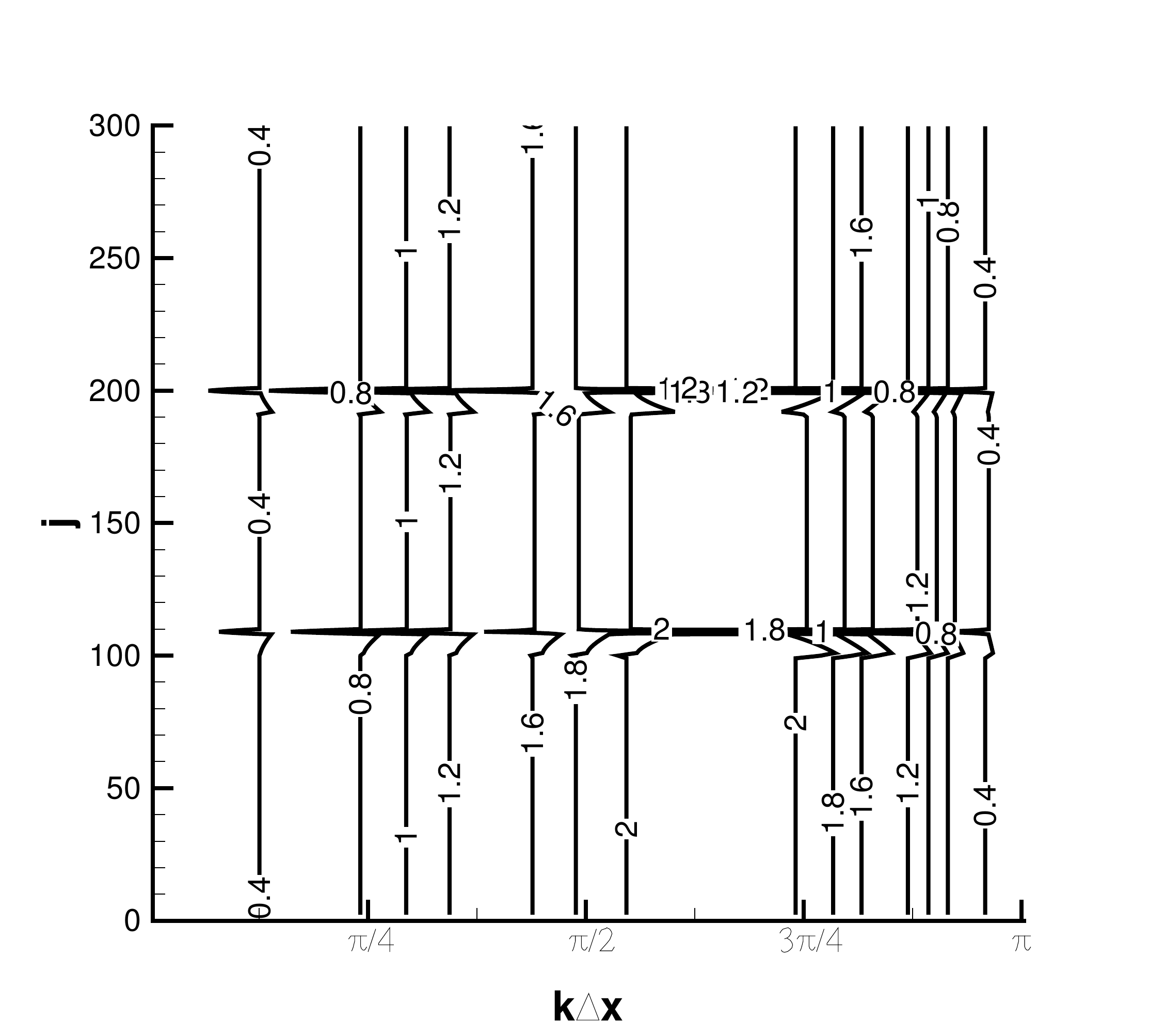}
\caption{Isocontours of the dispersion coefficient $\phi_j$ for the HCS1 scheme at CFL=0.1
\label{fig:HCS1_phase01}}
\end{center}
\end{figure}
\begin{figure}[!ht]
\begin{center}
\includegraphics [width=8cm]{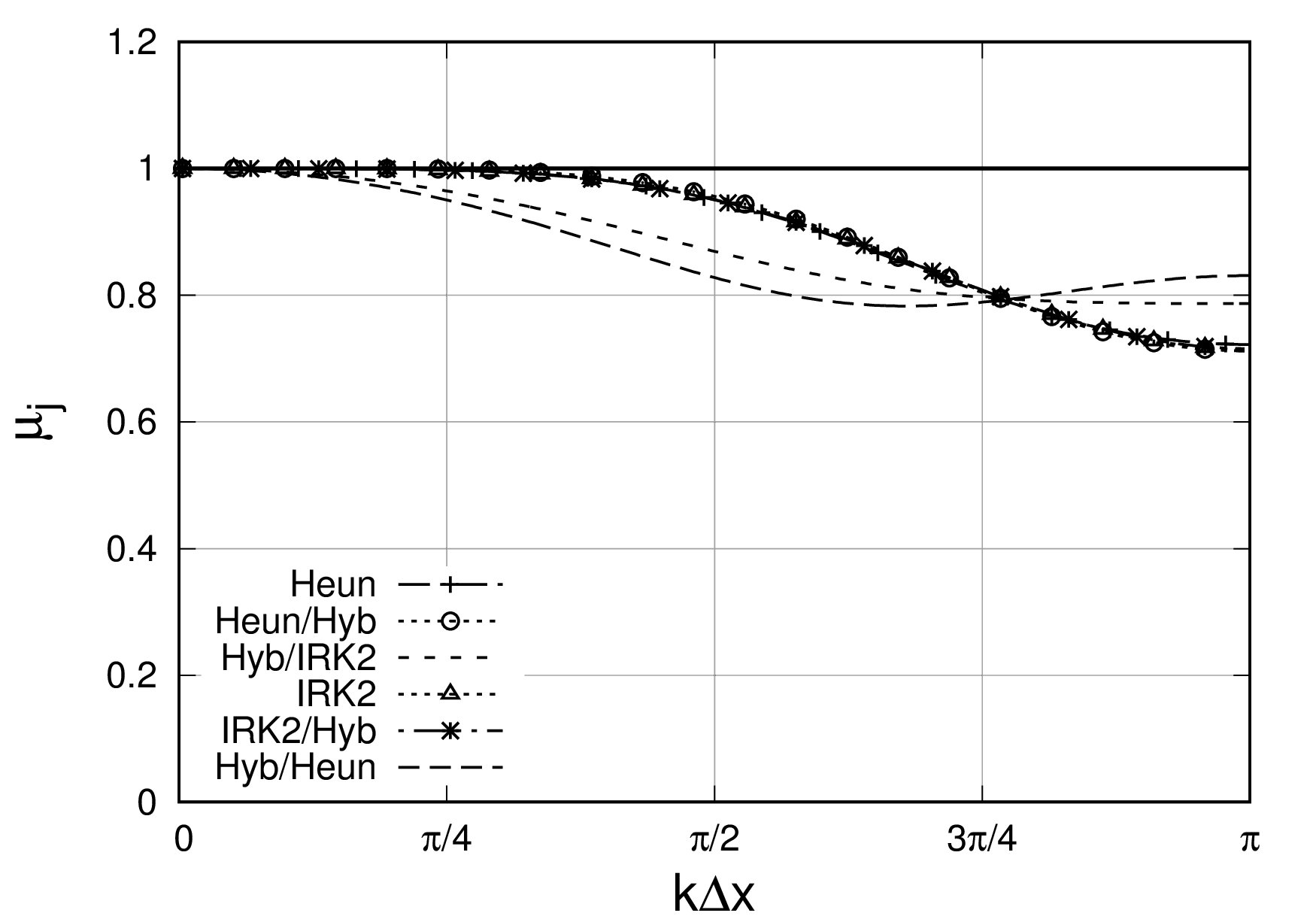}
\caption{Dissipation $\mu_j$ of the scheme HCS1 for all kinds of cells at CFL=0.1
\label{fig:dissipMenshovFig}}
\end{center}
\end{figure}
\begin{figure}[!ht]
\begin{center}
\includegraphics [width=8cm]{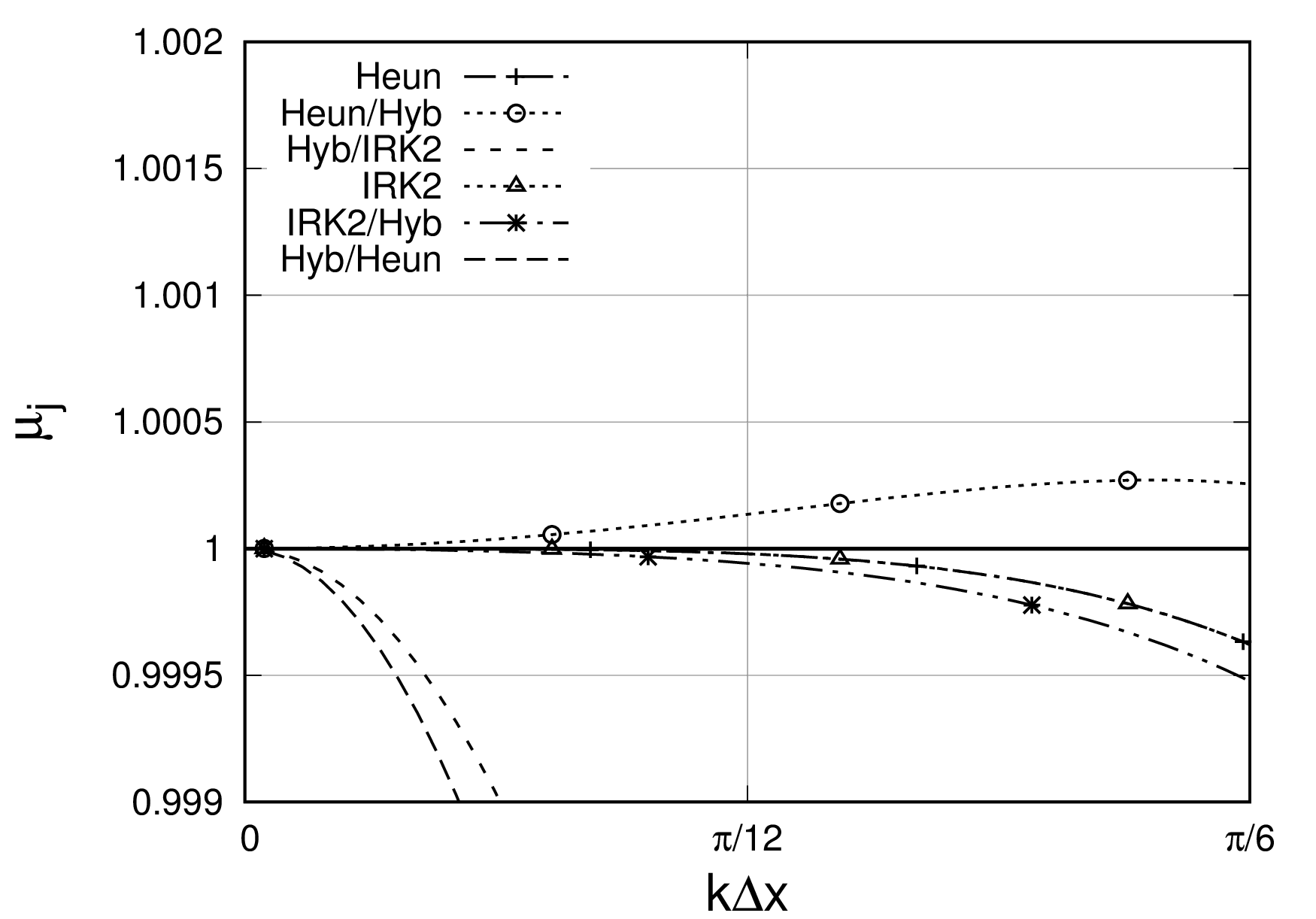}
\caption{Dissipation $\mu_j$ of the scheme HCS1 for all kinds of cells at CFL=0.1, zoom from Fig.~\ref{fig:dissipMenshovFig}
\label{fig:dissipMenshovZoomFig}}
\end{center}
\end{figure}
\begin{figure}[!ht]
\begin{center}
\includegraphics [width=8cm]{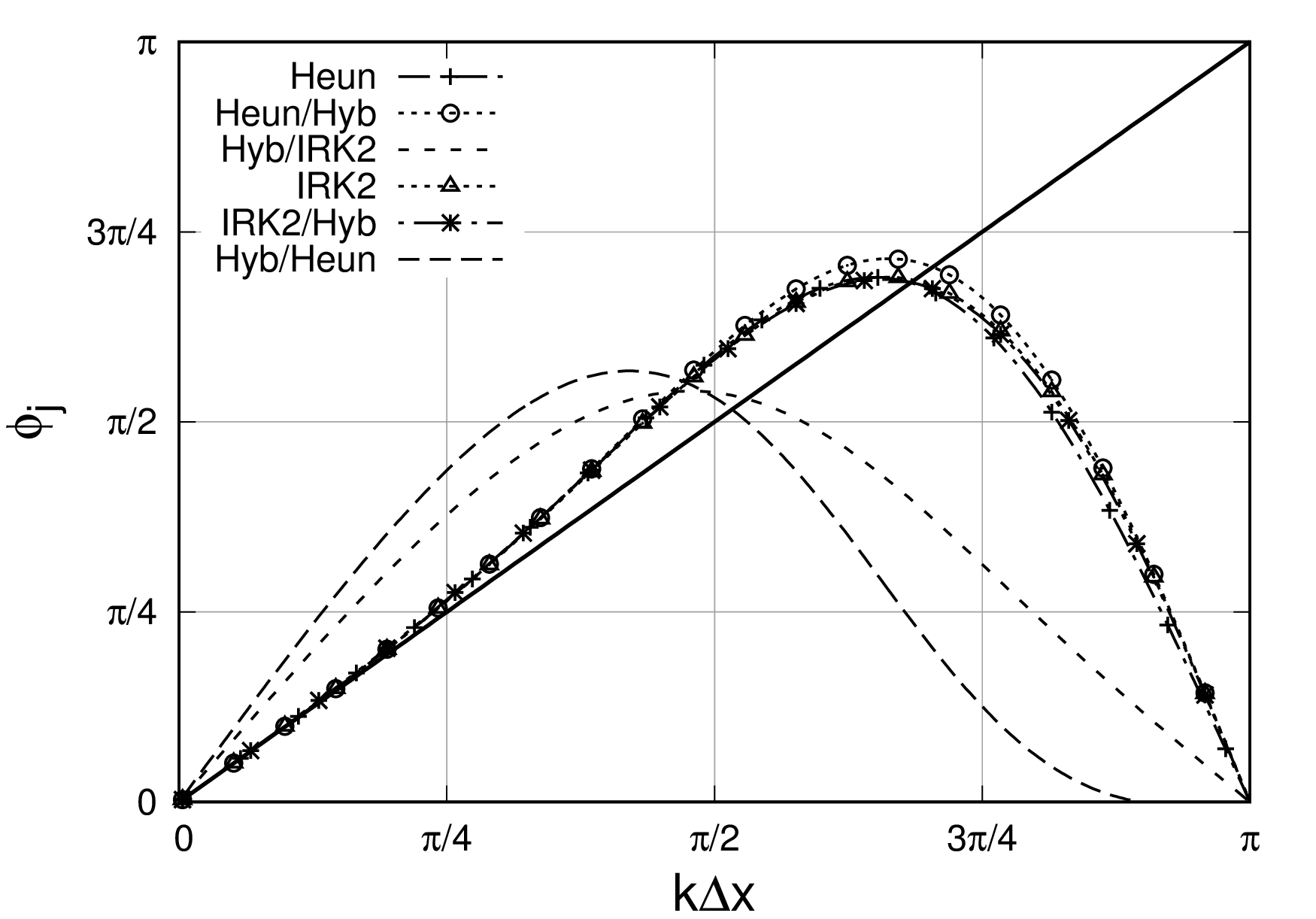}
\caption{Dispersion $\phi_j$ of the scheme HCS1 for all kinds of cells at CFL=0.1
\label{fig:phaseMenshovZoomFig}}
\end{center}
\end{figure}

Figures~\ref{fig:HCS1_abs01} and~\ref{fig:HCS1_phase01} give an overview of the dissipation and the dispersion of the
coupled time/space HCS1 scheme depending on the cell index $1 \le j \le N$ and the non dimensionalized wave number $ 0 \le k \Delta x \le \pi$ for CFL=0.1.
They clearly show that some phenomena occur at the transitions between the different temporal schemes.
As a consequence, the standard curves ($\mu_j$ and $\phi_j$ as
a function of $k \Delta x$) are introduced in Figs.~\ref{fig:dissipMenshovFig}-\ref{fig:phaseMenshovZoomFig}
for the specific cells at which there is a time integration transition. As indicated in Tab.~\ref{tab:cell-flux}, six
specific discretization cells are analysed. In addition, Tab.~\ref{tab:cell-flux} also gives the meaning of the legends
for Figs.~\ref{fig:dissipMenshovFig}-\ref{fig:phaseMenshovZoomFig}.

\begin{table}[!htbp]
\caption{Cell position, legend, and associated flux types. Note that left and right sides concern a 1D domain with a positive advection speed.\label{tab:cell-flux}}
\begin{center}
\begin{tabular}{c|c|c|c}
Legend & Flux type on left interface & Flux type on right interface & Cell position \\
\hline
Heun & explicit & explicit & $j=31$ \\
Heun/Hyb & explicit & hybrid & $j=100$\\
Hyb/IRK2 & hybrid & implicit & $j=104$\\
IRK2 & implicit & implicit & $j=150$\\
IRK2/Hyb & implicit & hybrid & $j=196$\\
Hyb/Heun & hybrid & explicit & $j=200$\\
\end{tabular}
\end{center}
\end{table}

Figure~\ref{fig:dissipMenshovFig} shows the dissipation coefficient. The focus for $k \Delta x \in [0,0.6]$ in Fig.~\ref{fig:dissipMenshovZoomFig}
shows that the scheme amplifies some wavenumbers for Heun/Hybrid cells.
The level of amplification is very low and one could assume that since the treatment is local (on one cell in 1D), the
amplification could not lead to global instability and failure of the computation. Fig.~\ref{fig:phaseMenshovZoomFig}
shows that cells with Hybrid/Heun and Hybrid/IRK2 fluxes produce the highest levels of dispersion, while dispersion is close to the reference
Heun's scheme for other choices.

\begin{figure}[!ht]
\begin{center}
\includegraphics [width=8cm]{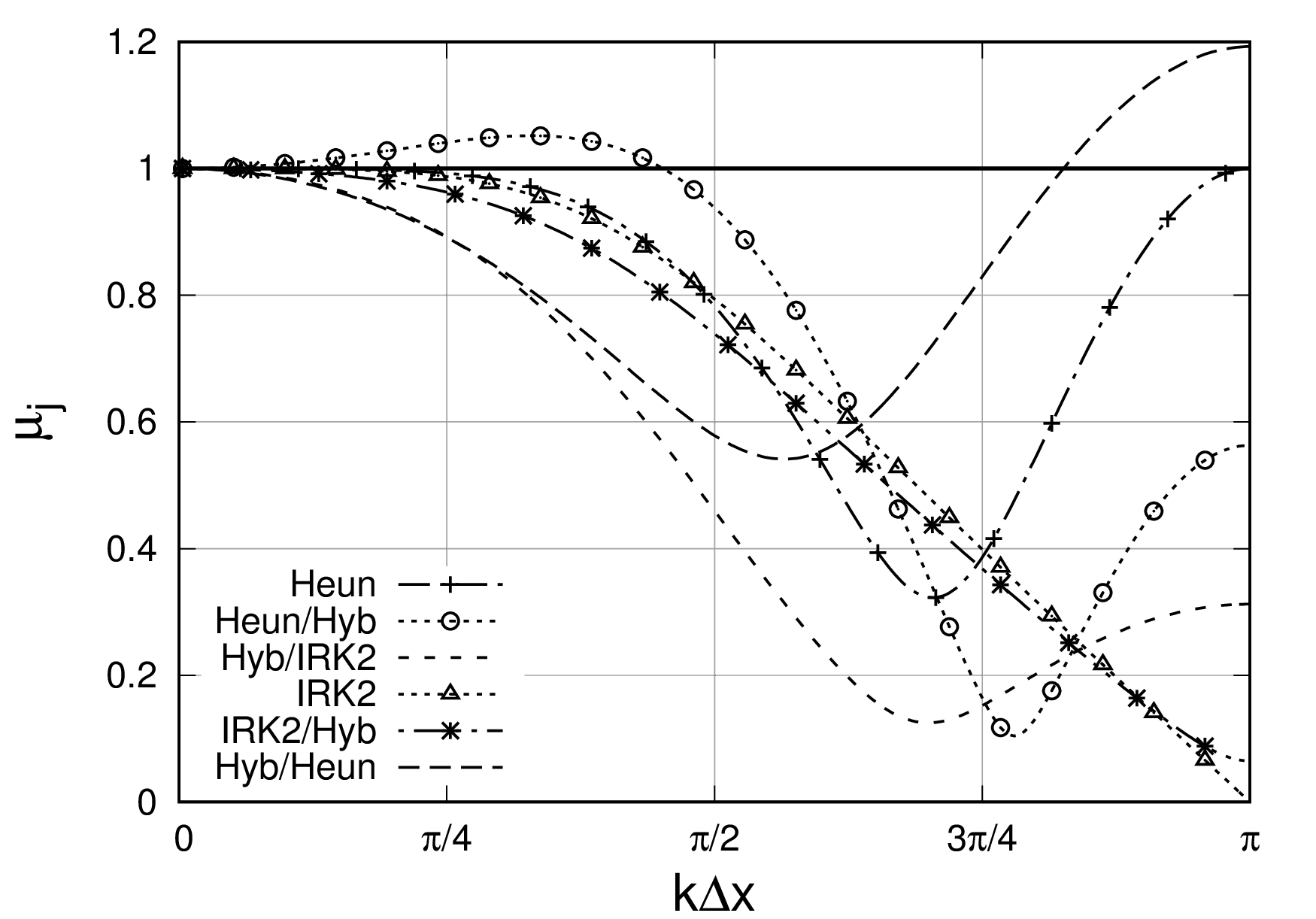}
\caption{Dissipation $\mu_j$ of the scheme HCS1 for all kinds of cells at CFL=0.6
\label{fig:dissipMenshov06Fig}}
\end{center}
\end{figure}

\begin{figure}[!ht]
\begin{center}
\includegraphics [width=8cm]{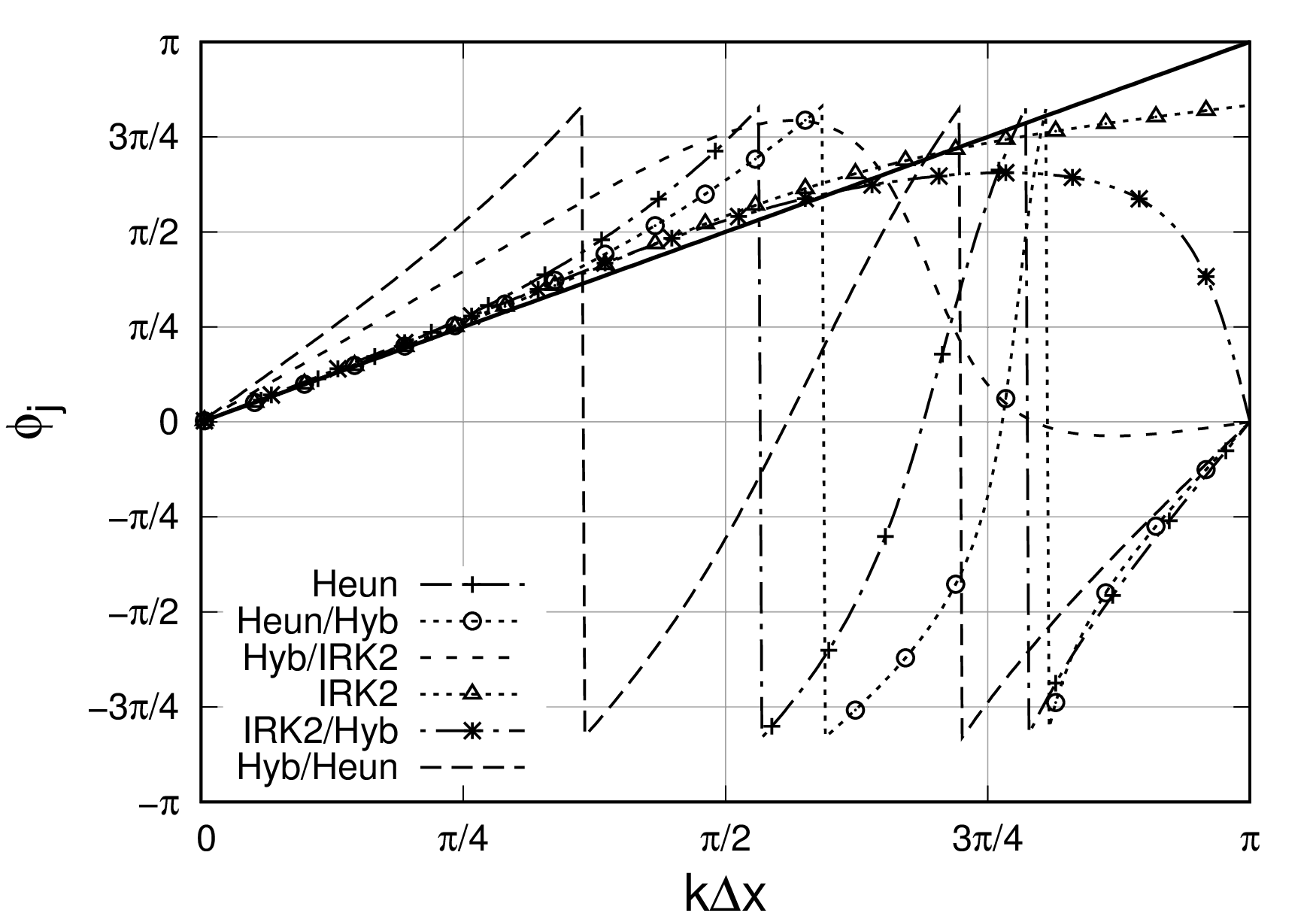}
\caption{Dispersion $\phi_j$ of the scheme HCS1 for all kinds of cells at CFL=0.6
\label{fig:phaseMenshov06Fig}}
\end{center}
\end{figure}

The same analysis is performed at CFL=0.6, {\it i.e.} at the limit without amplification at any wavenumber for the Heun's scheme. The dissipation and the dispersion curves are shown in Figs.~\ref{fig:dissipMenshov06Fig}-\ref{fig:phaseMenshov06Fig}. The dispersion curve for the Heun/Hybrid cell looks
like the one for Heun's scheme and presents a change of phase sign for $k \Delta x \simeq 1.8$.
The dissipation curve presents again amplification but the level of amplification is much higher. Since the
value of $\omega_j$ may be defined to follow some physical property, numerical amplification of a physical phenomena
could follow the phenomena itself, leading to an unacceptable flow or simulation break. Notice that the discrete jumps on
dispersion curves in Fig.~\ref{fig:phaseMenshov06Fig} at $k \Delta x \simeq 1.2, 1.6, 1.8, 2.2, 2.4$, and $2.6$.
These values of $k \Delta x$ are
maxima of dispersion error, and correspond to values of wave number $k \Delta x$ when $\mathcal{R}e(G_j)=0$.

The first coupled scheme HCS1 based on Heun and IRK2 schemes presents amplification. In order to remove it,
a new way to couple the desired explicit and implicit schemes is proposed in Sec.~\ref{sec:AION}.


\section{Another tIme integratiON (AION) scheme \label{sec:AION}}

In order to get better spectral characteristics than the HCS1 scheme,
a new hybrid scheme still based on both Heun and IRK2 schemes is designed.
This new hybrid time integrator is called AION (Another tIme integratiON).
It is applied to the Cauchy problem Eq.~\eqref{eq:compact_form}.

This hybrid scheme has a predictor-corrector formulation as TN's and Heun's
schemes. According to Tab.~\ref{tab:flux-faces}, for the 1D example on Fig.~\ref{fig:flux_conserv},
it is designed as:
\begin{equation}
\left\{
\begin{array}{ll}
\hbox{Predictor step:} & \displaystyle \widehat{W_j}=W_j^{n}+{\Delta t}R(W_j^{n})\\
\vspace*{-3mm} &  \\
\hbox{Corrector step:} & \displaystyle \left \{
    \begin{array}{ll}
    \displaystyle W^{n+1}_{j-1}=W^n_{j-1}+\frac{\Delta t}{2|\Omega_{j-1}|}( F_{j-\frac{1}{2}}^{n}+ \widehat{F}_{j-\frac{1}{2}} - F_{j-\frac{3}{2}}^{n}- \widehat{F}_{j-\frac{3}{2}} )  \\
    \vspace*{-3mm} &  \\
    \displaystyle W^{n+1}_j=W^n_j+\frac{\Delta t}{|\Omega_j|}( F_{j+\frac{1}{2}}^{Hybrid} - \frac{F_{j-\frac{1}{2}}^{n}+\widehat{F}_{j-\frac{1}{2}}}{2} )  \\
    \vspace*{-3mm} &  \\
    \displaystyle W^{n+1}_{j+1}=W^n_{j+1}+\frac{\Delta t}{|\Omega_{j+1}|}(\frac{F_{j+\frac{3}{2}}^{n}+ F_{j+\frac{3}{2}}^{n+1}}{2}-F_{j+\frac{1}{2}}^{Hybrid}     )  \\
    \vspace*{-3mm} &  \\
    \displaystyle W^{n+1}_{j+2}=W^n_{j+2}+\frac{\Delta t}{2|\Omega_{j+2}|}(F_{j+\frac{5}{2}}^{n}+ F_{j+\frac{5}{2}}^{n+1}- F_{j+\frac{3}{2}}^{n}- F_{j+\frac{3}{2}}^{n+1})  .
    \end{array}
\right.
\end{array}
\right.\label{eq:HCS2}
\end{equation}

{\bf Remark:} A situation not adressed in Tab.~\ref{tab:flux-faces} nor in Fig.~\ref{fig:flux_conserv} must be described. 
For any hybrid cell (according to the value $\omega_j$) with hybrid flux for all the faces, the predictor state $\widehat{W}_j$ is 
time-integrated by accounting for $\omega_j$: $\widehat{W}_j=W_j^{n}+\omega_j{\Delta t} R(W_j^{n})$.
For the mixed explicit / implicit regime, the hybrid flux differs due to the different reconstruction methods which is
 the key point to maintain stability without amplification.
The demonstration of the stability without amplification is the purpose of the next section.

The reconstructed states are defined as:
\begin{equation}
    \begin{array}{ll}
\LM{V}^L=& \omega_j  \displaystyle \bigg[ \frac{\LM{V}^n_j+\widehat{\LM{V}}_j}{2}\bigg]+\frac{1}{2}(\widetilde{\nabla \LM{V}})_j^{n}\cdot \overrightarrow{C_jC_f} + \bigg(\omega_j-\frac{1}{2}\bigg)(\widetilde{\widehat{\nabla \LM{V}}})_j \cdot \overrightarrow{C_jC_f}+\\
&(1-\omega_j)  \displaystyle \bigg[ \LM{V}^{n+1}_j+(\widetilde{\nabla \LM{V}})_j^{n+1}\cdot \overrightarrow{C_jC_f} - \frac{1-\omega_j}{2}(\widetilde{\Delta_t \LM{V}})^{n+1}_j \bigg], \\
\LM{V}^R=& \omega_i  \displaystyle \bigg[ \frac{\LM{V}^n_i+\widehat{\LM{V}}_i}{2}\bigg]+\frac{1}{2}(\widetilde{\nabla \LM{V}})_i^{n}\cdot \overrightarrow{C_iC_f}
+ \bigg(\omega_i - \frac{1}{2}\bigg)(\widetilde{\widehat{\nabla \LM{V}}})_i \cdot \overrightarrow{C_iC_f}+\\
&(1-\omega_i)  \displaystyle \bigg[ \LM{V}^{n+1}_i+(\widetilde{\nabla \LM{V}})_i^{n+1}\cdot \overrightarrow{C_iC_f} - \frac{1-\omega_i}{2}(\widetilde{\Delta_t \LM{V}})^{n+1}_i \bigg].\\
\end{array}
\label{eq:AION_NHRM}
\end{equation}
For $\omega_j=1$ and $\omega_i=1$, the following reconstruction is obtained:
\begin{equation}
\begin{aligned}
\LM{V}^L=& \bigg[ \frac{\LM{V}^n_j+\widehat{\LM{V}}_j}{2}\bigg]+\frac{1}{2}(\widetilde{\nabla \LM{V}})_j^{n}\cdot \overrightarrow{C_jC_f} + \frac{1}{2}(\widetilde{\widehat{\nabla \LM{V}}})_j \cdot \overrightarrow{C_jC_f},\\
\LM{V}^R=& \bigg[ \frac{\LM{V}^n_i+\widehat{\LM{V}}_i}{2}\bigg]+\frac{1}{2}(\widetilde{\nabla \LM{V}})_i^{n}\cdot \overrightarrow{C_iC_f} + \frac{1}{2}(\widetilde{\widehat{\nabla \LM{V}}})_i \cdot \overrightarrow{C_iC_f},\\
\end{aligned}
\end{equation}
and the hybrid part leads to the Heun's scheme for the one-dimension linear advection equation Eq.~\eqref{conv}.

\subsection{Space-Time Stability Analysis}
\label{sec:VonNeumannAION}
\begin{figure}[!htbp]
\begin{center}
\includegraphics [width=8cm]{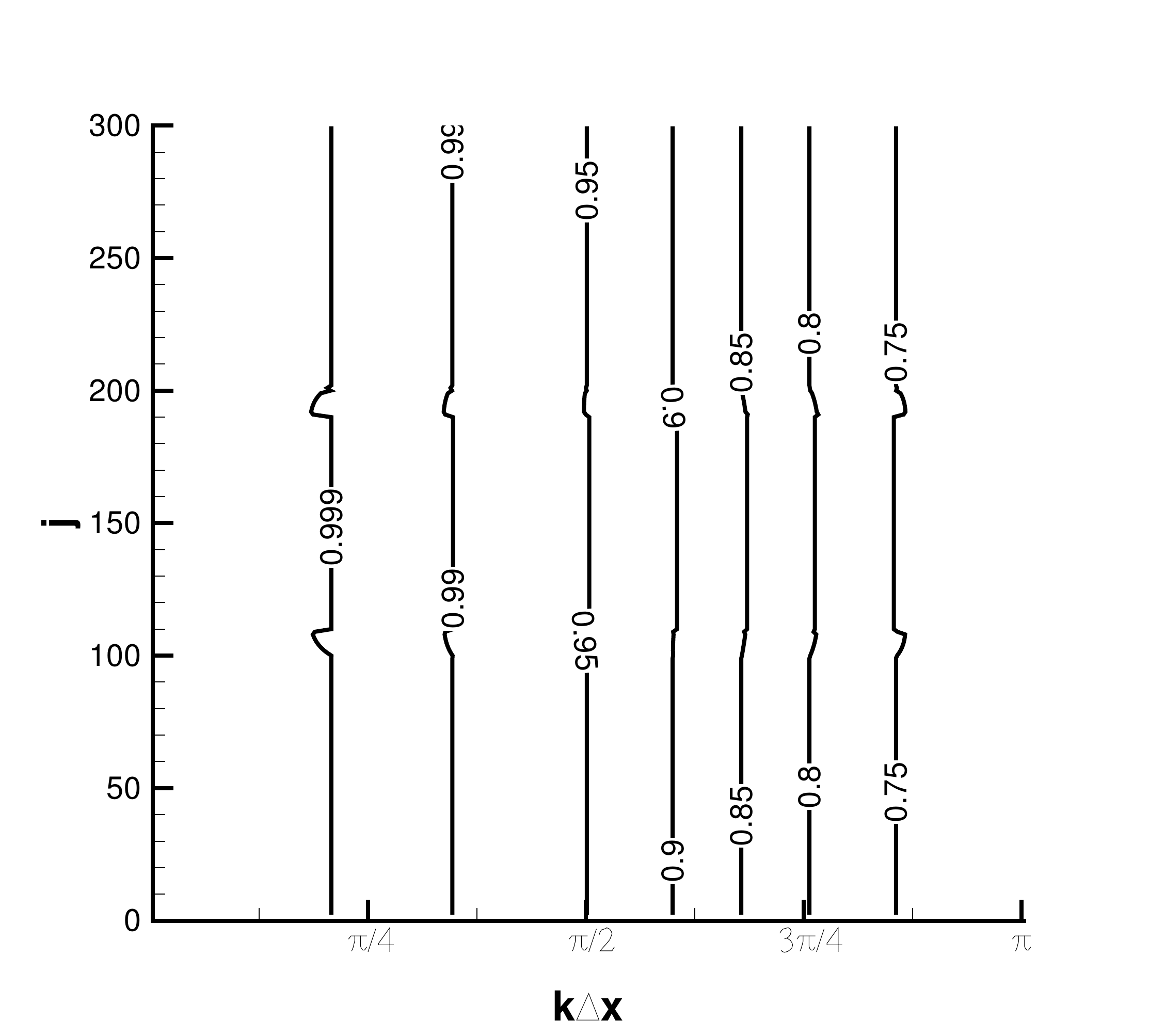}
\caption{Isocontours of the dissipation $\mu_j$ for the AION scheme and $1 \le j \le N$ at CFL=0.1
\label{fig:HCS2_abs01}}
\end{center}
\end{figure}

\begin{figure}[!htbp]
\begin{center}
\includegraphics [width=8cm]{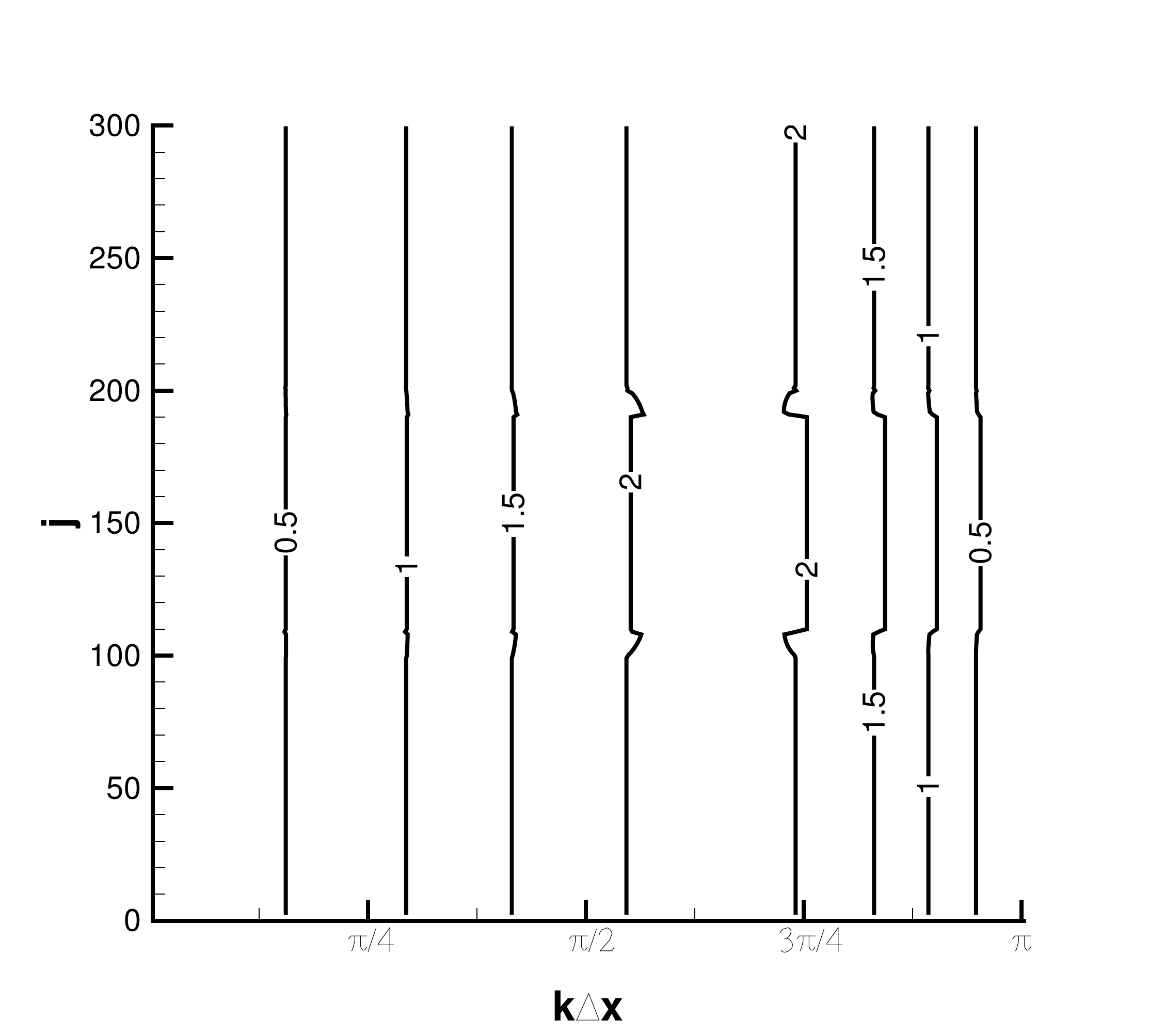}
\caption{Isocontours of the dispersion $\phi_j$ for the AION scheme and $1 \le j \le N$ at CFL=0.1
\label{fig:HCS2_phase01}}
\end{center}
\end{figure}

The analysis introduced in Sec.~\ref{sec:VonNeumannHCS1} is now applied to the AION scheme.
The spectral behaviour is shown in Figs.~\ref{fig:HCS2_abs01} and~\ref{fig:HCS2_phase01} for each cell
index $1 \le j \le N$ and wave number $k \Delta x \in [0,\pi]$.
Using the isocontours of iso-$\mu_j$ and iso-$\phi_j$, it is clear that the spectral behaviour of the scheme depends on the cell
index $j$ {\it via} the cell status $\omega_j$.

\begin{figure}[!htbp]
\begin{center}
\includegraphics [width=8cm]{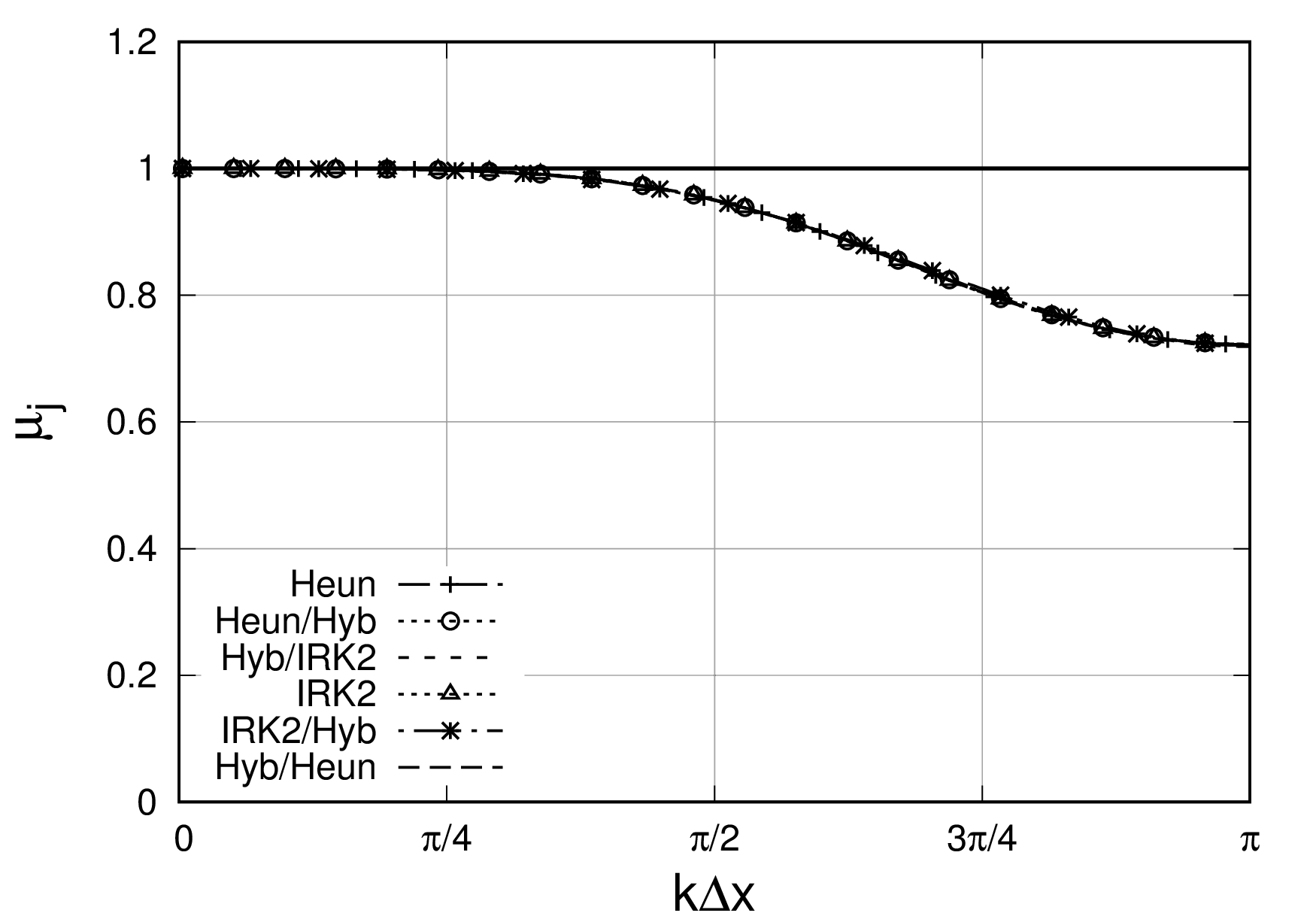}
\caption{Dissipation $\mu_j$ of the scheme AION depending on the kind of cell at CFL=0.1
\label{fig:dissipAIONFig}}
\end{center}
\end{figure}

\begin{figure}[!htbp]
\begin{center}
\includegraphics [width=8cm]{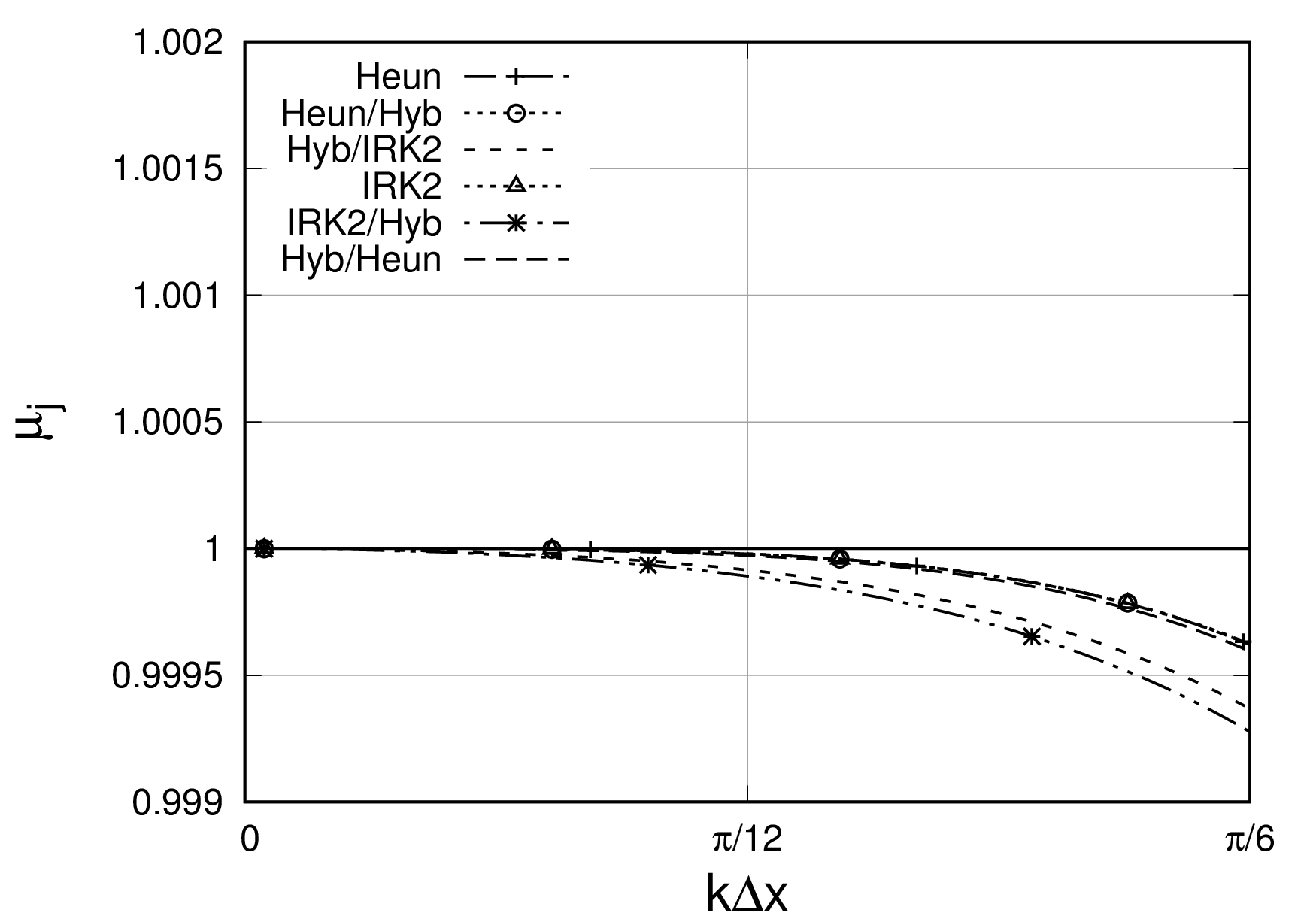}
\caption{Dissipation $\mu_j$ of the scheme AION depending on the kind of cell at CFL=0.1, zoom from Fig.~\ref{fig:dissipAIONFig}
\label{fig:dissipAIONZoomFig}}
\end{center}

\end{figure}

\begin{figure}[!htbp]
\begin{center}
\includegraphics [width=8cm]{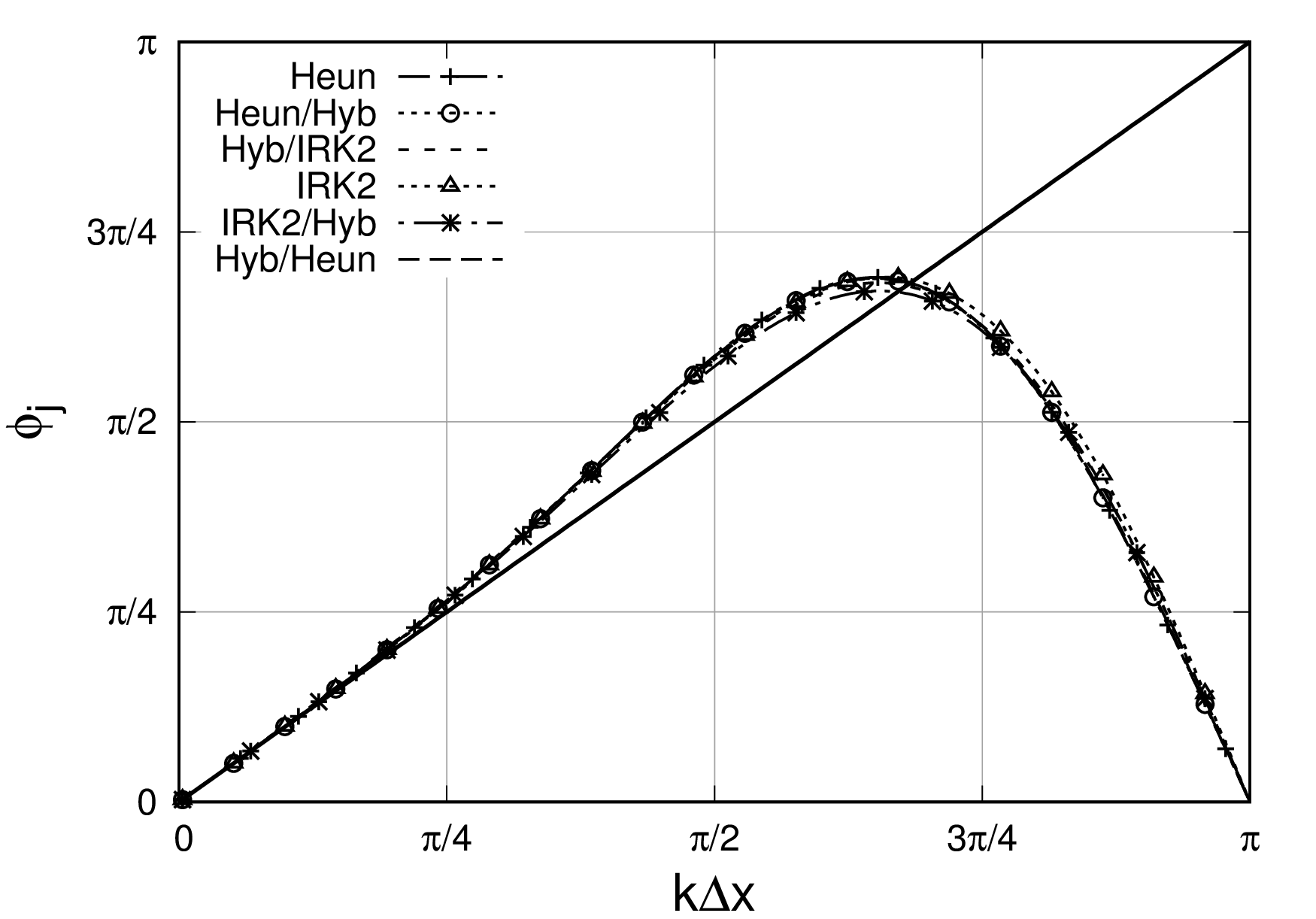}
\caption{Dispersion $\phi_j$ of the scheme AION depending on the kind of cell at CFL=0.1
\label{fig:phaseAIONZoomFig}}
\end{center}
\end{figure}

\begin{figure}[!htbp]
\begin{center}
\includegraphics [width=8cm]{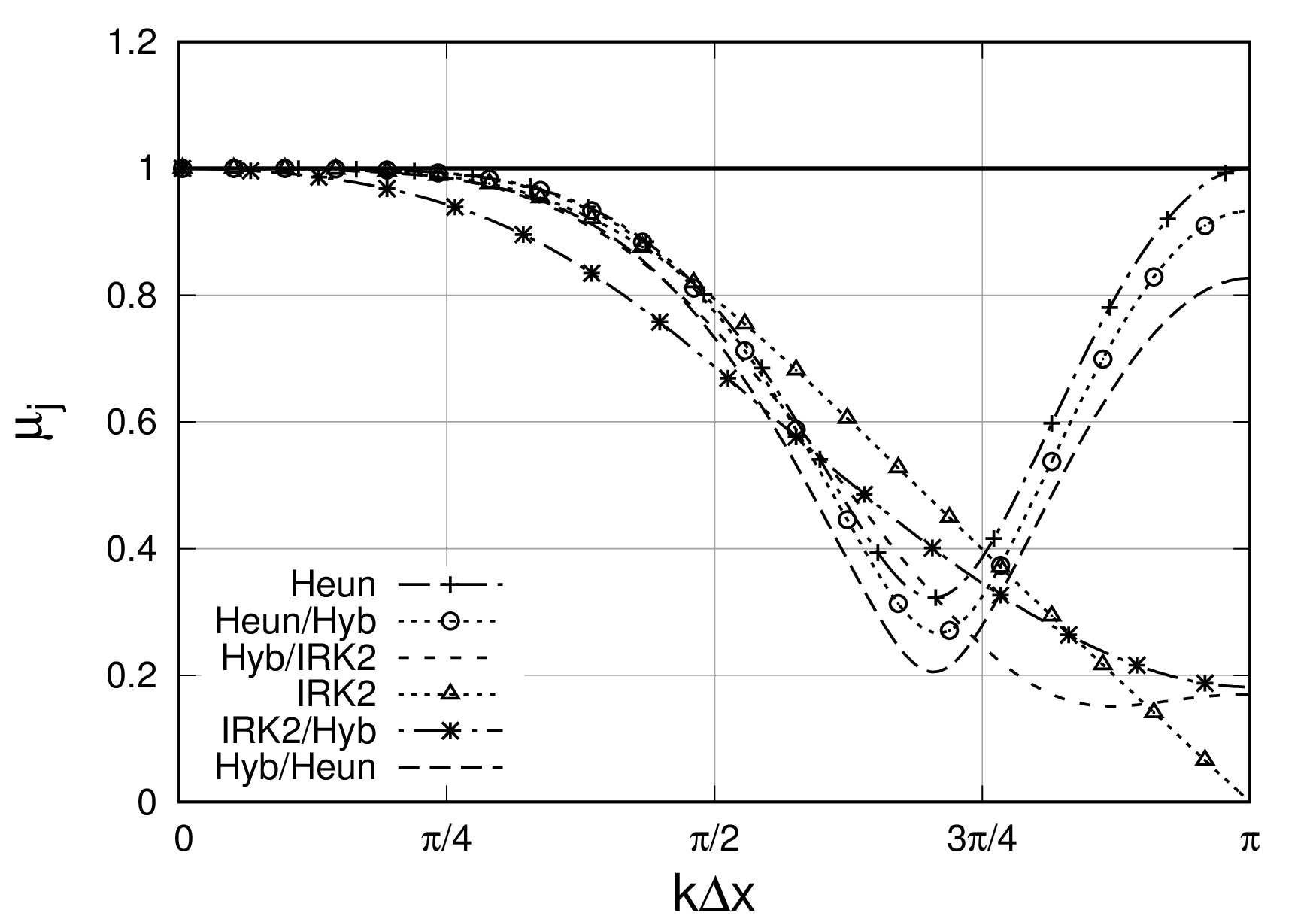}
\caption{Dissipation $\mu_j$ of the scheme AION depending on the kind of cell at CFL=0.6
\label{fig:dissipAIONFig06}}
\end{center}
\end{figure}

\begin{figure}[!htbp]
\begin{center}
\includegraphics [width=8cm]{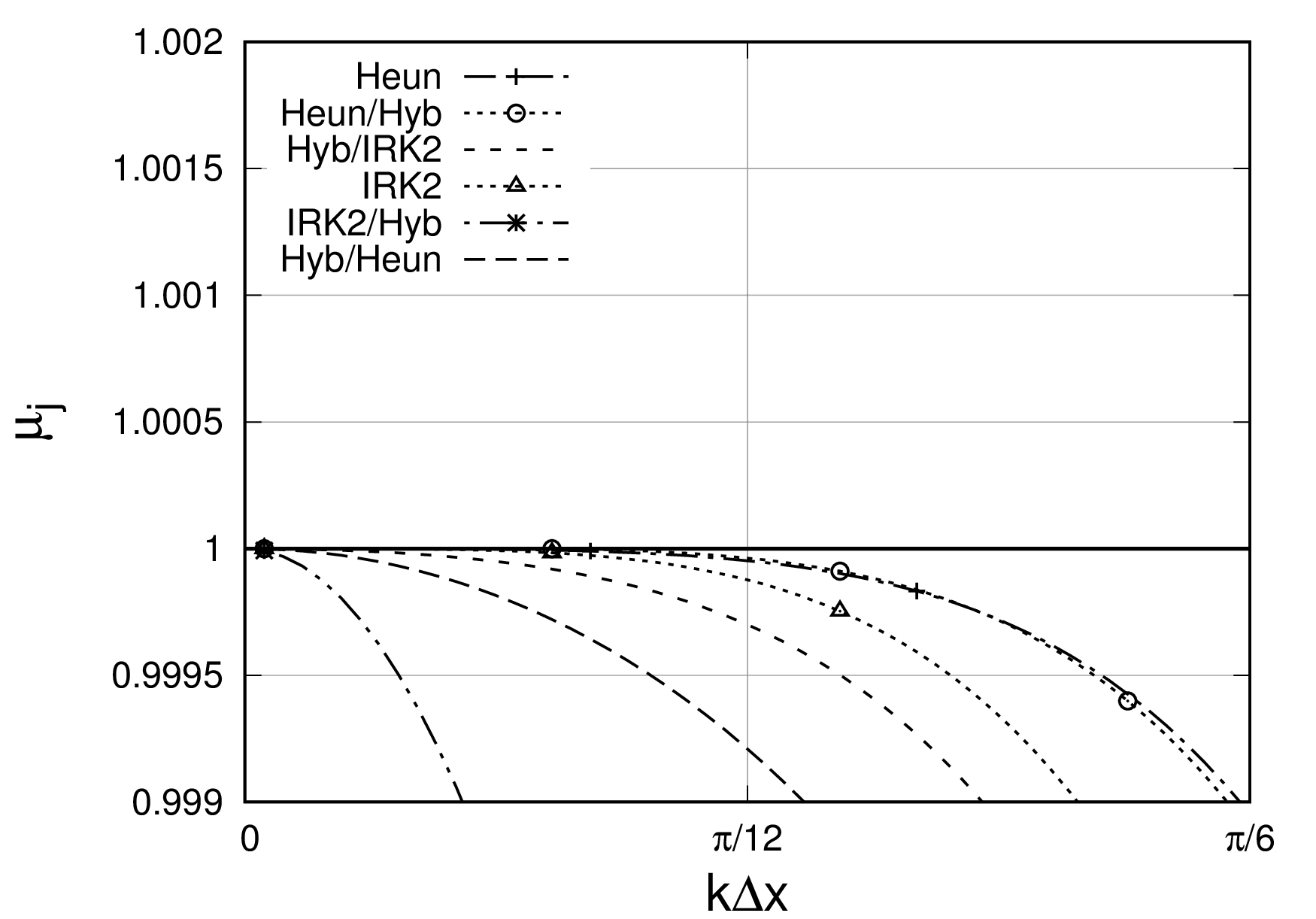}
\caption{Dissipation $\mu_j$ of the scheme AION depending on the kind of cell at CFL=0.6, zoom from Fig.~\ref{fig:dissipAIONFig06}
\label{fig:dissipAIONZoomFig06}}
\end{center}
\end{figure}

\begin{figure}[!htbp]
\begin{center}
\includegraphics [width=8cm]{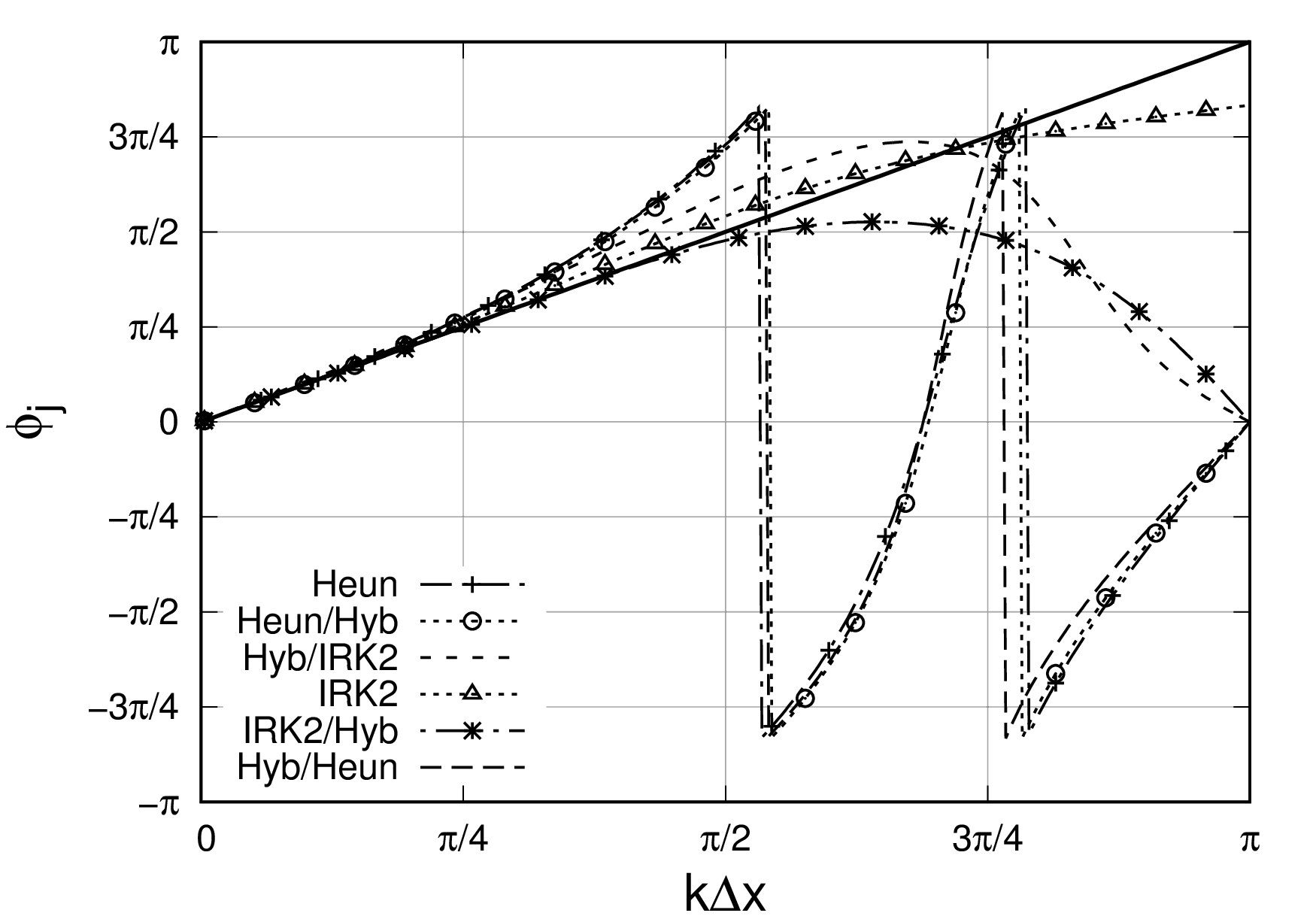}
\caption{Dispersion $\phi_j$ of the scheme AION depending on the kind of cell at CFL=0.6
\label{fig:phaseAIONFig06}}
\end{center}
\end{figure}

As before, the standard curves are now presented for the 6 values of $j$ introduced in Tab.~\ref{tab:cell-flux} at CFL=0.1
and CFL=0.6.
At CFL=0.1, the overall view of the dissipation coefficient in Fig.~\ref{fig:dissipAIONFig} shows that the AION scheme
seems to have the same behaviour for any cell index.
In addition, a zoom for $k \Delta x \in [0,0.6]$ does not show any amplification of waves
and stability is kept for any kind of cell (Fig.~\ref{fig:dissipAIONZoomFig}). Finally, all schemes present the same kind of dispersion in
Fig.~\ref{fig:phaseAIONZoomFig}. This shows that the limitation of the initial HCS1 scheme is now removed by our new coupling technique at
CFL=0.1. In addition, as shown in Figs.~\ref{fig:dissipAIONFig06}-\ref{fig:phaseAIONFig06}, the same conclusion is 
obtained at CFL=0.6, which is the limit without amplification at any wavenumber for the pure explicit scheme.


For this example of linear advection, the proposed scheme is shown to be stable. The value of $\omega_j$ is set to vary between 0 and 1. For $\omega_j=1$, it was shown that the reconstructed states
at the interface enable to recover the standard explicit Heun's scheme for the linear equation.
So, the idea is to keep a smooth transition from explicit to hybrid regions. However, it would be convenient to switch to implicit regions
as fast as possible. In the next section, attention is paid on the transition between implicit
and hybrid cells.

\subsection{Analysis of the Hyb/IRK2 and IRK2/Hyb Transitions}

This analysis is motivated by the desire to switch as fast as possible to the implicit formulation, while maintaining stability.
The computational domain with 300 cells is split into two parts.
Half of the domain is dedicated to the hybrid part of the AION scheme while the rest of the domain is dedicated to the IRK2 scheme.
The spectral analysis is performed at the transition between the chosen schemes.
Two configurations are analysed, one for the wave going from the hybrid domain to the implicit domain, and one
for the wave going from the implicit domain to the hybrid domain.

\begin{figure}[!htbp]
\begin{center}
\includegraphics [width=8cm]{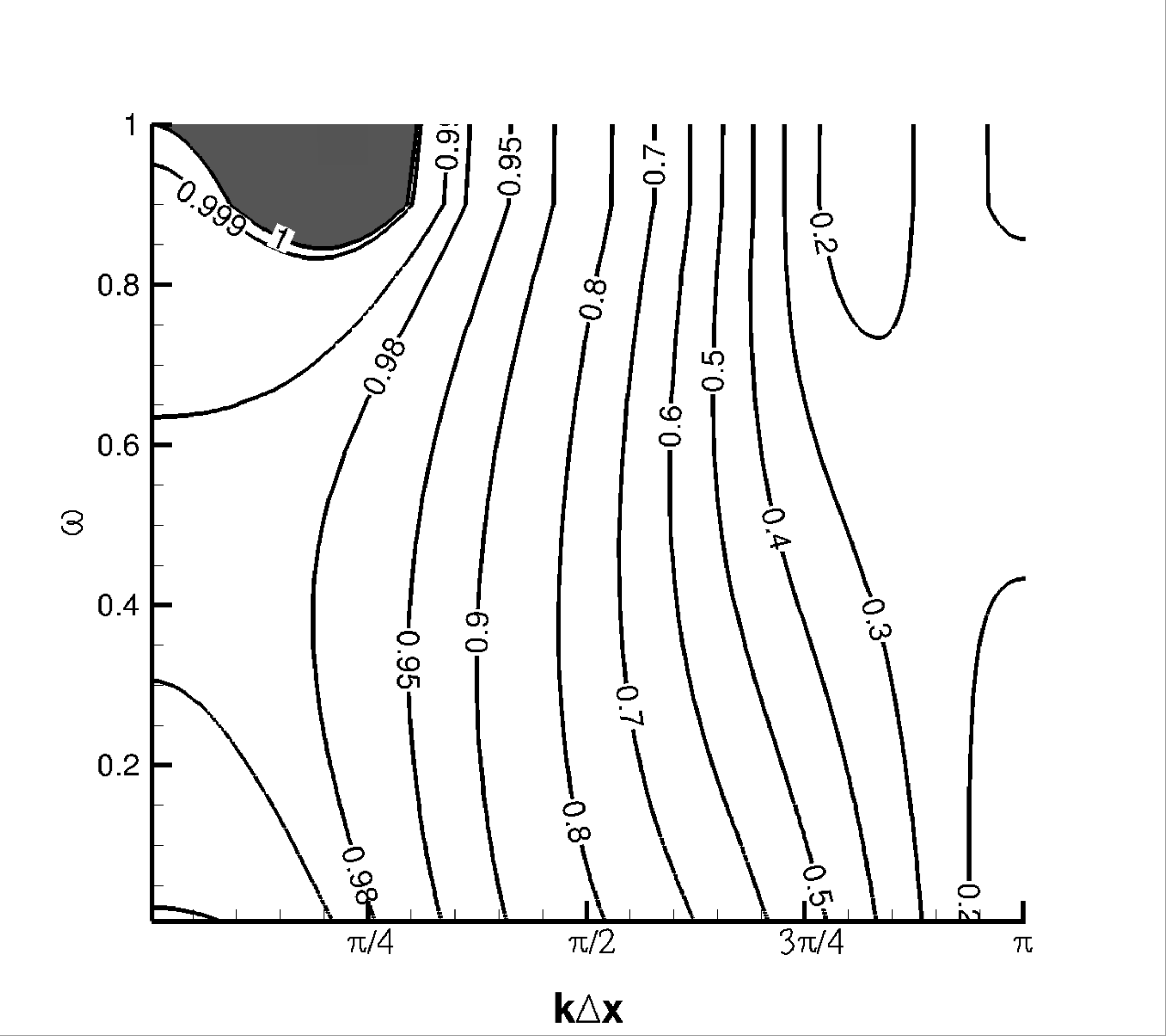}
\caption{Isocontours of the dissipation $\mu_j$ for Hyb/IRK2 transition at CFL=0.5 as a function of $\omega_j$ and $k \Delta x$. The grey area represents the area
of amplification. \label{fig:absAION_IRK2_omega05}}
\end{center}
\end{figure}

\begin{figure}[!htbp]
\begin{center}
\includegraphics [width=8cm]{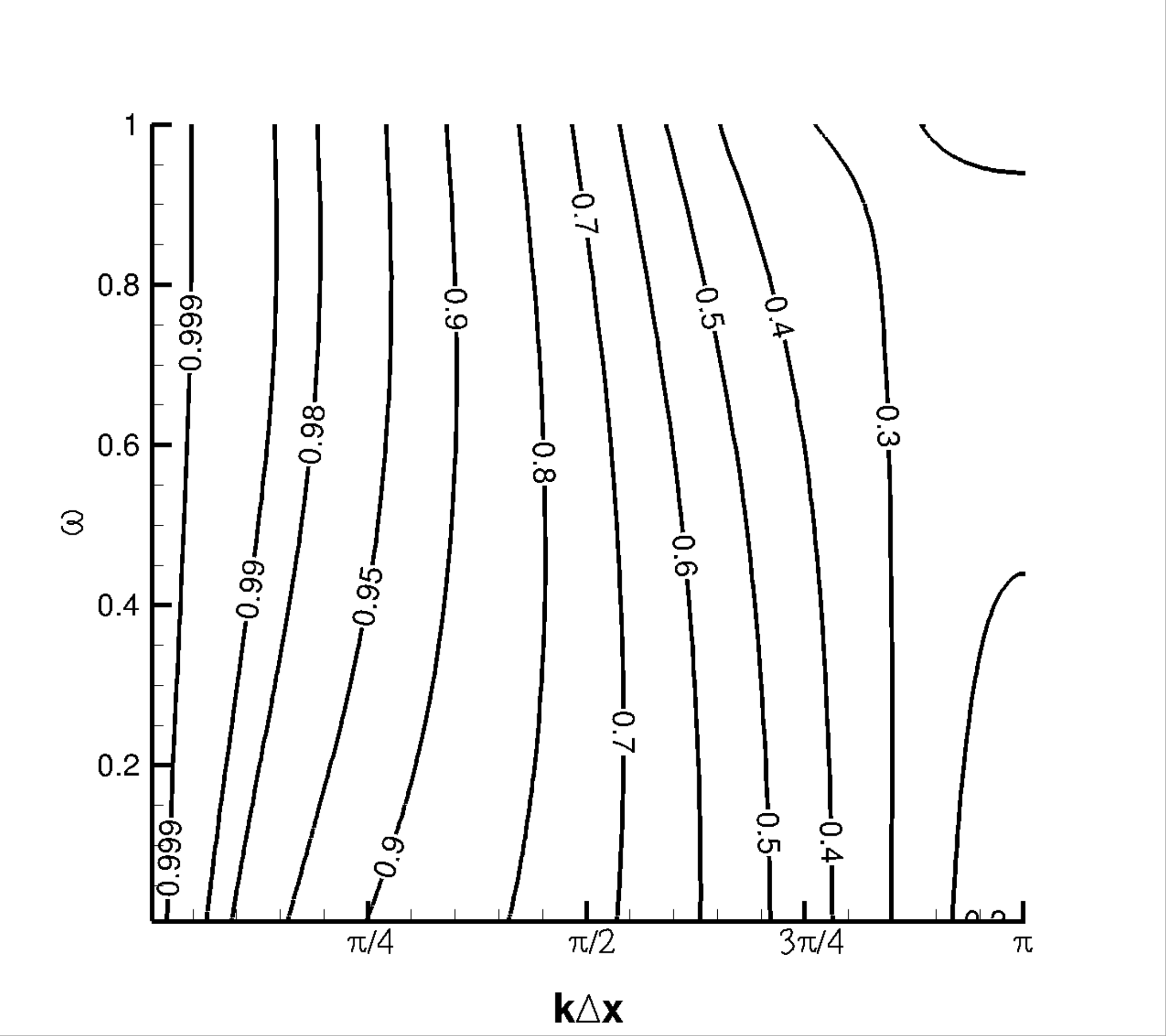}
\caption{Isocontours of the dissipation $\mu_j$ for Hyb/IRK2 transition at CFL=0.5 as a function of $\omega_j$. The coupled approach is
always stable\label{fig:absIRK2_AION_omega05}}
\end{center}
\end{figure}

\begin{figure}[!htbp]
\begin{center}
\includegraphics [width=8cm]{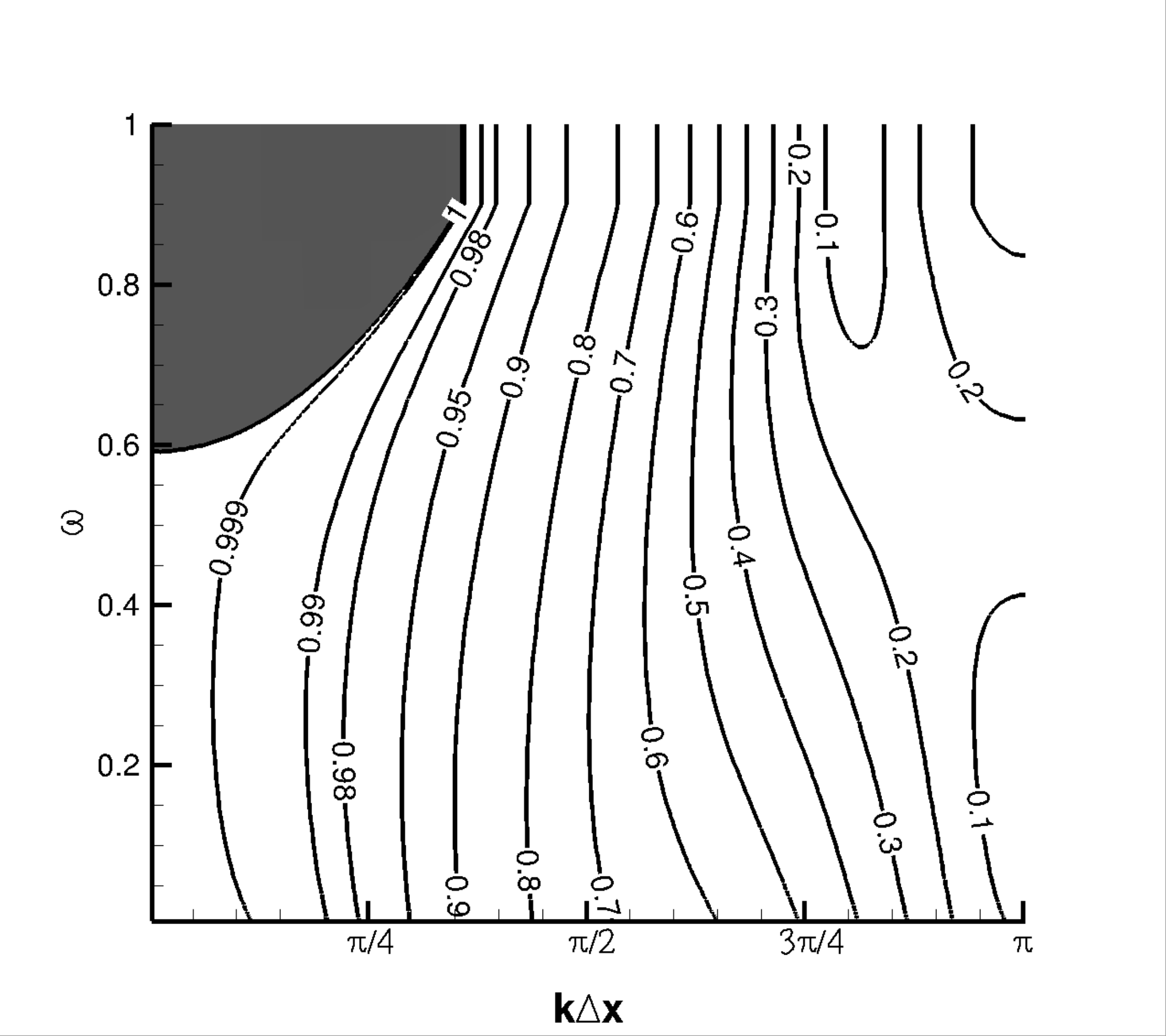}
\caption{Isocontours of the dissipation $\mu_j$ for Hyb/IRK2 transition at CFL=0.6 as a function of $\omega_j$. The grey area represents the area
of amplification. \label{fig:absAION_IRK2_omega06}}
\end{center}
\end{figure}

\begin{figure}[!htbp]
\begin{center}
\includegraphics [width=8cm]{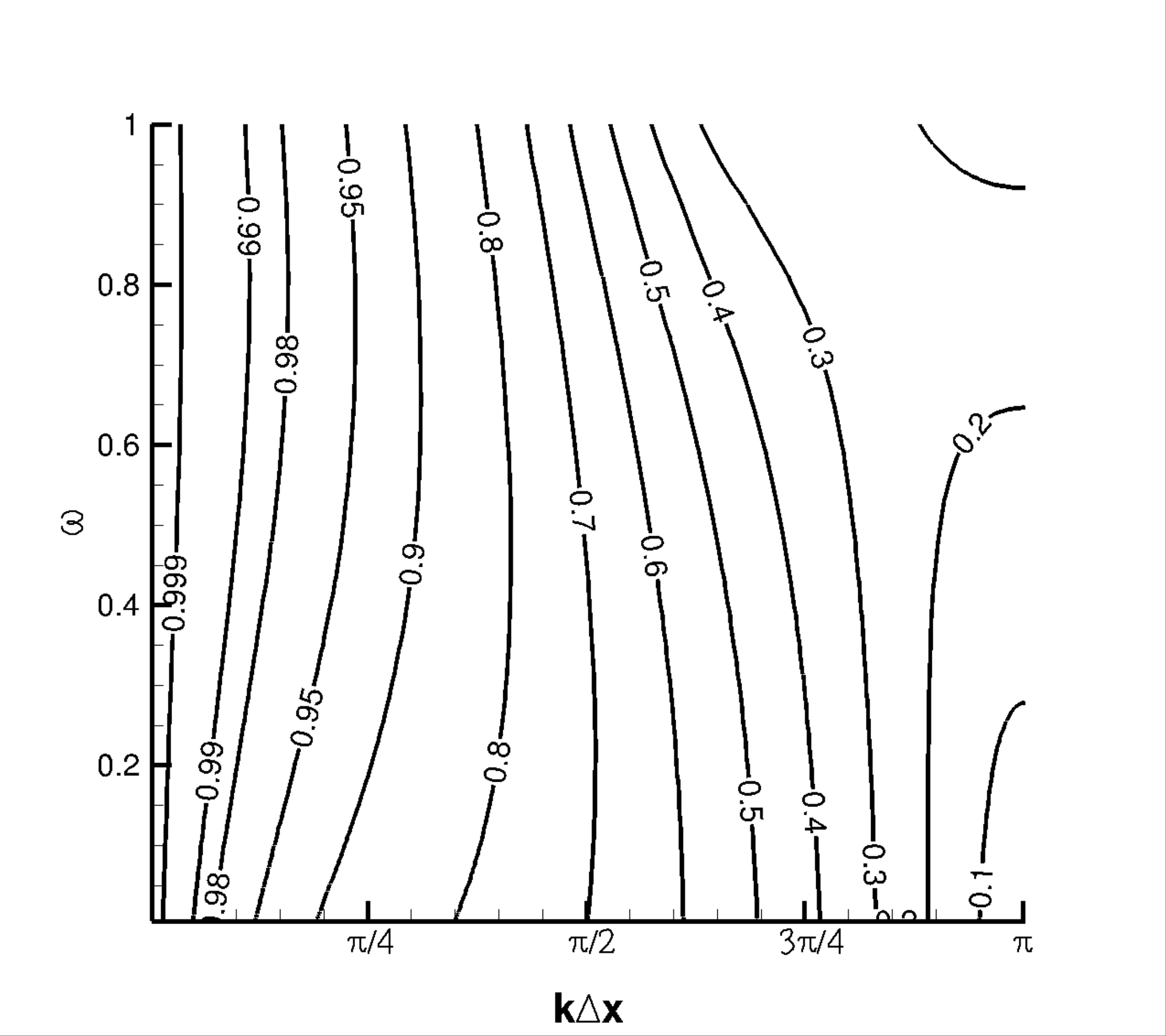}
\caption{Isocontours of the dissipation $\mu_j$ for IRK2/Hyb transition at CFL=0.5 as a function of $\omega_j$. The coupled approach is
always stable.\label{fig:absIRK2_AION_omega06}}
\end{center}
\end{figure}

Figs.~\ref{fig:absAION_IRK2_omega05} and~\ref{fig:absAION_IRK2_omega06} show the dissipation coefficient for CFL=0.5 and for CFL=0.6.
For the transition between implicit and hybrid schemes. Amplification (positive dissipation) is represented by a grey area.
Amplification occurs for the transition Hyb/IRK2.
In the reverse mode, amplification does not occur, as highlighted by Figs.~\ref{fig:absIRK2_AION_omega05} and~\ref{fig:absIRK2_AION_omega06}.
Moreover, a stable formulation is obtained for $\omega_j\le0.6$, and this
motivates our final choice to switch at $\omega_j=0.6$ between the hybrid flux computation and the implicit Runge-Kutta scheme. As
a consequence and for the following applications, $\omega_j = 1$ will lead to the Heun's scheme, $\omega_j < 0.6$ will lead to the
IRK2 scheme. The new hybrid reconstruction method based on the reconstructed states Eq.~\eqref{eq:AION_NHRM}
is applied for $0.6 \le \omega_j \le 1$.

\subsection{Analysis of $q-$waves}


Focusing on the one-dimensional linear advection equation with a constant velocity $c$, several authors mentioned and explained the existence
of numerical waves travelling in the direction opposite to $c$, introducing the notion of $q-$waves. Indeed,
a $p$-wave is a wave travelling in the same direction as $c$, eventually not at the same speed. $q-$waves
are travelling in the opposite direction. Existence of $q$-waves
was shown to be independent of the stability analysis of numerical scheme and was highlighted for instance in the seminal work of
Vichnevetsky, summarised in \cite{Vichnevetsky_1982_book}, by Poinsot and Veynante for combustion \cite{Poinsot_2005_book} or by
Trefethen \cite{Trefethen_1992_SIAM}. In this context, Sengupta {\it et al.} \cite{Sengupta_2012_AMC} linked the numerical group velocity
with the existence of $q$-waves. Following the definition proposed in \cite{Sengupta_2012_AMC}, new quantities
related to the one-dimensional linear advection equation Eq.~\eqref{conv} are introduced. The numerical phase speed is defined as
 \begin{equation}
\begin{aligned}
c^N_j=\frac{arg(G_j)}{k\,\Delta t},\label{eq:def_cn}
\end{aligned}
\end{equation}
from which is deduced the numerical group velocity,
\begin{equation}
\begin{aligned}
V^{gN}_j=\frac{d \big(arg(G_j)\big)}{d k} \frac{c}{CFL\,\Delta x}.\label{eq:def_vgn}
\end{aligned}
\end{equation}
Indeed, for a positive advection velocity $c>0$, $q$-waves are propagating upstream and their group velocity $V^{gN}$ is negative
\cite{Sengupta_2012_AMC}. It is important to notice that the negative group velocity is only a necessary condition for apparition of
$q$-waves. Indeed, observation of $q$-waves also depends upon the real and imaginary part of the amplification factor of the space-time discretization and with an excessive dissipation, filtering or damping, the $q$-waves can be removed very quickly. Even if
it was mentioned in \cite{Sengupta_2012_AMC} that initial condition, grid resolution and multi-dimensional case have effects on $q$-waves,
the analysis presented in the following is restricted to the one-dimensional case. 

\begin{figure}[!htbp]
\begin{center}
\includegraphics [width=9cm]{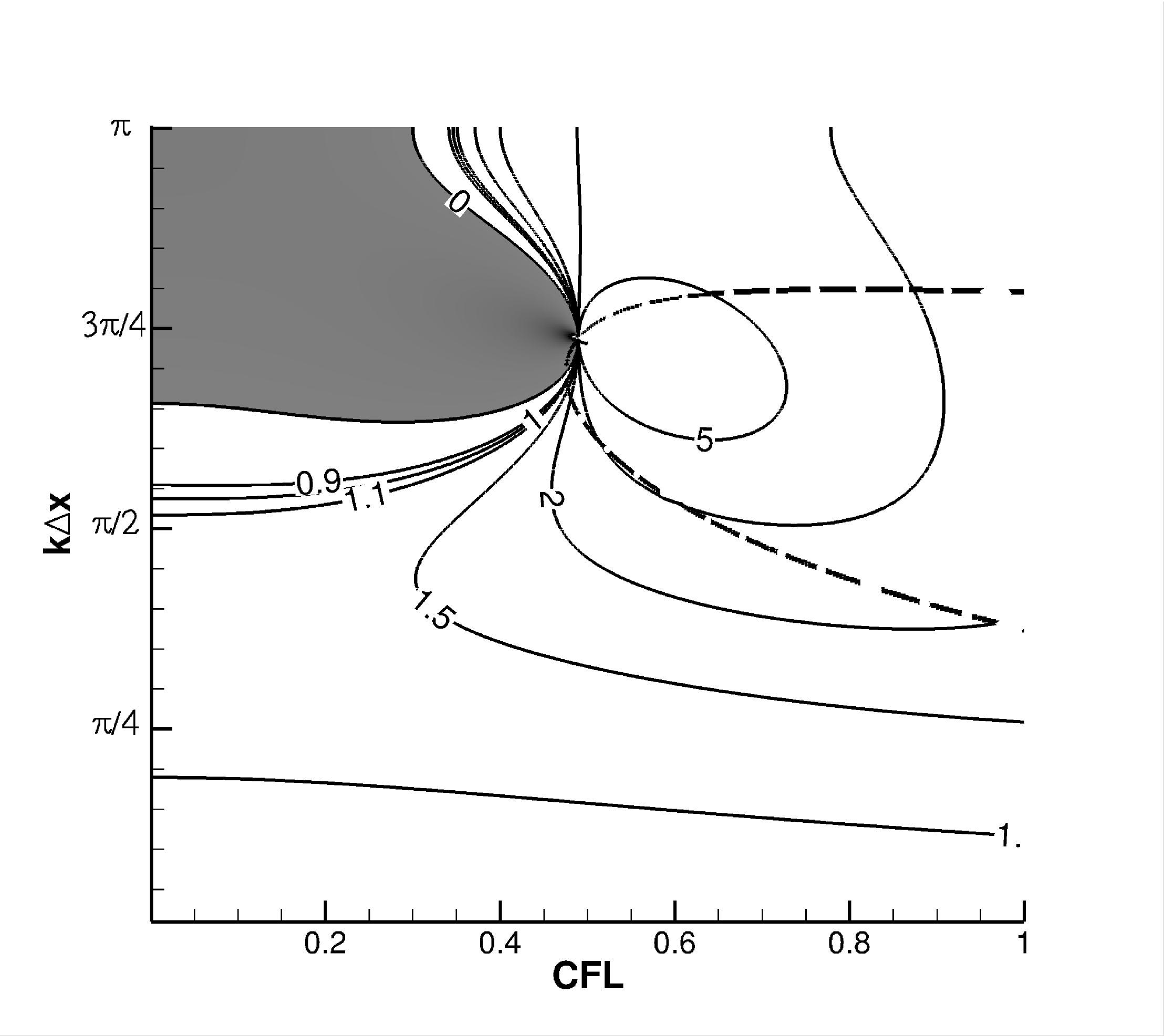}
\caption{Isocontours of $V_{gN}$ for full Heun scheme}
\label{fig:qwavesHeun}
\end{center}
\end{figure}

\begin{figure}[!htbp]
\begin{center}
\includegraphics [width=9cm]{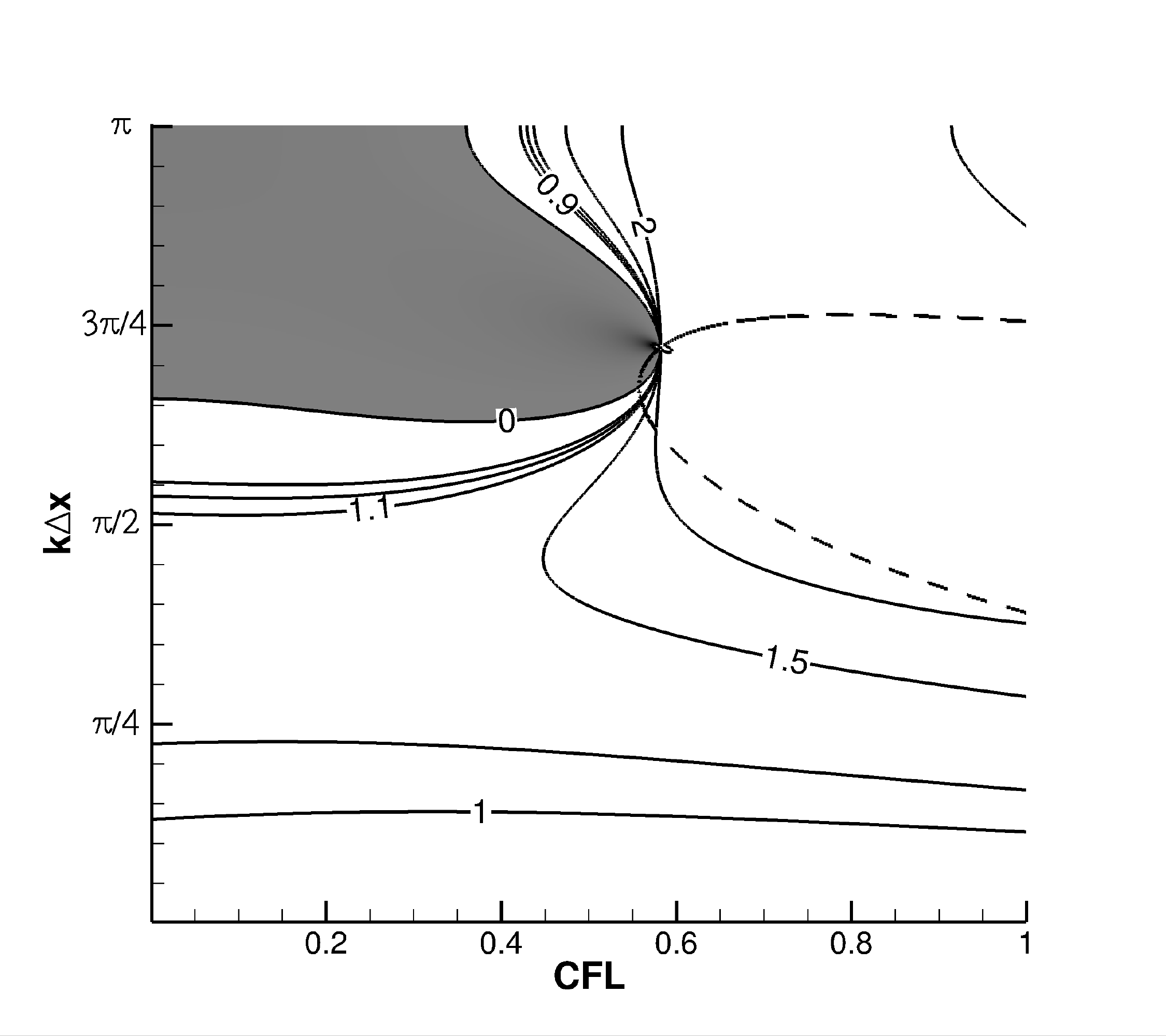}
\caption{Isocontours of $V_{gN}$ for intersection Heun/Hyb scheme}
\label{fig:qwavesHeunMuscat}
\end{center}
\end{figure}

\begin{figure}[!htbp]
\begin{center}
\includegraphics [width=9cm]{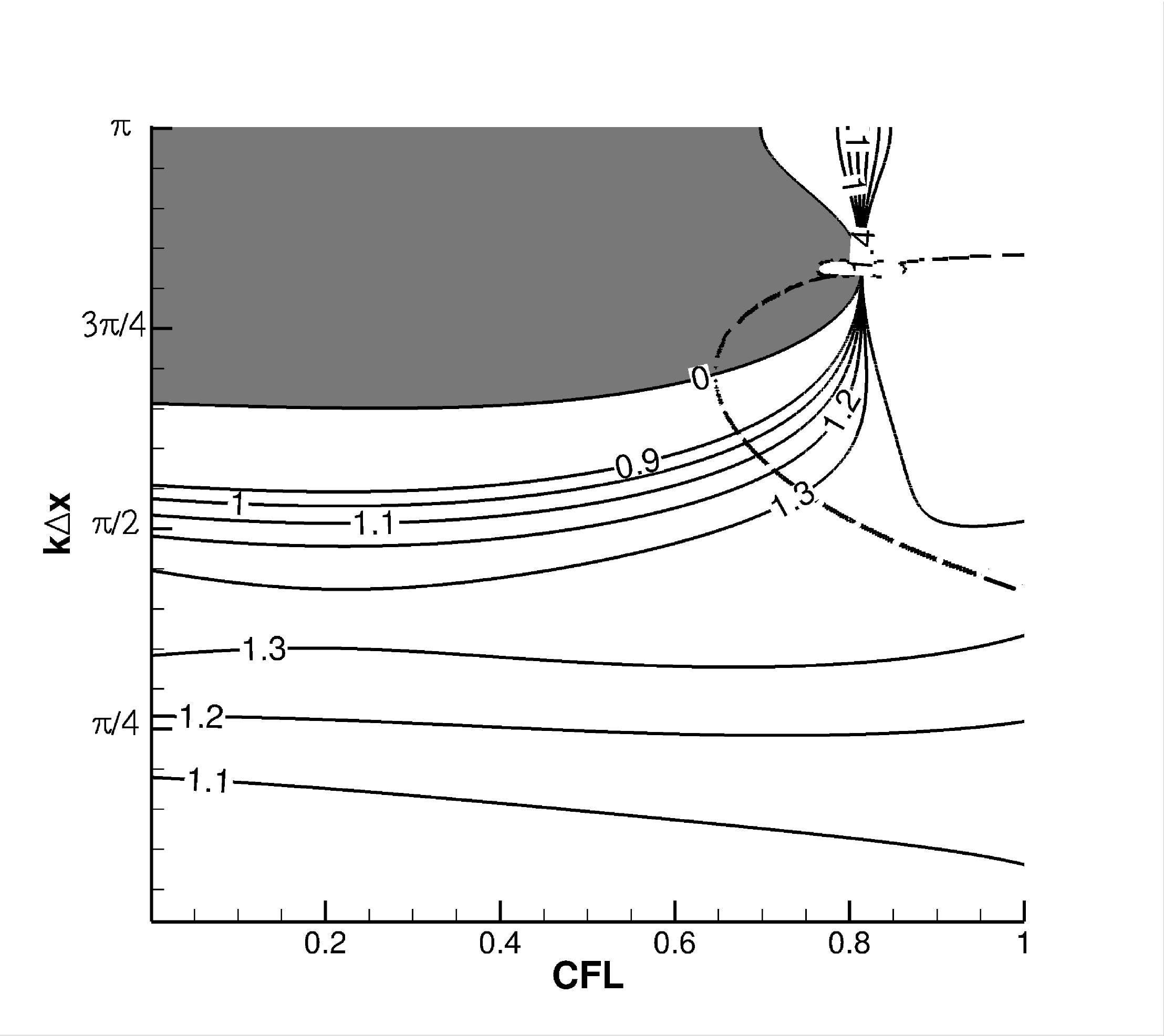}
\caption{Isocontours of $V_{gN}$ for intersection Hyb/IRK2 scheme}
\label{fig:qwavesMuscatIRK2}
\end{center}
\end{figure}

\begin{figure}[!htbp]
\begin{center}
\includegraphics [width=9cm]{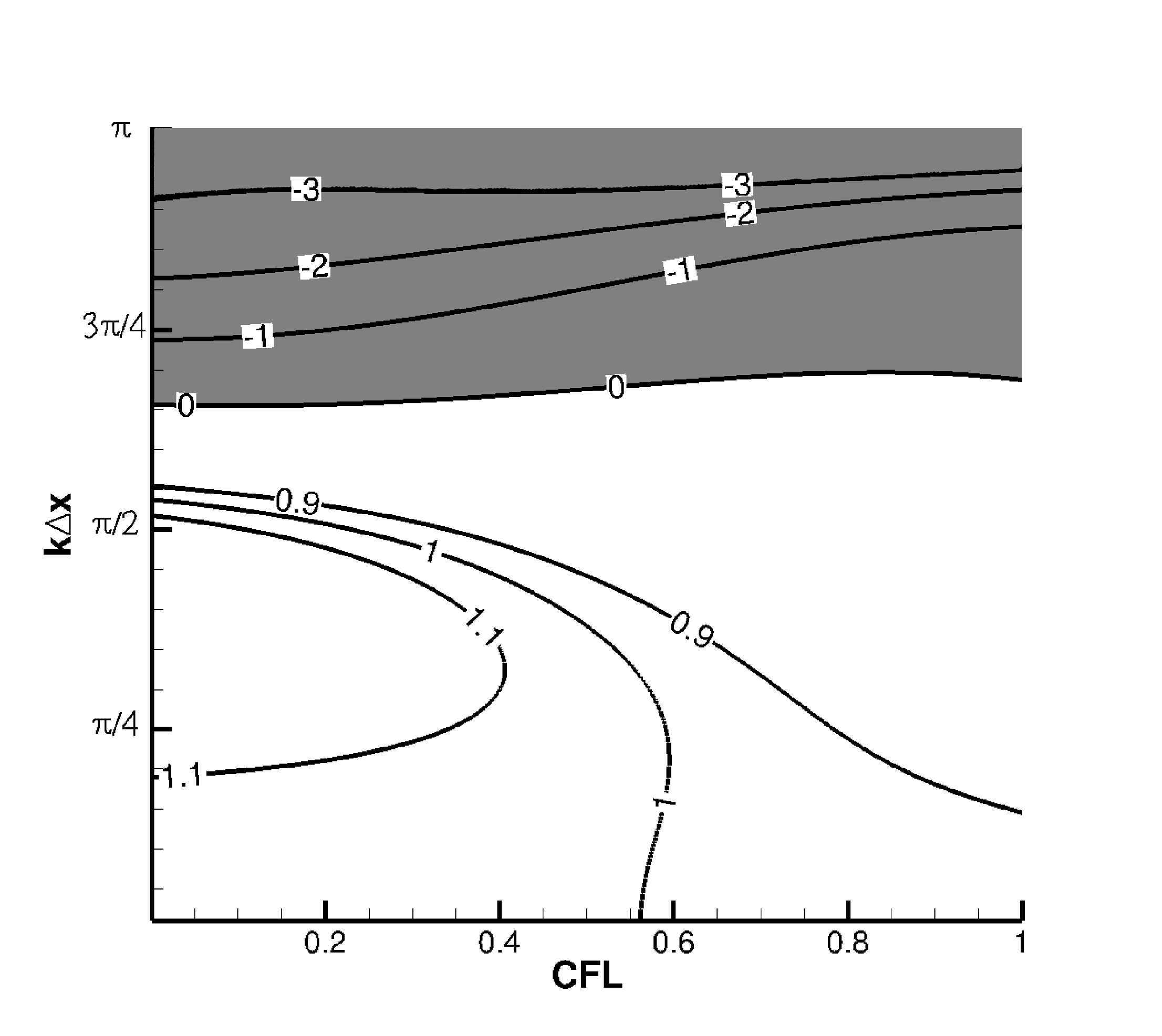}
\caption{Isocontours of $V_{gN}$ for intersection IRK2/hyb scheme}
\label{fig:qwavesIRK2Muscat}
\end{center}
\end{figure}

A mesh composed of 300 cells is considered with the manufactured choice of $\omega_j$ introduced in
Sec.~\ref{sec:VonNeumannHCS1} and used again in Sec.~\ref{sec:VonNeumannAION}. The numerical phase speed $c_j^N$ and
the numerical group velocity $V^{gN}_j$ are computed from $G_j$ (Eq.~\ref{eq:defGj}) for any cell $j$ (and therefore for a given $\omega_j$).
The analysis of $q-$waves is performed for the AION scheme and for CFL $\in [0,1]$. Attention is paid on the computation
of the group velocity $V^{gN}_j$ for the different intersection of time integrators in AION scheme.
In Figs.~\ref{fig:qwavesHeun}-\ref{fig:qwavesIRK2Muscat}, dashed curves correspond to $|V^{gN}|/|c|>>1$,
which represents discontinuity of
dispersion $\phi_j$, and grey zones correspond to negative group velocity $V^{gN}_j<0$.

{\bf Remark:} The choice of the CFL domain is dictated by the desire to compare several schemes. Of course, amplification always occurs
for the Heun's scheme at CFL$>0.6$ but the analysis can still be performed for CFL$>0.6$.

\begin{figure}[!htbp]
\begin{center}
\includegraphics [width=9cm]{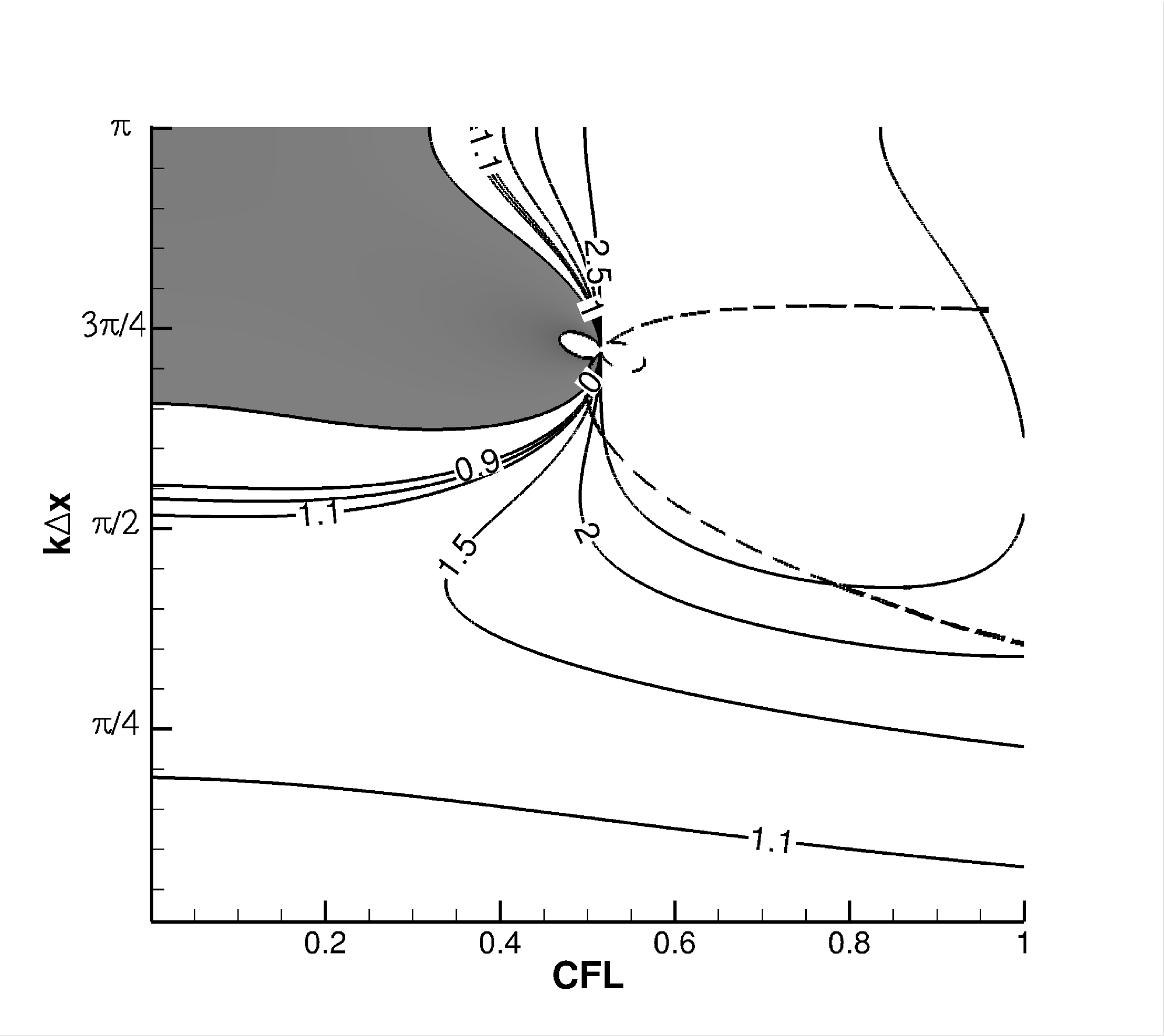}
\caption{Isocontours of $V_{gN}$ for intersection Hyb/Heun scheme}
\label{fig:qwavesMuscatHeun}
\end{center}
\end{figure}

\begin{figure}[!htbp]
\begin{center}
\includegraphics [width=9cm]{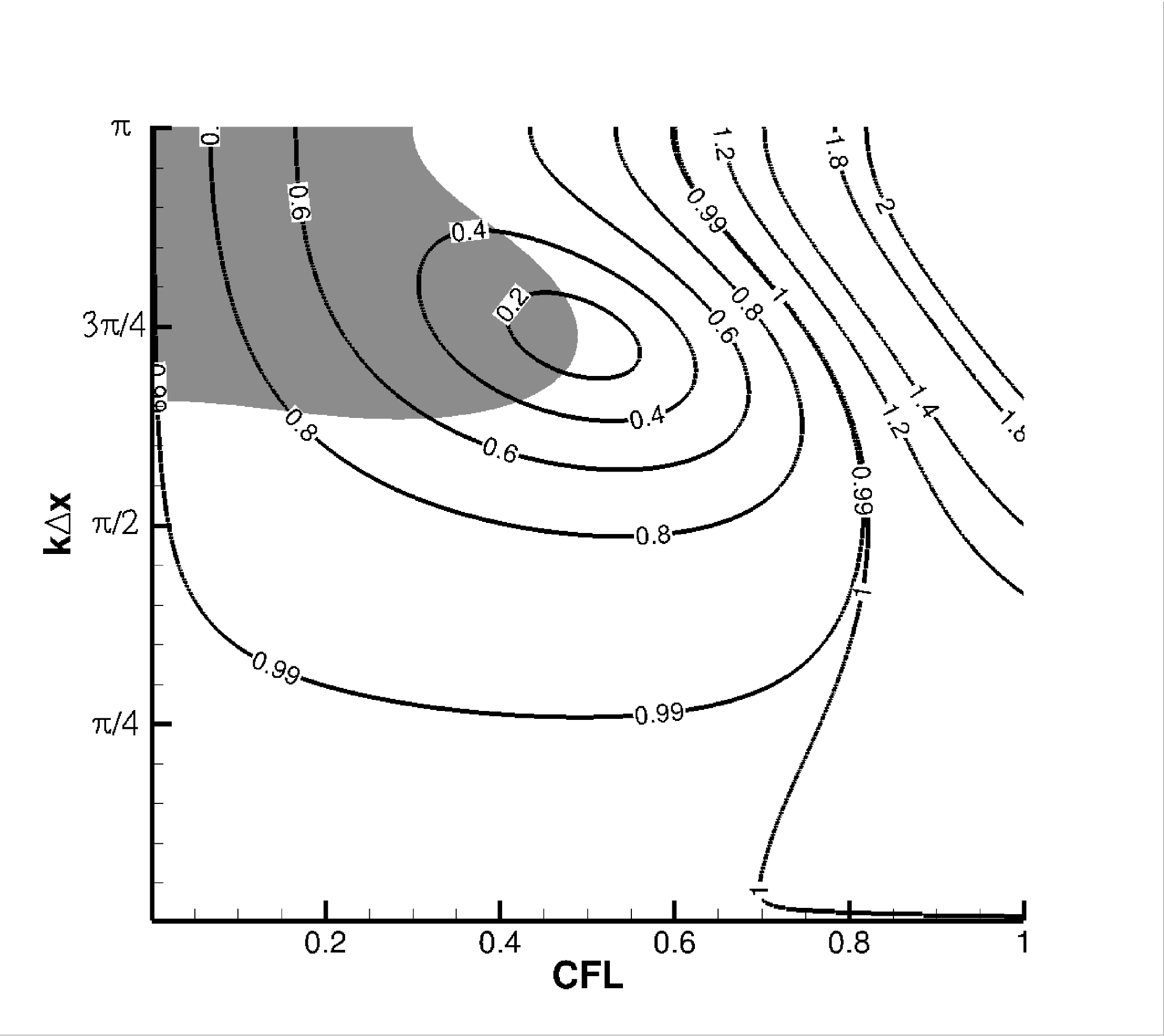}
\caption{Isocontours of the dissipation $\mu_j$ for the standard Heun's scheme and grey zone for negative group velocity
\label{fig:qwaves_absG_Heun}}
\end{center}
\end{figure}

\begin{figure}[!htbp]
\begin{center}
\includegraphics [width=9cm]{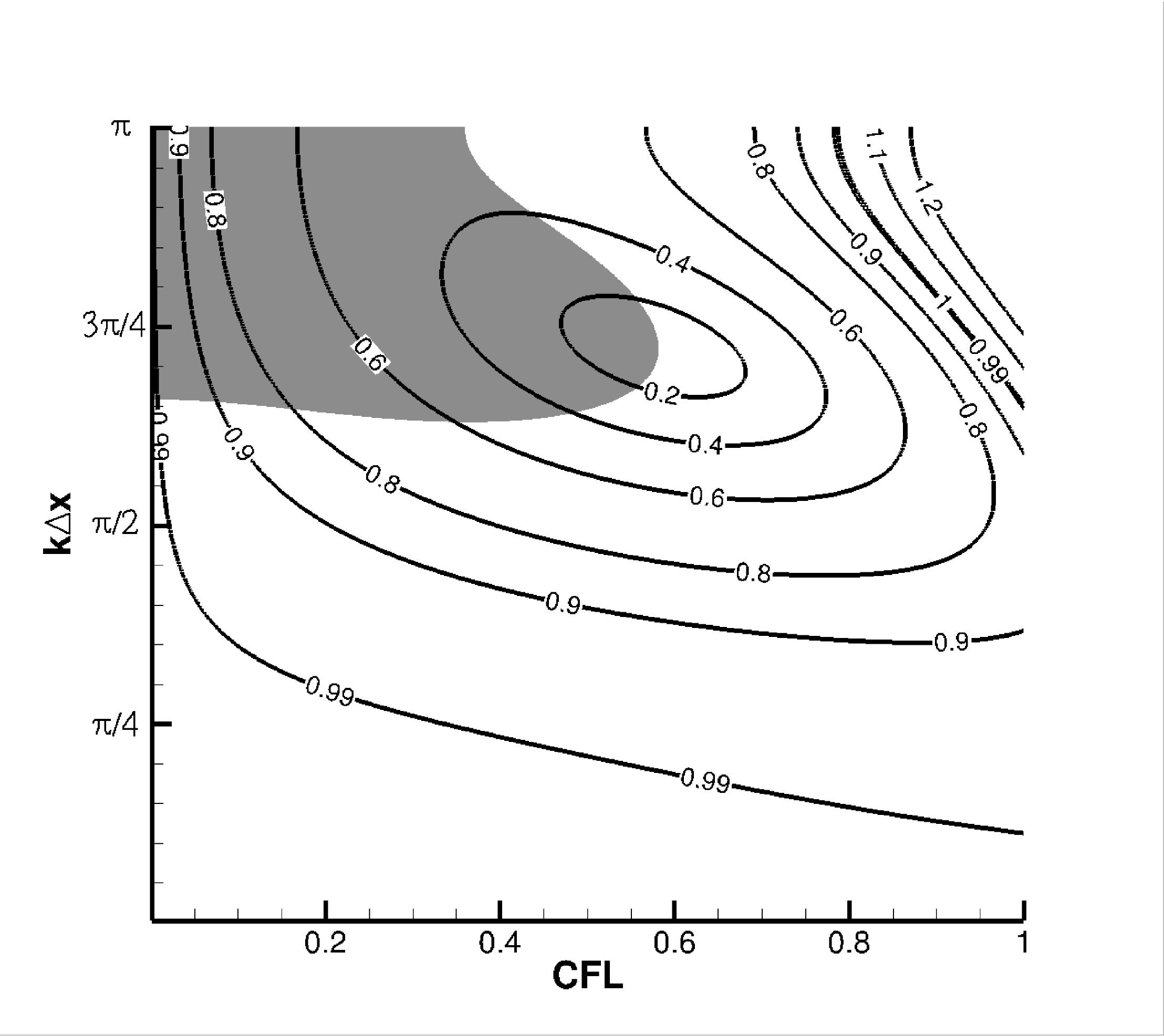}
\caption{Isocontours of the dissipation $\mu_j$ for a cell with Heun/Hyb reconstruction and grey zone for negative group velocity
\label{fig:qwaves_absG_HeunMuscat}}
\end{center}
\end{figure}

\begin{figure}[!htbp]
\begin{center}
\includegraphics [width=9cm]{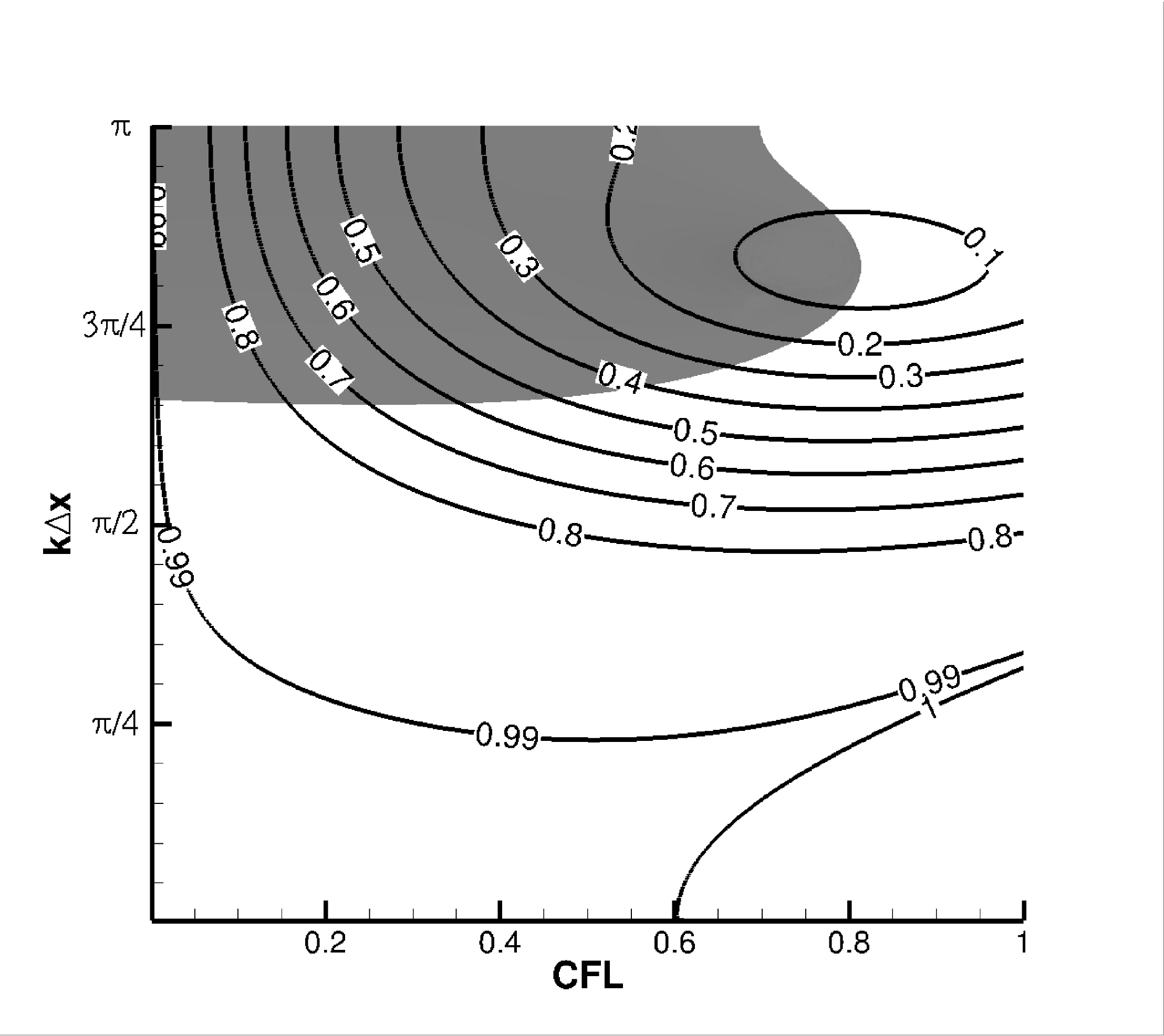}
\caption{Isocontours of the dissipation $\mu_j$ for a cell with Hyb/IRK2 reconstruction and grey zone for negative group velocity
\label{fig:qwaves_absG_MuscatIRK2}}
\end{center}
\end{figure}

\begin{figure}[!htbp]
\begin{center}
\includegraphics [width=9cm]{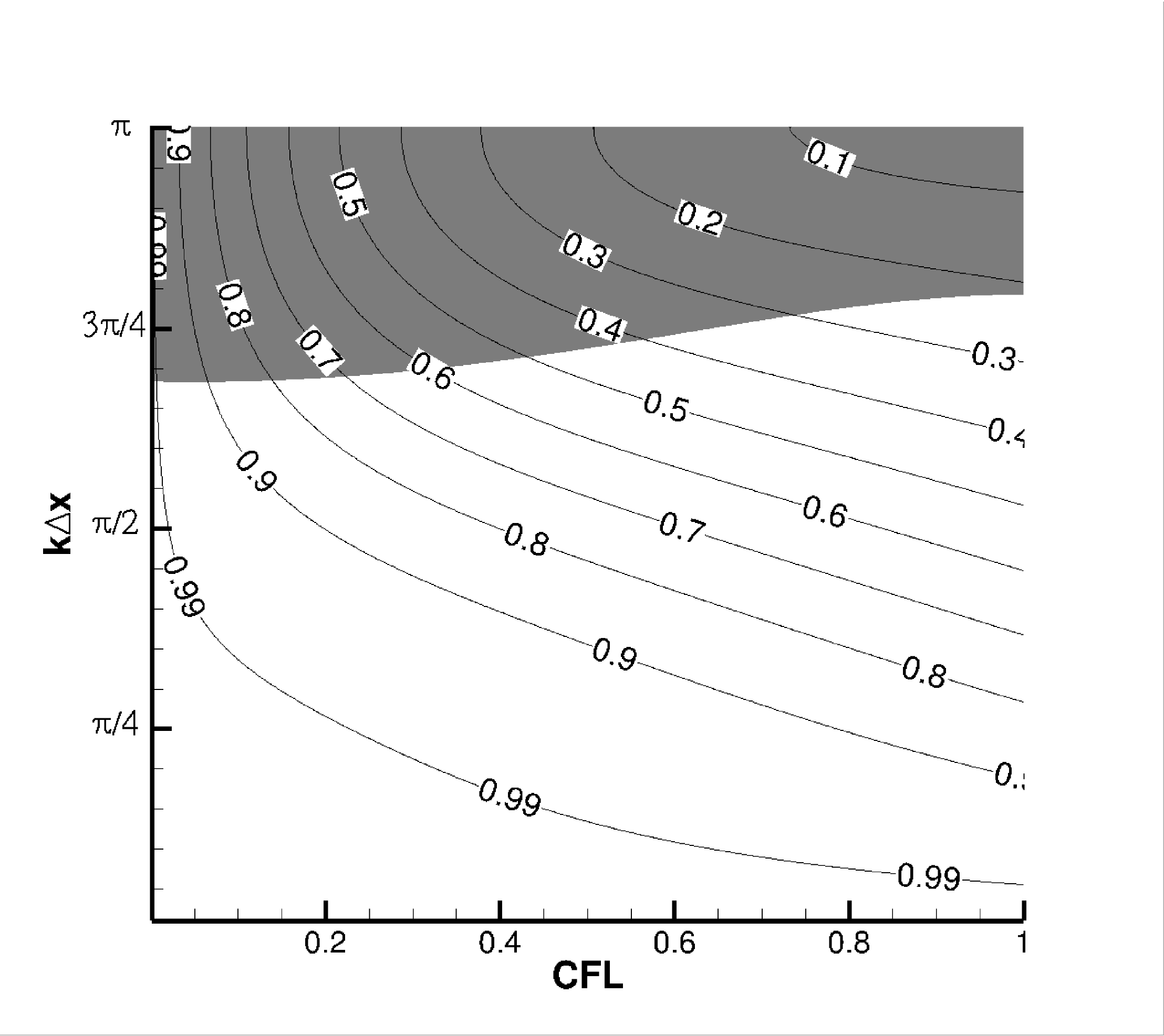}
\caption{Isocontours of the dissipation $\mu_j$ for a cell with IRK2/Hyb reconstruction and grey zone for negative group velocity
\label{fig:qwaves_absG_IRK2Muscat}}
\end{center}
\end{figure}

\begin{figure}[!htbp]
\begin{center}
\includegraphics [width=9cm]{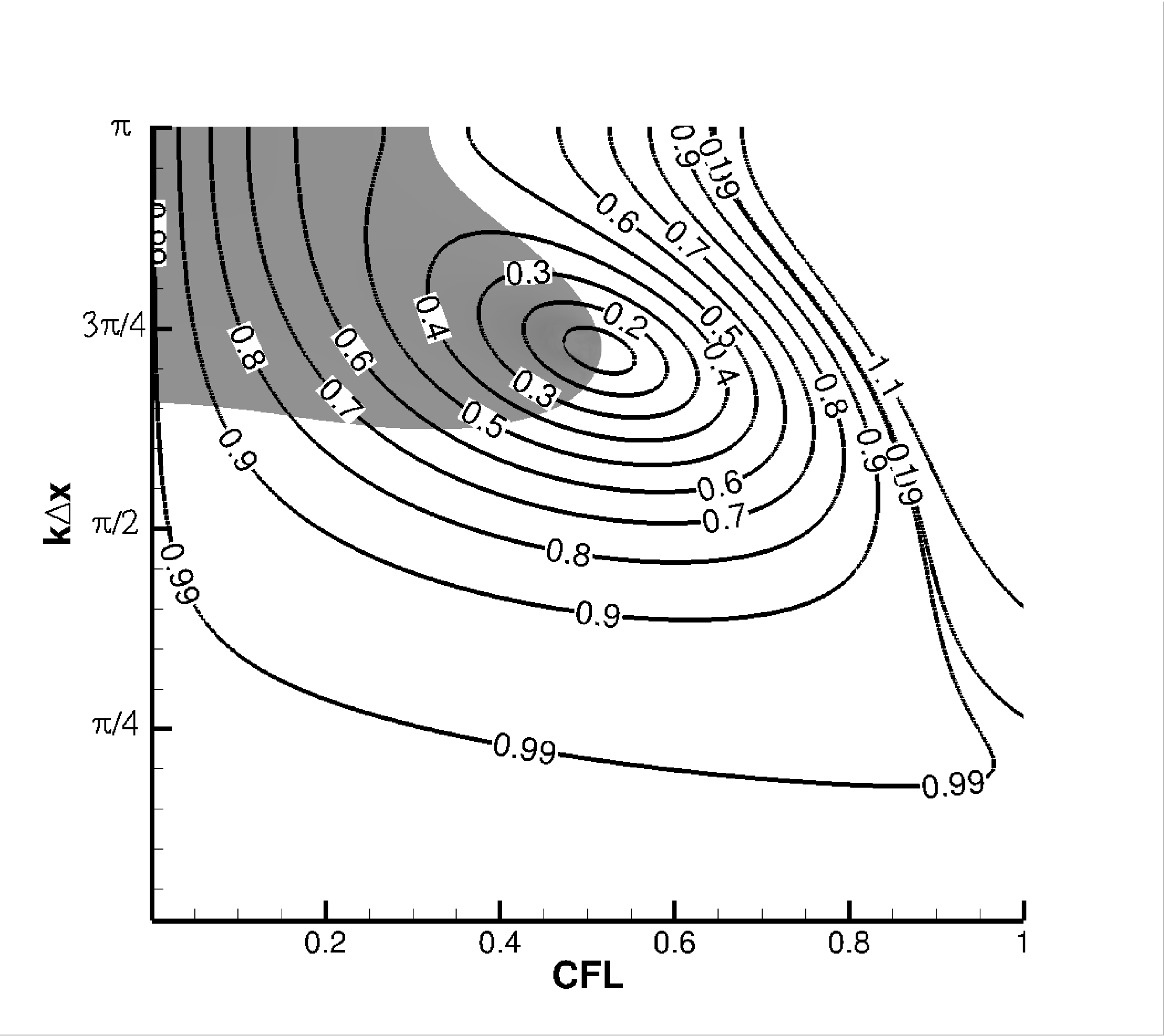}
\caption{Isocontours of the dissipation $\mu_j$ for a cell with Hyb/Heun reconstruction and grey zone for negative group velocity
\label{fig:qwaves_absG_MuscatHeun}}
\end{center}
\end{figure}

Concerning the areas of negative-$V^{gN}_j$ group velocity, it can be noticed that the zone is slightly more expanded for Heun/Hyb cells
than for the pure Heun scheme, according to Figs.~\ref{fig:qwavesHeun} and~\ref{fig:qwavesHeunMuscat}. Indeed,
negative group velocities can be found up to CFL=0.57 at Heun/Hyb cells, but the full Heun's scheme negative-group-velocity zone is present
up to CFL=0.47. In addition, this negative group-velocity zone is larger for the other transitions according to
Figs.~\ref{fig:qwavesMuscatIRK2} and~\ref{fig:qwavesIRK2Muscat}. So, the presence of $q$-waves is possible in a larger
domain of CFL for AION scheme than for the full Heun's scheme.
The last question concerns the capability of the AION scheme to damp $q-$waves.
To answer it, it is mandatory to couple the observations on negative group velocity with the dissipation property of the AION scheme, as
shown in Figs.~\ref{fig:qwaves_absG_Heun}-\ref{fig:qwaves_absG_MuscatHeun}.

\newpage
According to Fig.~\ref{fig:qwaves_absG_Heun}, the CFL limit of the negative-$V_{gN}$ zone (CFL=0.47) due to Heun time integration corresponds to $\mu_j<0.2$.
In the case of the time integrator transitions of AION scheme, the CFL limit of their negative-$V_{gN}$ zone corresponds to $\mu_j<0.3$ for cell Heun/Hyb (Fig.~\ref{fig:qwaves_absG_HeunMuscat}), $\mu_j<0.5$ for hyb/IRK2 and IRK2/Hyb cells
(Fig.~\ref{fig:qwaves_absG_MuscatIRK2}~\ref{fig:qwaves_absG_IRK2Muscat}), and finally $\mu_j<0.2$ for cell Hyb/Heun cells
(Fig.~\ref{fig:qwaves_absG_MuscatHeun}). Fortunately, it seems that the effect of extension of the $q$-waves zone due to our coupling scheme can be attenuated by dissipation.
All this information allows us to conclude that the AION scheme seems to be as stable as the full Heun scheme on uniform mesh.
Moreover, the possible additional $q$-waves that could propagate would be quite dissipated.
The good numerical behaviour obtained through the proposed theoretical analysis must now be confirmed by numerical simulations,
which is the goal of the next sections.

\section{Validation\label{sec:Validation}}

This section is devoted to the validation of the AION scheme using several test cases of increasing complexity,
starting from 1D advection solution and Euler solutions to 3D Navier-Stokes computations.

\subsection{Advection of a Gaussian hump}
\GP{
The first test case is the linear advection of a Gaussian hump (Eq.~\ref{eq:euler} with $F(W)=W$ and $G(W,\nabla W)=0$ 
in a periodic domain) and serves to assess both stability and accuracy of the AION, Heun and IRK2 schemes. 
Introducing the periodic domain $[0,1]$ and the parameters of the initial hump $x_0=0.5m$, $\sigma=0.1$,
the analytic theoretical initial solution is simply: 
\begin{equation}
\begin{aligned}
y=\frac{1}{\sqrt{2\pi}\sigma}.\exp{-\frac{(x-x_0)^2}{2\sigma^2}}.
\end{aligned}
\end{equation}
The initial solution must be recovered at any time $t=k\,s$ with $k \in \mathbb{N}$ and here, the analysis is performed 
at $t=10s$. The mesh is composed of $N=333$ cells and is irregular but manufactured  using the following rule:
\begin{equation}
\begin{aligned}
\Delta x_j=\alpha\,\Delta x_{j-1}
&\text{~for~} \frac{N}{2}-50 \leq j\leq \frac{N}{2},  \\
\Delta x_j=\frac{1}{\alpha}.\Delta x_{j-1}
&\text{~for~} \frac{N}{2}+1 \leq j \leq \frac{N}{2}+50. \\
\Delta x_j=\Delta x_{max}
&\text{~elsewhere~} 
\end{aligned}
\end{equation}
with $\alpha=0.9$. Then the parameter $\omega_j$ is controlled as :
\begin{equation}
\begin{aligned}
\omega_j=\frac{\Delta x_j}{\Delta x_{max}}
\end{aligned}
\end{equation}
Such a variation in mesh size enables a strong control of the maximum CFL value authorized for a computation using 
an explicit time integration. In addition, the coupling scheme is fully explicit in the regular part of the 
grid ({\it ie.} when $\Delta x_j=\Delta x_{max}$).
The spatial scheme used is a second order spatial scheme for regular and irregular grids.
The time step is fixed in order to obtain a CFL constant in regular grid ($=\hbox{CFL}_{min}$) and leads to higher values 
in the smaller cells of the irregular grid. 

The first computations are performed using IRK2, Heun and AION scheme at $\hbox{CFL}_{min}=0.005$ in the explicit part. 
The largest CFL value in the implicit part is 1. Fig.~\ref{fig:gauss005} illustrates the fact that the AION scheme conserves 
the time accuracy of Heun's scheme but the IRK2 scheme involves more dissipation. At this level, it must be mentionned 
that the Newton algorithm for the convergence of IRK2 and AION scheme is guaranteed: the final error (at any time step) 
is very low. This is to avoid any confidence in the results due to a bad convergence of the Newton algorithm.

\begin{figure}[!ht]
\begin{center}
\includegraphics [width=10cm]{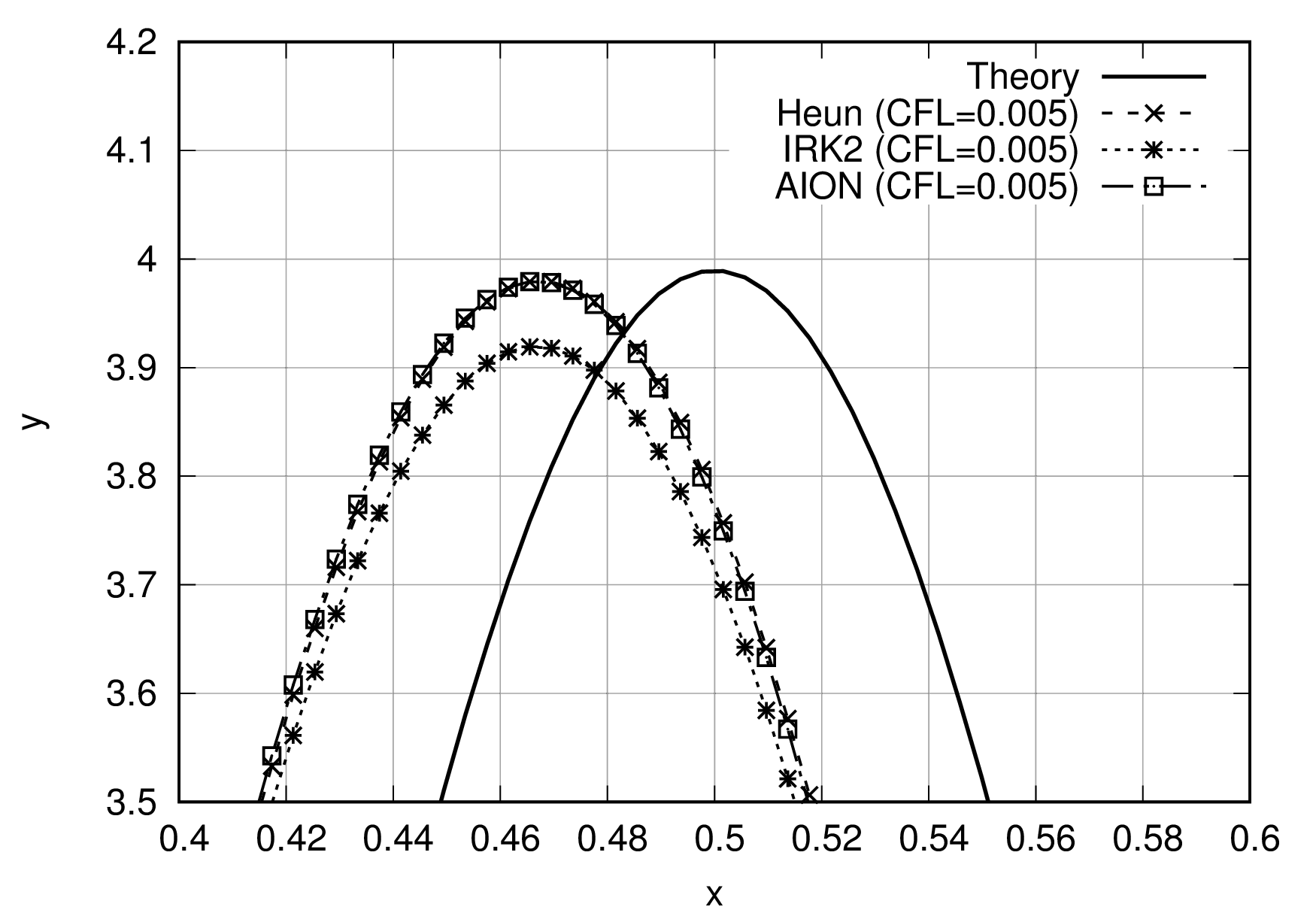}
\caption{Advection of a Gaussian hump at $CFL_{min}=0.005$, effect of the time integration scheme
\label{fig:gauss005}}
\end{center}
\end{figure}

Now, the same test case is performed with a higher $\hbox{CFL}_{min}=0.1$. In that case, the maximum CFL number in the implicit part 
reaches $20$, out of the stability region of Heun's scheme. In this configuration, both AION and IRK2 schemes are stable. Here again, 
the IRK2 scheme is more dissipative than the AION scheme. Moreover, for the present case, the convergence of the implicit schemes (AION and IRK2) is performed until 
a low error on the solution. The number of Newton steps is much larger for the IRK2 
scheme than for the AION scheme.

\begin{figure}[!ht]
\begin{center}
\includegraphics [width=10cm]{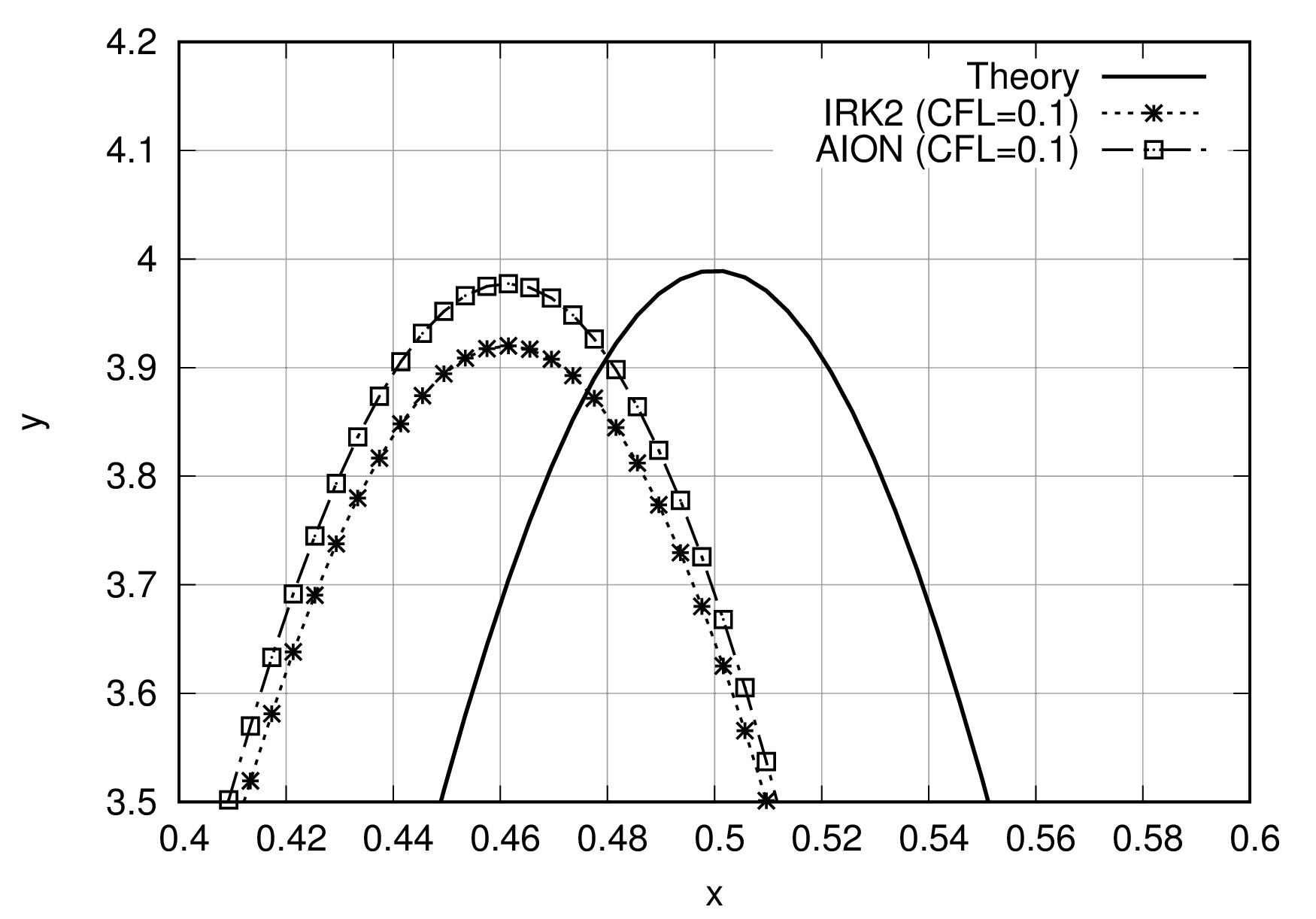}
\caption{Gaussien advection at $CFL_{min}=0.1$
\label{fig:gauss01}}
\end{center}
\end{figure}

The last question of importance concerns the interest of using the AION scheme 
at moderate CFL numbers versus the IRK2 scheme at larger CFL value. 
The computation is now performed with the IRK2 scheme at an associated $\hbox{CFL}_{min}=1$ in the explicit part. At this CFL number, the AION computation cannot be 
performed since Heun's scheme is now unstable. 
As expected (Fig.~\ref{fig:gauss05}), the IRK2 scheme is usable in this case 
but the number of iterations 
to converge  Newton's algorithm has a strong influence on the solution accuracy. 
In our example, the reference solution needs 90 steps for Newton's algorithm 
and the solution is still dissipated and dispersed. In order to try 
to represent an industrial computation, the same simulation is now performed 
but only 60 steps of Newton's algorithm are allowed. The solution is more dispersive. For the sake of consiceness, the AION solution at 
$\hbox{CFL}_{min}=0.1$ is recalled. For the implicit part of the scheme, 
the convergence of Newton's algorithm does not require more than 10 steps.

\begin{figure}[!ht]
\begin{center}
\includegraphics [width=10cm]{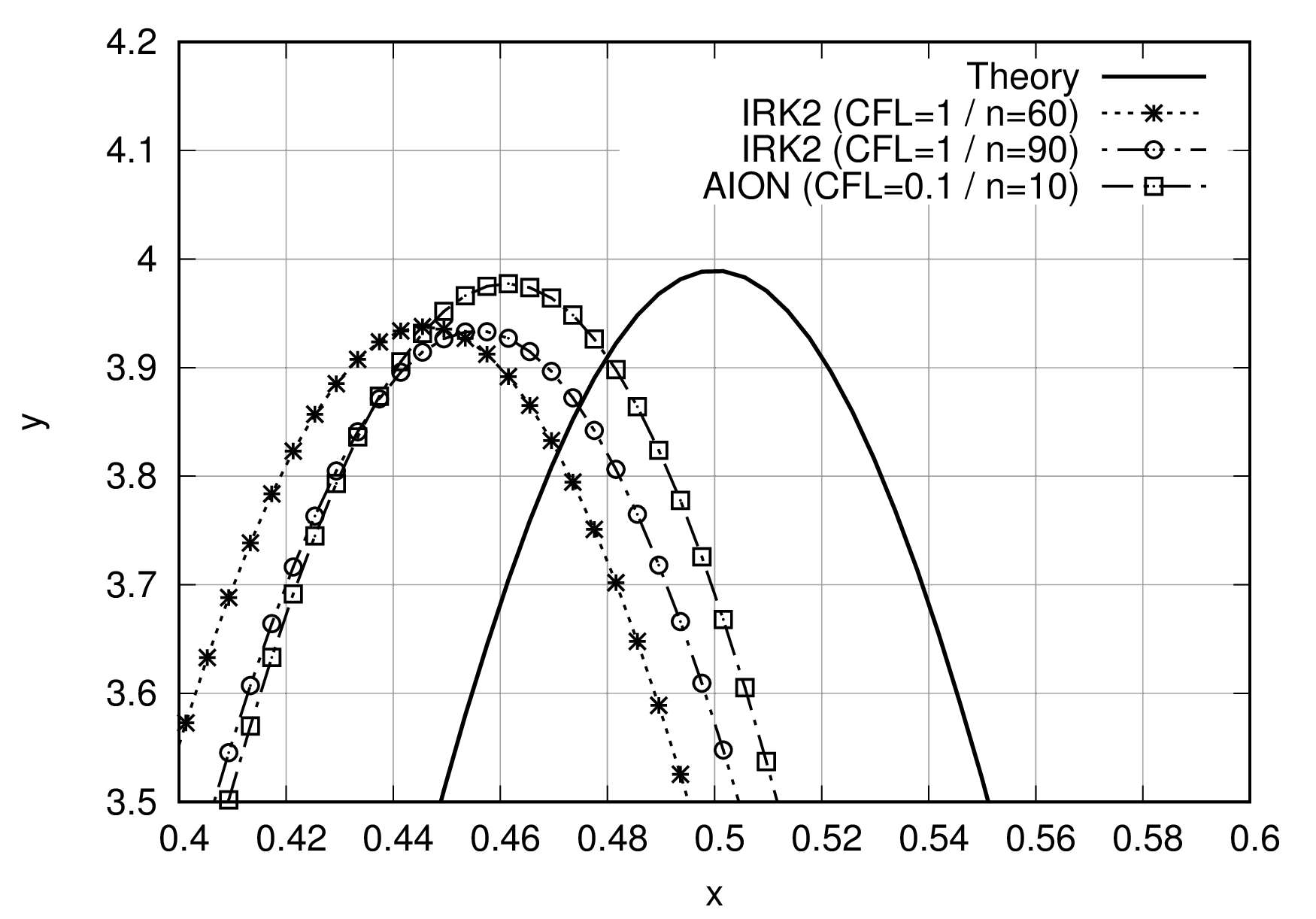}
\caption{Comparaison with IRK2 scheme at $CFL_{min}=1$
\label{fig:gauss05}}
\end{center}
\end{figure}

Indeed several articles mentionned that the time integration of LES and DNS simulation by implicit time integrators involve a damping of the high-frequency solution modes for large time steps \cite{Cinnella_2016_JCP,MARTIN_2006_JCP}. Furthermore the number of Newton subiteration is not easy to control in order to obtain convergence of the error. It remains more interesting to perform implicit time integration in stiff and/or stationnary zone of the numerical domain and perform explicit time integration elsewhere.
}

\subsection{Sod's tube}

The Sod's tube is an unsteady inviscid one-dimensional problem used to characterize the properties of both time integration and
convection schemes. The computational domain of length $L_x = 1~m$ is split into two parts separated by a membrane located
initially at $x=0.5L_x$. The initial flow is defined by
\begin{equation}
\begin{aligned}
\begin{pmatrix}
\rho_L \\[3mm]
p_L \\[3mm]
U_L \\
\end{pmatrix}=\begin{pmatrix}
1.0\\[3mm]
1.0 \\[3mm]
0.0 \\
\end{pmatrix}
, \begin{pmatrix}
\rho_R \\[3mm]
p_R \\[3mm]
U_R \\
\end{pmatrix}=\begin{pmatrix}
0.125\\[3mm]
0.1 \\[3mm]
0.0 \\
\end{pmatrix},
\end{aligned}
\end{equation}
where $L$ refers to the left side and $R$ to the right side of the membrane. At $t=0$, the membrane blows up and waves are travelling inside the computational domain. At the final time $t=0.2s$, the solution is composed of a rarefaction wave, a contact discontinuity and a shock.

\begin{figure}[!htbp]\begin{center}
\begin{tabular}{cc}
\includegraphics[width=8cm]{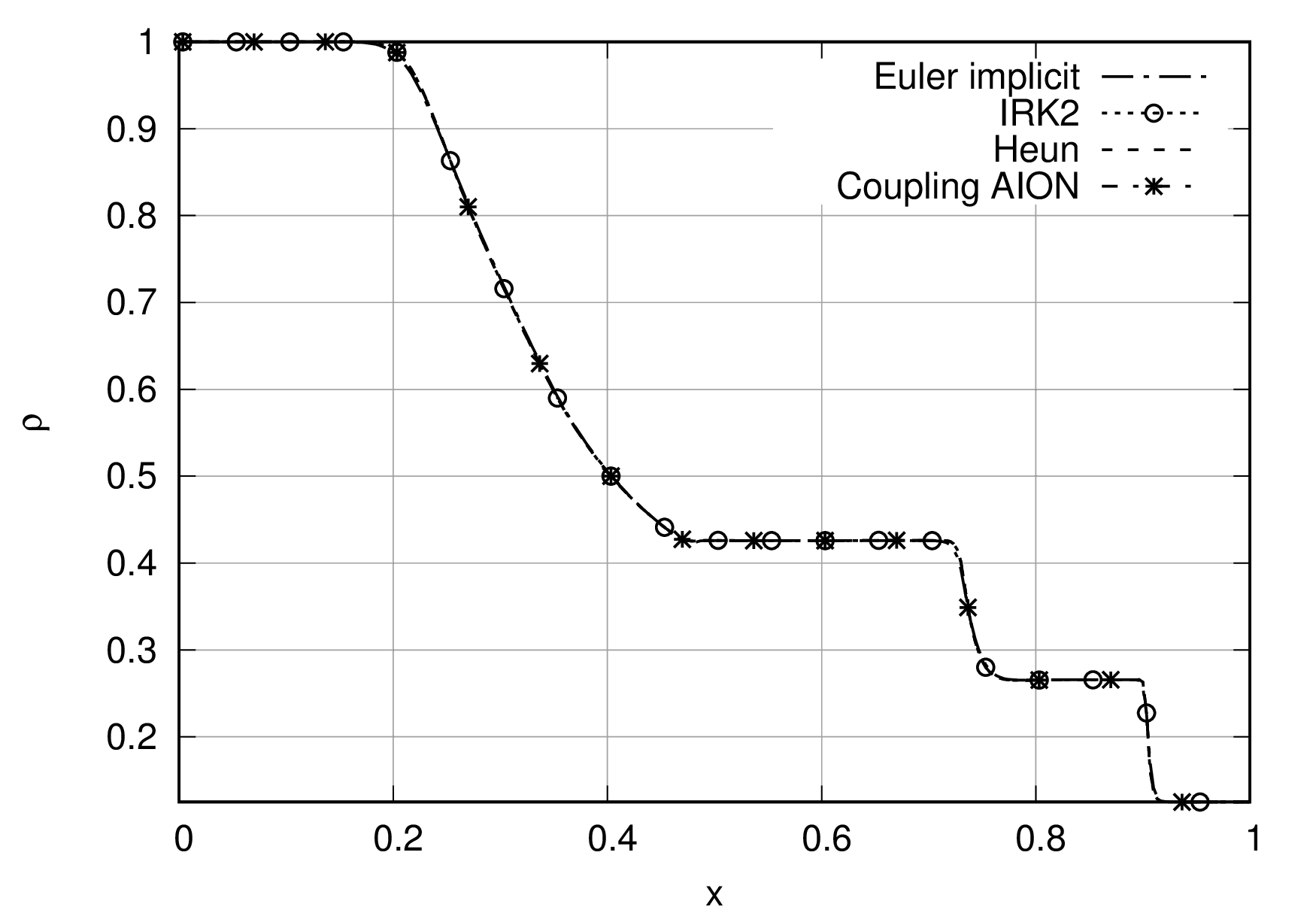} & \includegraphics [width=8cm]{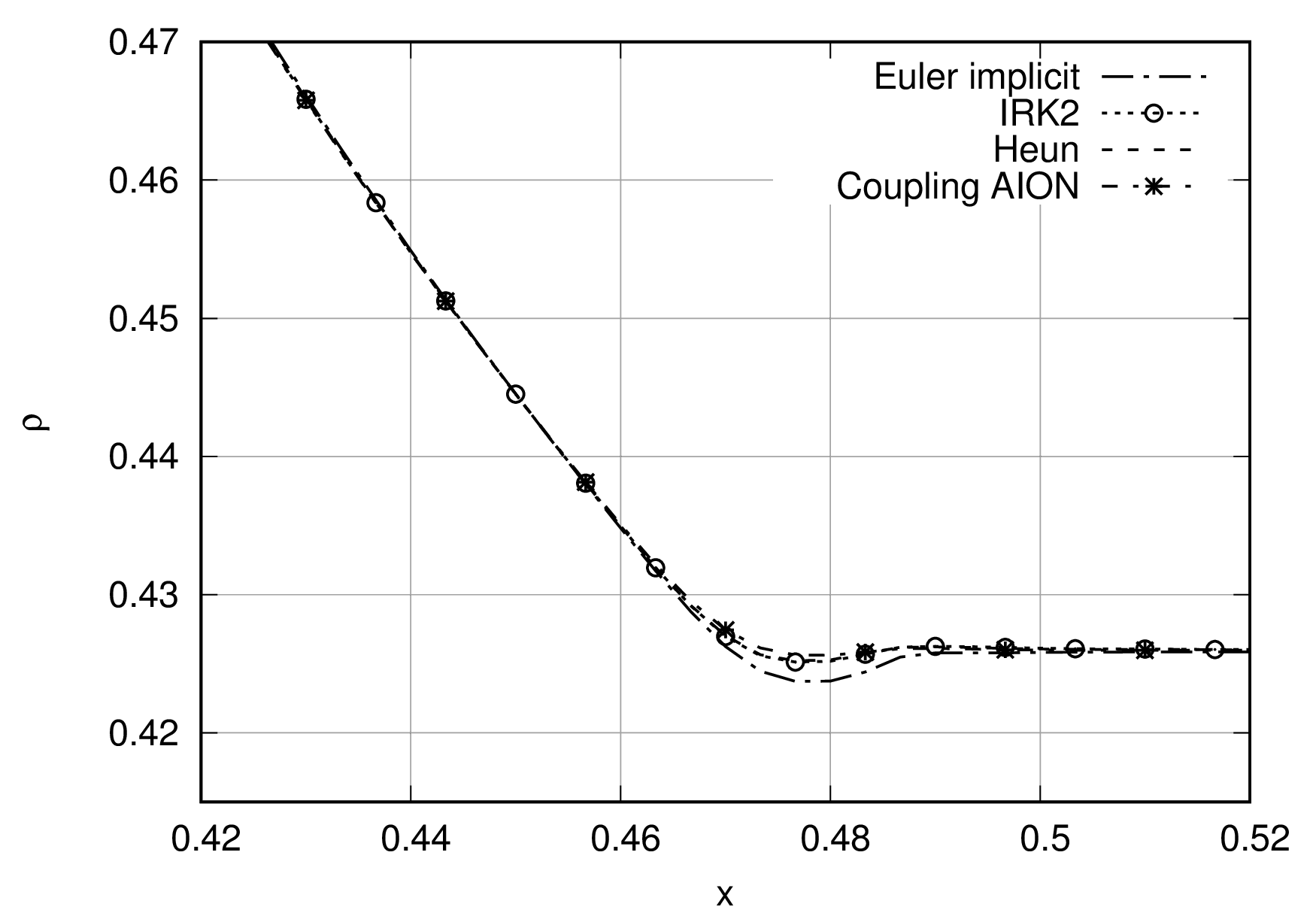}\\
\end{tabular}
\includegraphics [width=8cm]{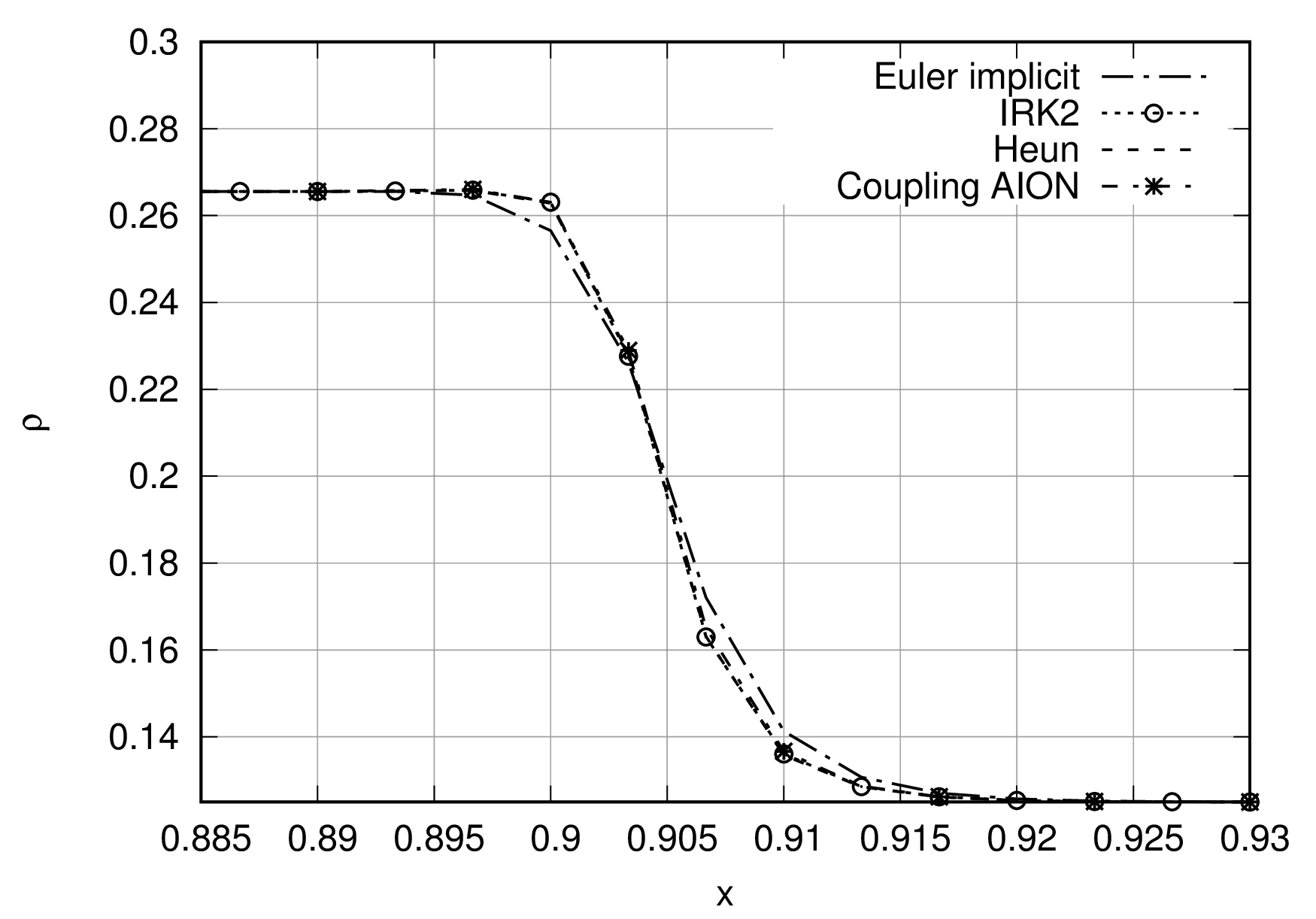}
\caption{Sod's shock tube at $\nu_{min}=0.1$. Global view of the density profile at $t=2s$ and close-up views
near the rarefaction wave and near the shock. \label{fig:rho_Sod_CFL0.1}}
\end{center}
\end{figure}

\begin{figure}[!htbp]\begin{center}
\begin{tabular}{cc}
\includegraphics[width=8cm]{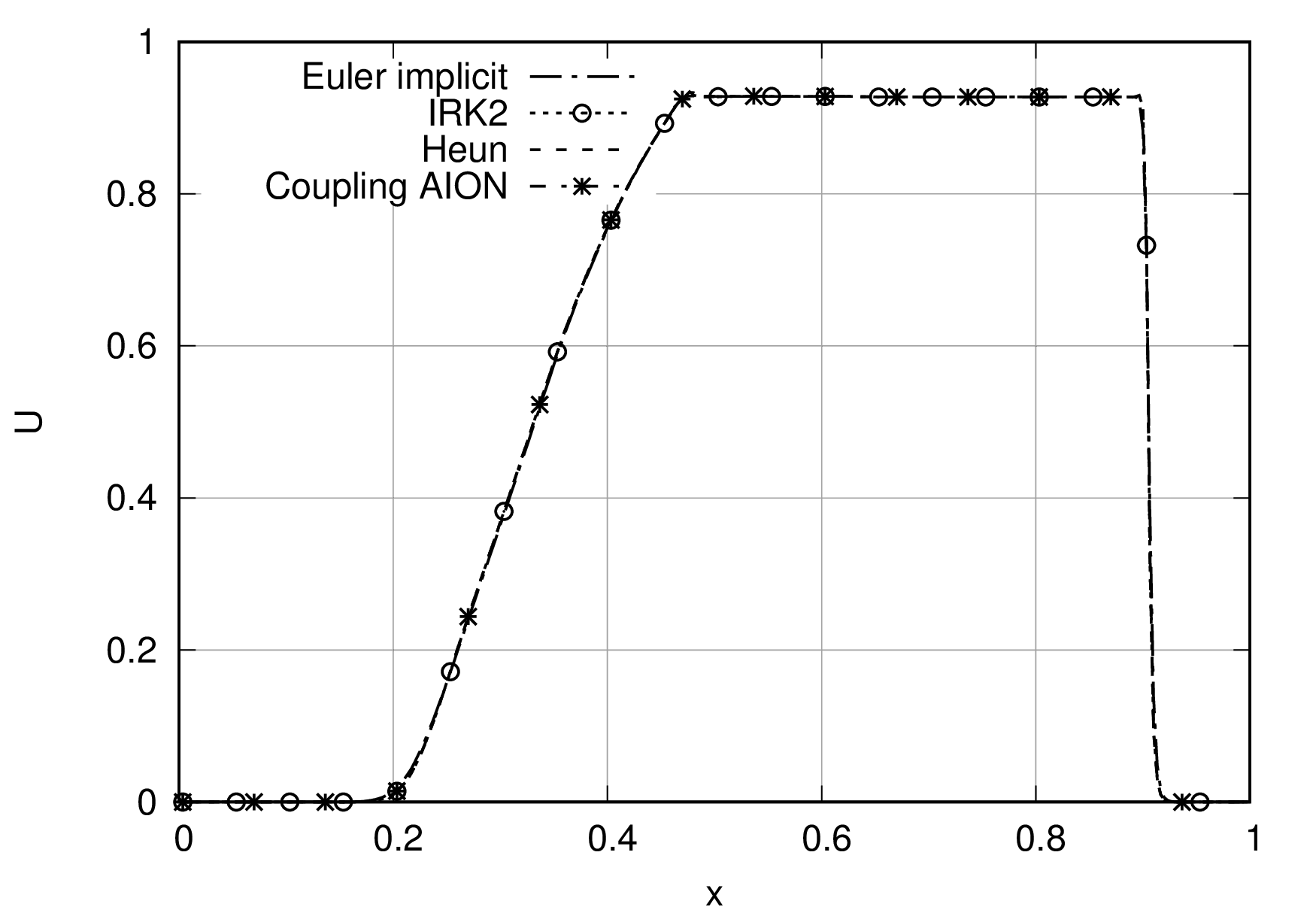} & \includegraphics [width=8cm]{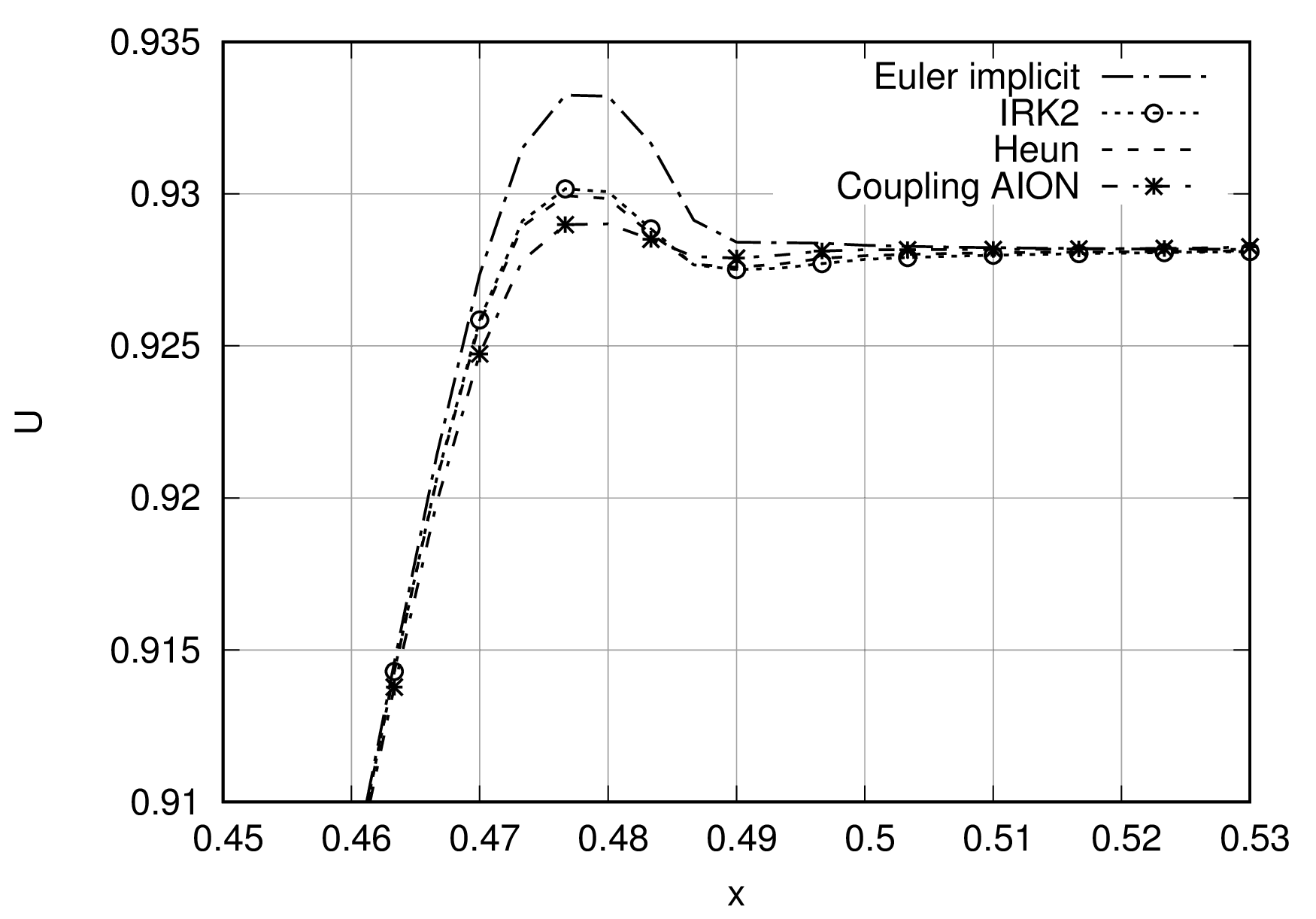}\\
\end{tabular}
\includegraphics [width=8cm]{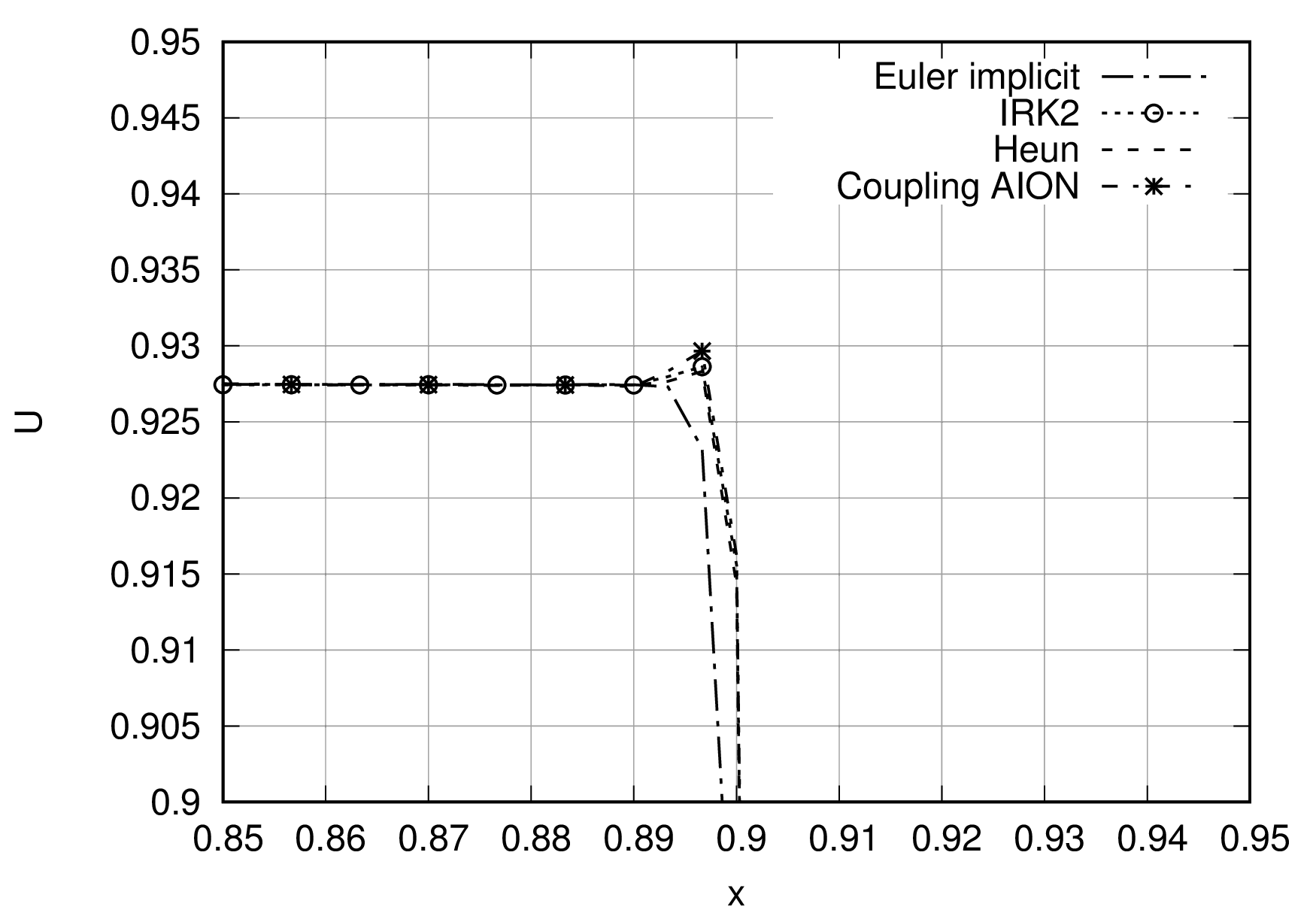}
\caption{Sod's shock tube at $\nu_{min}=0.1$. Global view of the \LM{velocity} profile at $t=2s$ and close-up views
near the rarefaction wave and near the shock. \label{fig:U_Sod_CFL0.1}}
\end{center}
\end{figure}

The Euler equations are solved using a 1-exact (second order) upwind scheme using the Roe approximate Riemann solver. 
In order to avoid spurious oscillation and to compute TVD solution, our limiting strategy is provided with the minmod slope limiter \cite{Roe_1986_ARFM}.
The computational domain is composed of $N=300$ cells located in regular parts with a uniform mesh size and
in irregular parts with non-uniform mesh size such that
\begin{equation}
\begin{aligned}
\text{for} \text{ } \text{ } \frac{N}{2}-50 \leq j \leq \frac{N}{2} \text{ }: \Delta x_j=\alpha.\Delta x_{j-1} \\
\text{for} \text{ } \text{ } \frac{N}{2}+1 \leq j \leq \frac{N}{2}+50 \text{ }: \Delta x_j=\frac{1}{\alpha}.\Delta x_{j-1} \\
\end{aligned}
\end{equation}
with $\alpha=0.98$. Then the parameter $\omega_j$ is controlled as:
\begin{equation}
\begin{aligned}
\omega_j=\frac{\Delta x_j}{\Delta x_{max}}.
\end{aligned}
\end{equation}
This definition of $\omega_j$ leads to a fully explicit time integration on the regular part of the grid, where
$\Delta x_j=\Delta x_{max}$. In the following, $\nu_j$ is defined by $\nu_j=\frac{\Delta t}{\Delta x_j}$.
For $\nu_{min}=0.1$ in the explicit part, the maximum value of $\nu$ in the implicit part reaches $0.3$.
Figures~\ref{fig:rho_Sod_CFL0.1} and~\ref{fig:U_Sod_CFL0.1} show the density and velocity profiles, with zooms in the region of the
rarefaction wave and near the contact discontinuity. The AION scheme appears to be as accurate as the second order time
accurate standard schemes and in agreement with the solution obtained using the Euler explicit scheme.

\begin{figure}[!htbp]\begin{center}
\begin{tabular}{cc}
\includegraphics[width=8cm]{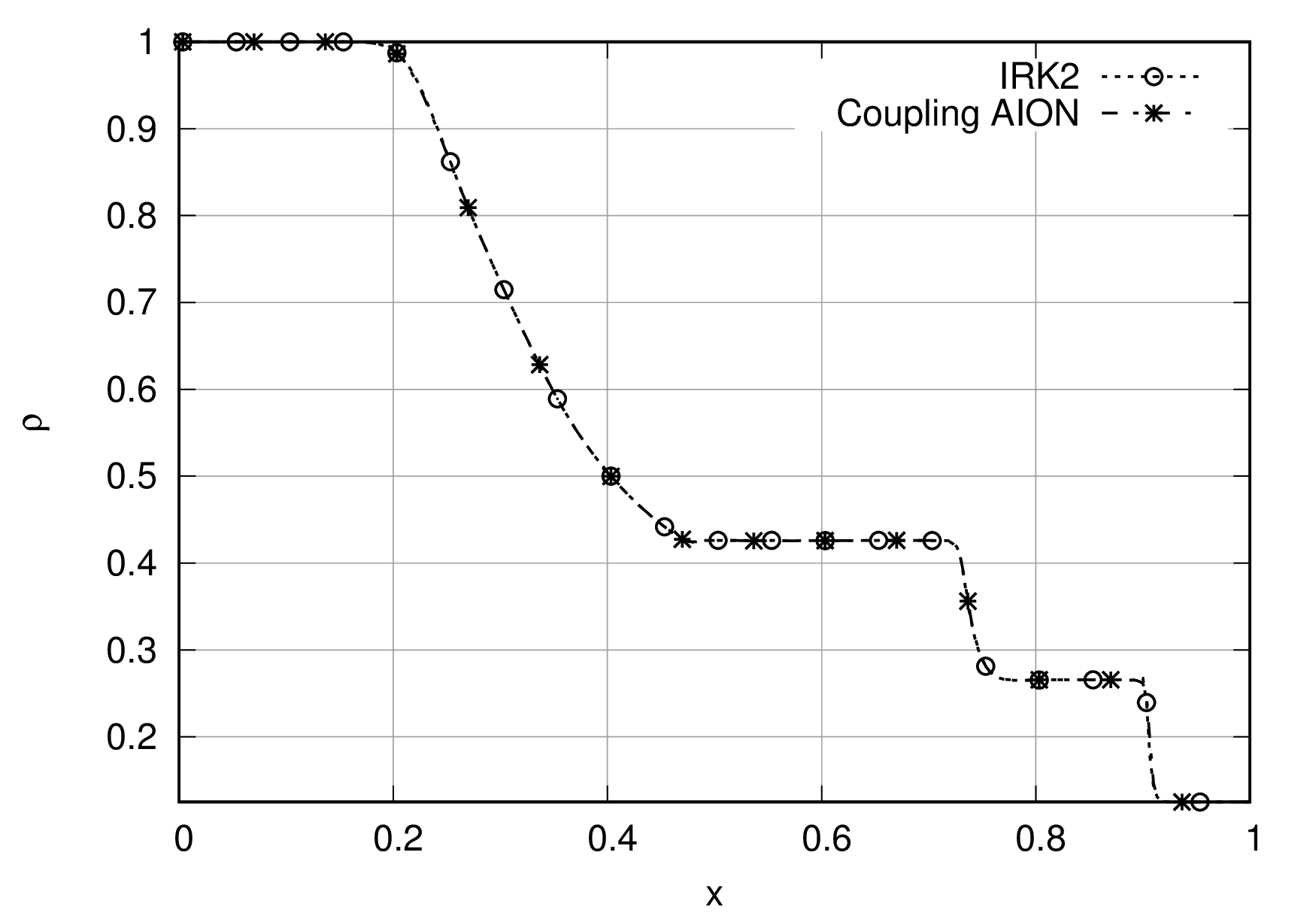} & \includegraphics [width=8cm]{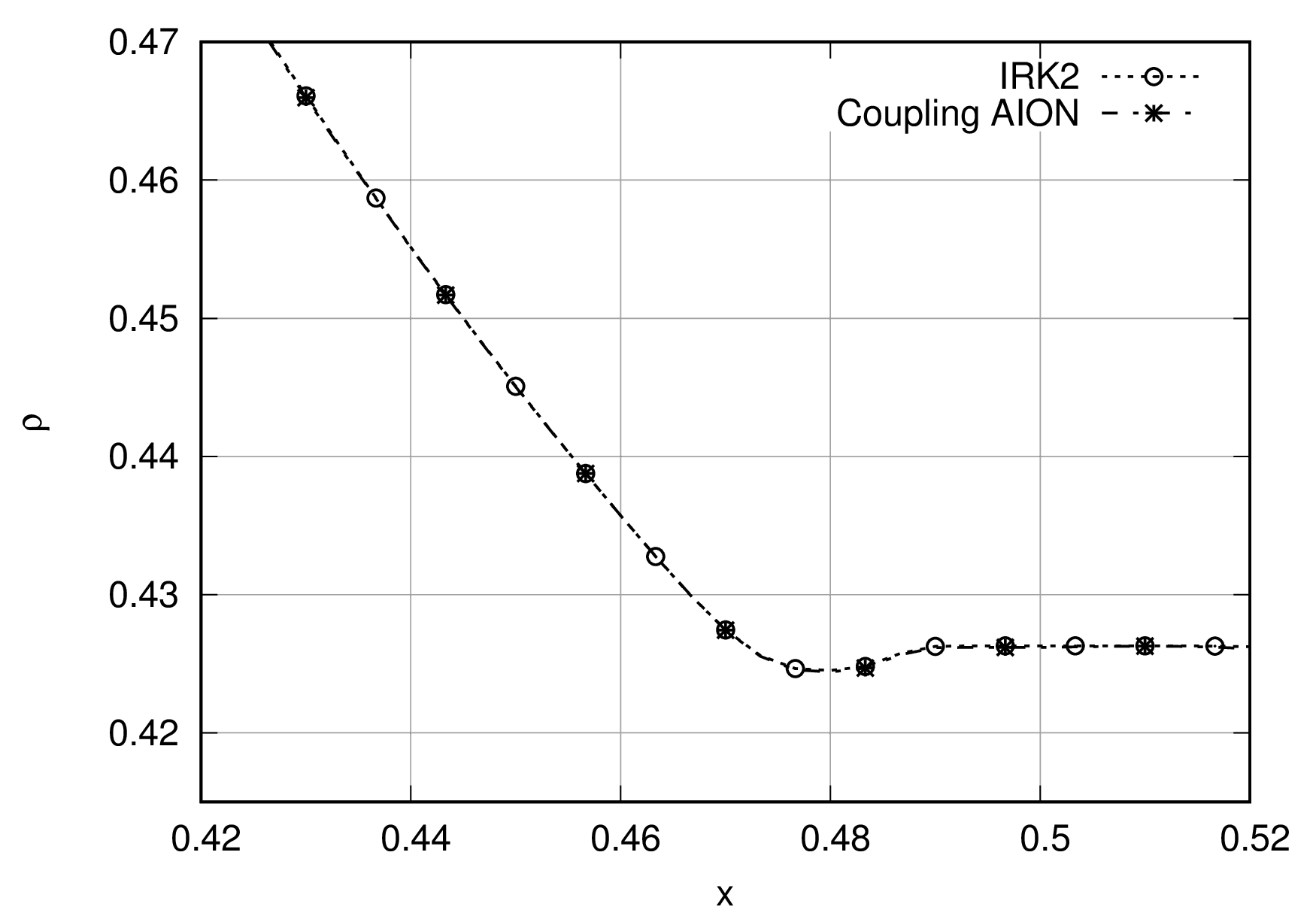}\\
\end{tabular}
\includegraphics [width=8cm]{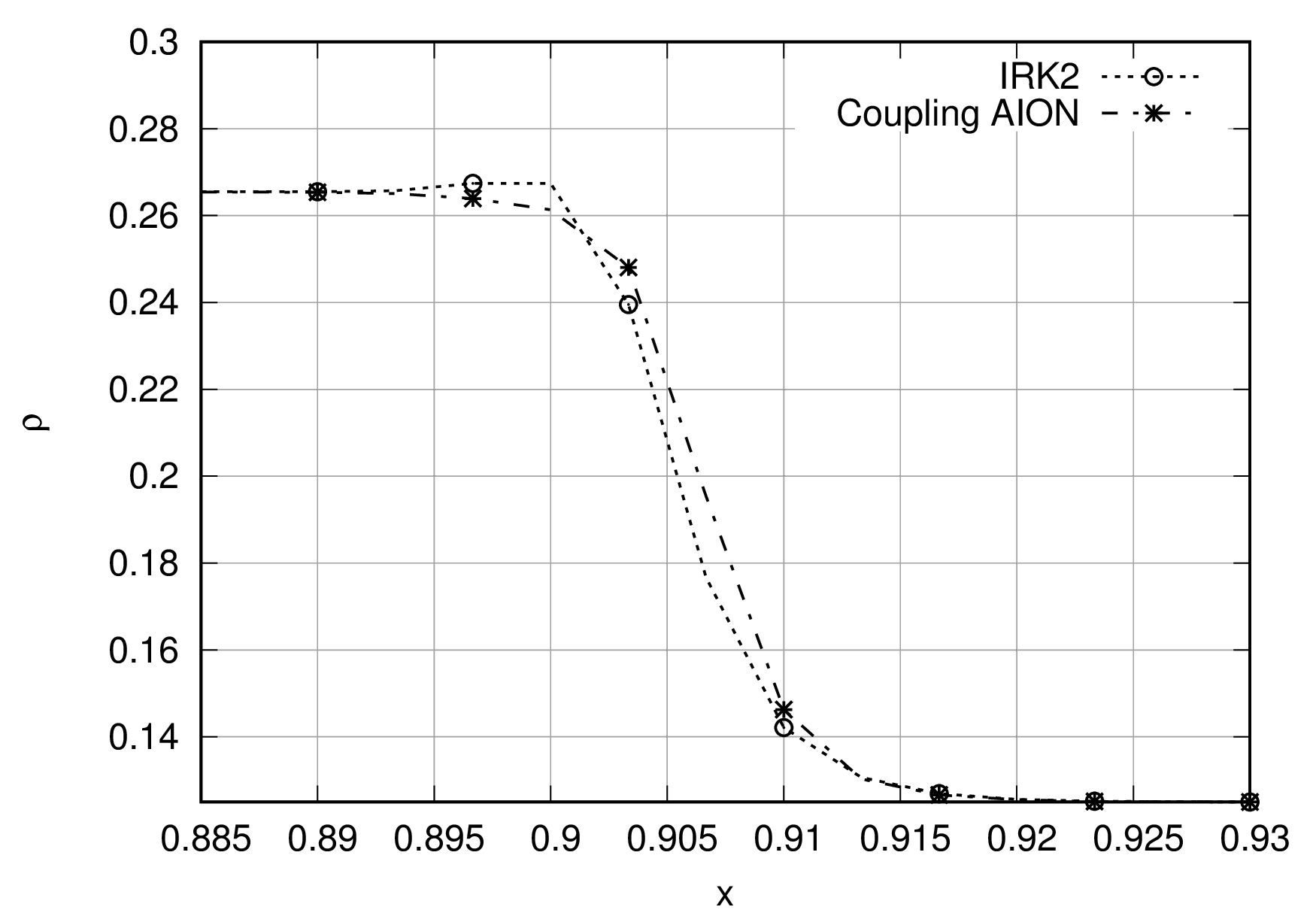}
\caption{Sod's shock tube at $\nu_{min}=0.4$. Global view of the density profile at $t=2s$ and close-up views
near the rarefaction wave and near the shock. \label{fig:rho_Sod_CFL0.4}}
\end{center}
\end{figure}

\begin{figure}[!htbp]\begin{center}
\begin{tabular}{cc}
\includegraphics[width=8cm]{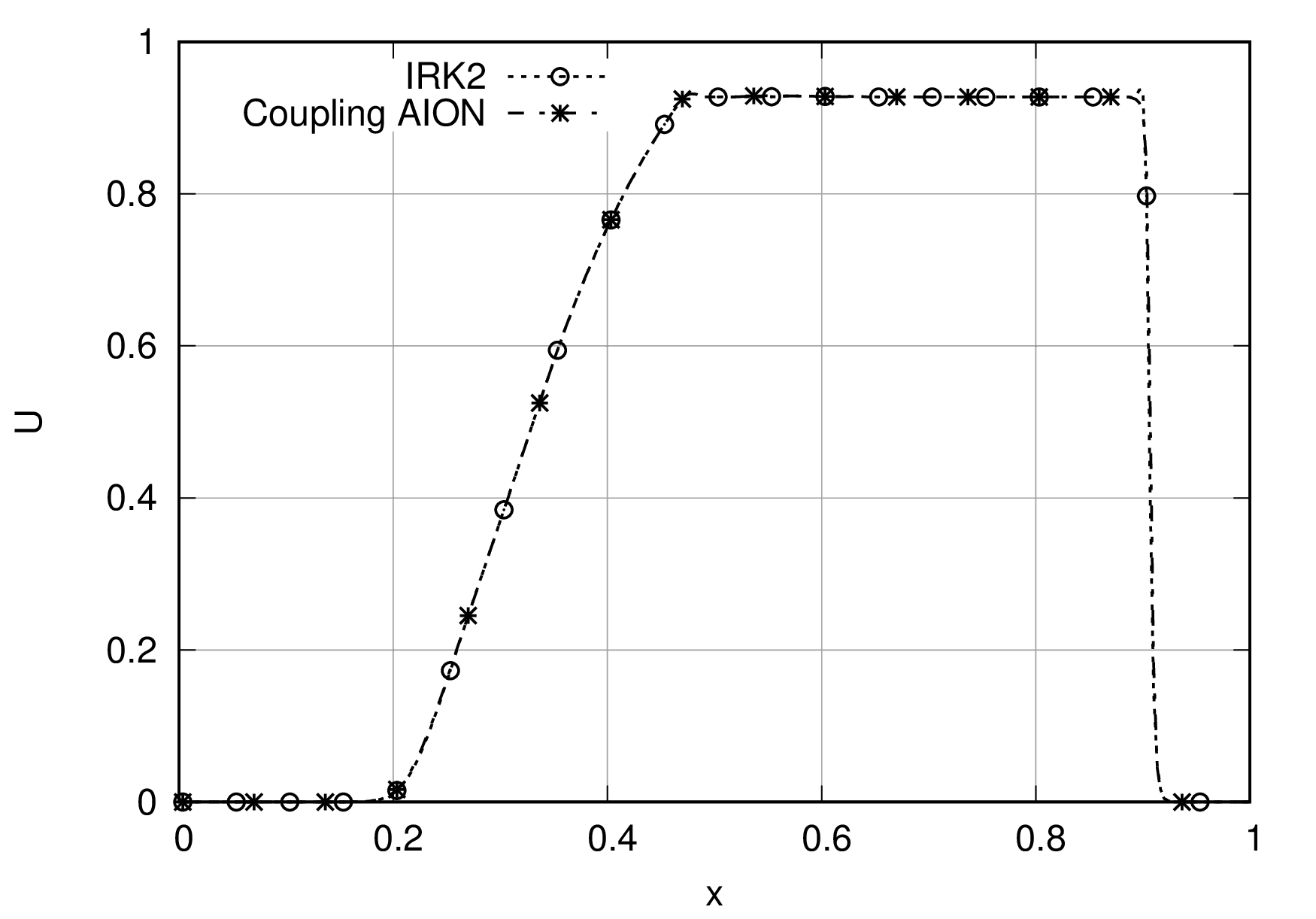} & \includegraphics [width=8cm]{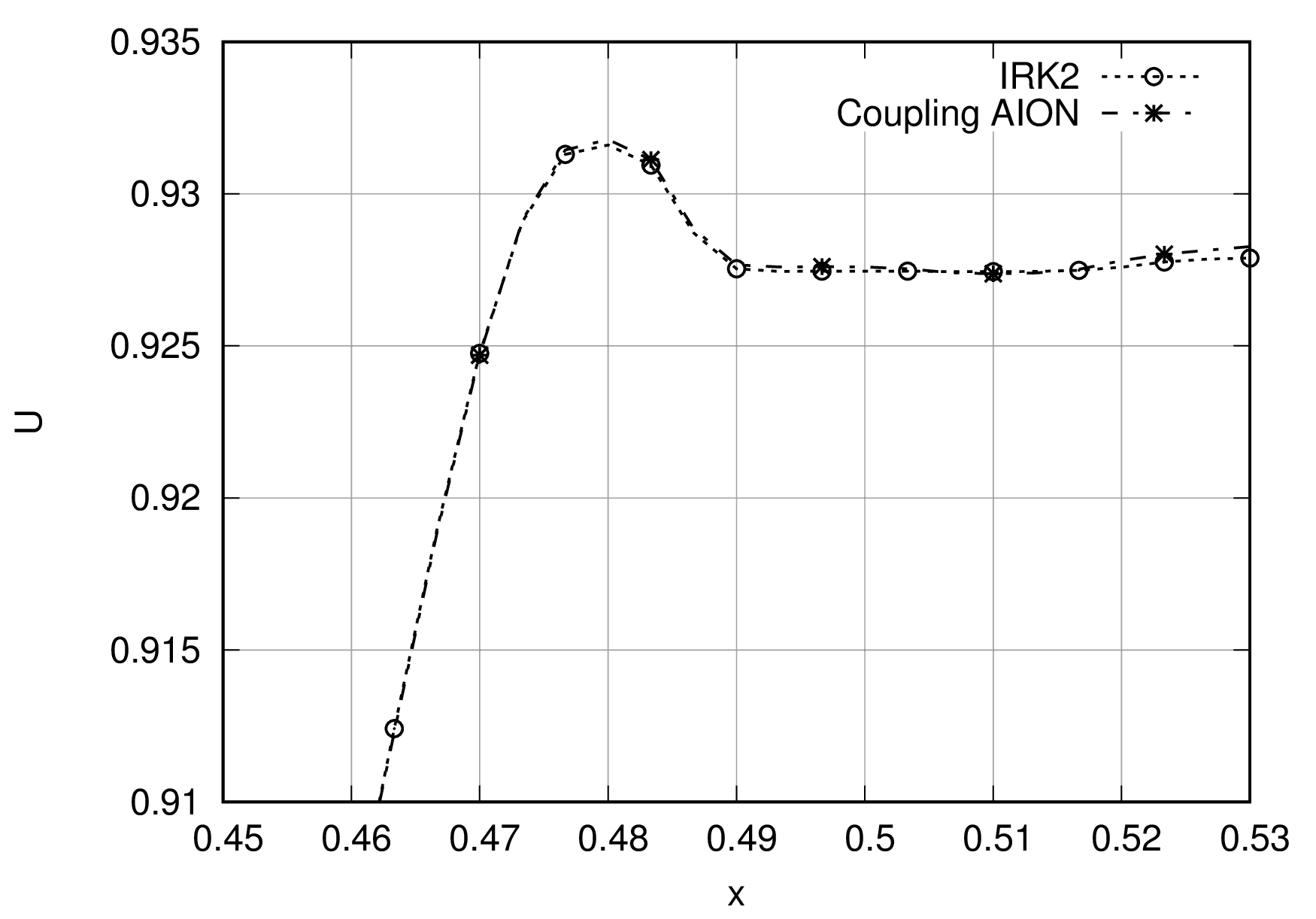}\\
\end{tabular}
\includegraphics [width=8cm]{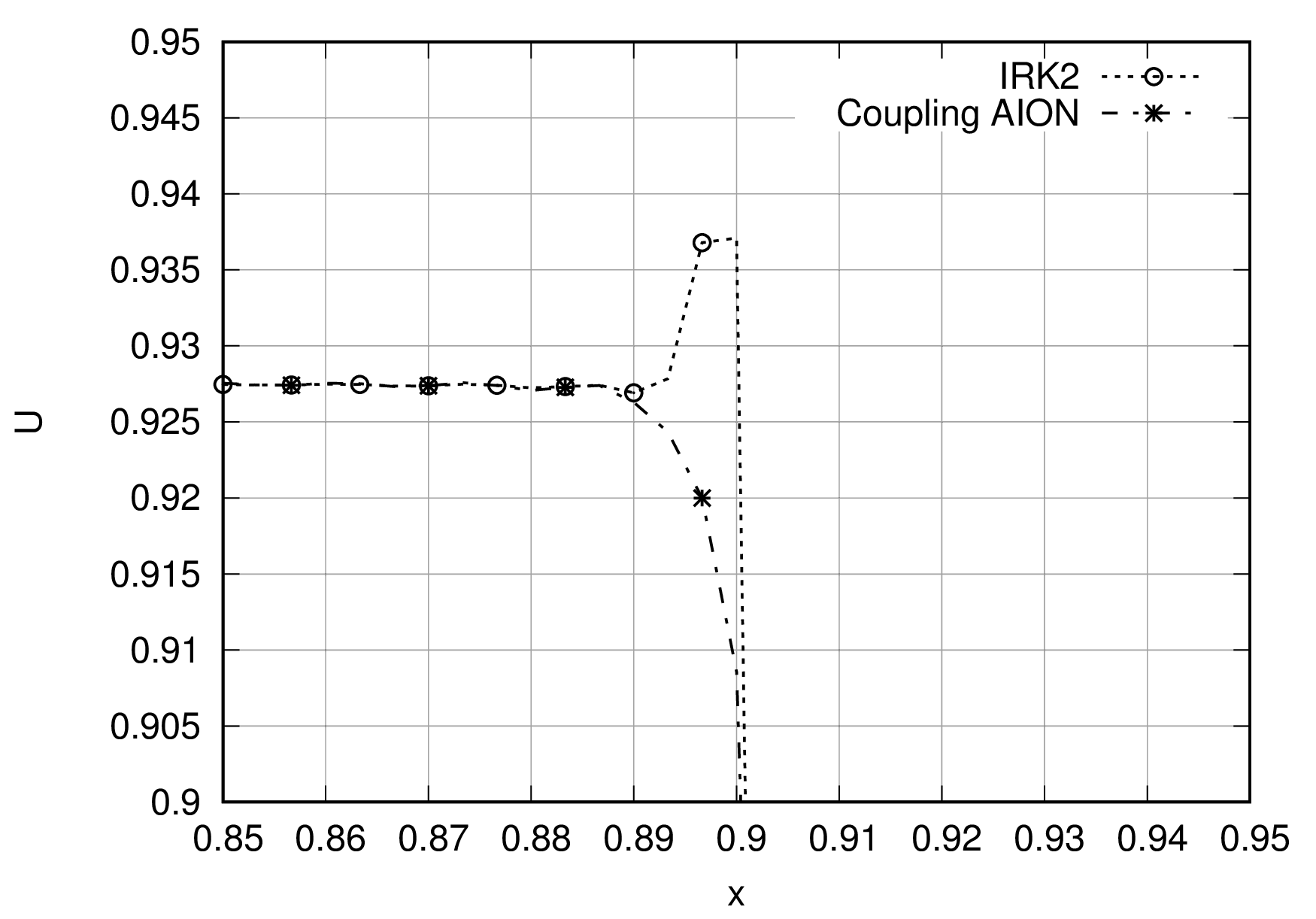}
\caption{Sod's shock tube at $\nu_{min}=0.4$. Global view of the \LM{velocity} profile at $t=2s$ and close-up views
near the rarefaction wave and near the shock. \label{fig:U_Sod_CFL0.4}}
\end{center}
\end{figure}

A second set of computations is performed for $\nu_{min}=0.4$ in the explicit part, which leads to
a maximum value of $\nu$ in the implicit part of $1.01$. The Heun' scheme
is unstable, but Figs~\ref{fig:rho_Sod_CFL0.4} and~\ref{fig:U_Sod_CFL0.4} show that the AION and IRK2 schemes are still stable.
Both solutions are very close.
Paying attention to the velocity and density near the shock, it seems that the AION scheme slightly dissipates
the overshoots given by the IRK2 implicit scheme.

From this case, it can be concluded that the reconstruction procedure to define the AION scheme is in agreement 
with the requirements to handle shock, contact discontinuity and rarefaction wave. Moreover, the AION scheme improves 
the explicit scheme with an enhanced stability and is more computationally efficient than the fully-implicit scheme. 
The next test case is dedicated to the analysis of the scheme accuracy.


\subsection{Two-dimensional linear advection of an isentropic vortex}

The advection of an isentropic vortex is a famous test case of the High Order Workshop. This problem is designed in order
to test the solver capability to preserve vorticity in an unsteady simulation and this verification case is suitable for
verifying total order of accuracy. The total error $E_{total}$ may be expressed as:
\begin{equation}
\begin{aligned}
E_{total}=Ah^p+B\Delta t^q +O(h^{p+1},\Delta t^{q+1})
\end{aligned}
\end{equation}
with $(A,B)\in \mathbb{R}^2$.
The total order of accuracy is estimated as $min(p,q)$.
For self-consistence, the test case definition is summarized below.
The computational domain is the square domain $[-\frac{L}{2}, \frac{L}{2}]^2$ ($L=0.1$)
with periodic boundary conditions. On a uniform flow of pressure $P_{\infty}$, temperature $T_{\infty}$ and
Mach number $M_{\infty}$ is superimposed an isentropic vortex defined by its characteristic radius $R$ and strength $\beta$.
The vortex is initialized in the center of the computational domain $(x_c,y_c)=(0,0)$. The initial state is defined by:
\begin{equation}
\begin{aligned}
\delta u&=- U_{\infty} \, \beta \, \frac{(y-y_c)}{R}e^{-\frac{r^2}{2}}, \\
\delta v&= U_{\infty} \, \beta \, \frac{(x-x_c)}{R}e^{-\frac{r^2}{2}},\\
\delta T&=\frac{(U_{\infty} \, \beta)^2}{2}.e^{-\frac{r}{2}}, \\
u_0&=U_{\infty}+\delta u, \\
v_0&=\delta v,
\end{aligned}
\end{equation}
with:
\begin{equation}
\begin{aligned}
r&=\frac{\sqrt{(x-x_c)^2+(y-y_c)^2}}{R}, \\
U_{\infty}&=M_{\infty} \, \sqrt{\gamma \, R_{gas} \, T_{\infty}}.
\end{aligned}
\end{equation}
with $R_{gas}=287.15 \text{  }J/kg/K$ the gas constant and a constant ratio of specific heats $\gamma $ equal to $1.4$.
The isentropic relation leads to the complete set of unknowns
 \begin{equation}
\begin{aligned}
T_0&=T_{\infty}-\delta T, \\
\rho_0&=\frac{P_{\infty}}{R_{gas} \, T_{\infty}} \, \big(\frac{T_0}{T_{\infty}}\big)^{\frac{1}{\gamma-1}}, \\
P_0&=\rho_0 \, R_{gas} \, T_0
\end{aligned}
\end{equation}
Here, the "fast vortex" test case is considered and defined by
 \begin{equation}
P_{\infty} = 10^5 \text{  } N/m^2,
T_{\infty} = 300\text{  }K,
M_{\infty} = 0.5,
\beta =\frac{1}{5},
R = 0.005 .
\end{equation}

\begin{figure}[!htbp]
\begin{center}
\includegraphics [width=10cm]{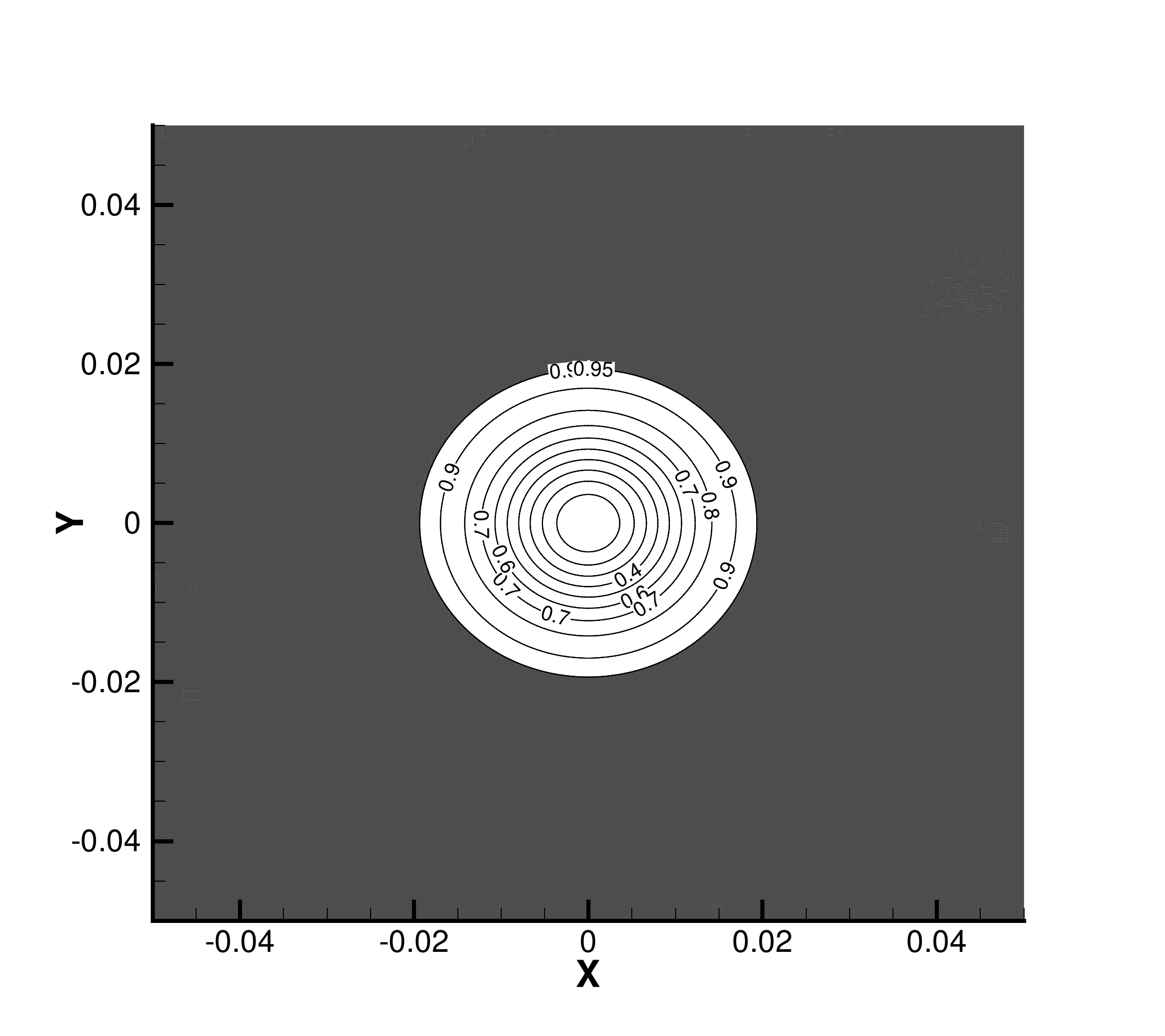}
\caption{Behaviour of $\omega_j$ for the manufactured solution ; the grey zone corresponds to $\omega_j=1$
\label{fig:omega_j_256}}
\end{center}
\end{figure}

The solution is time-marched during three rotations inside the periodic box.
The computation is performed with a Successive-Correction 2-exact formulation for the spatial scheme (order three)
and time integrated by Heun, IRK2 and AION schemes on the baseline Cartesian grids of $64^2$, $128^2$ and $256^2$
degrees of freedom (DOF). The time integration of theses \LM{computations} was performed at CFL=0.1.
For the AION scheme, it is mandatory to define the hybrid parameter $\omega_j$ ; it is configured according to following equation:
 \begin{equation}
\begin{aligned}
\omega_{j}=1-e^{-\frac{r^2}{2}} \hbox{ with }r^2=\frac{(x_j-x_c)^2+(y_j-y_c)^2}{R^2}.
\end{aligned}
\end{equation}
Fig.~\ref{fig:omega_j_256} represents the distribution of $\omega_j$ on the $256^2$ mesh.

\begin{figure}[!ht]
\begin{center}
\includegraphics [width=10cm]{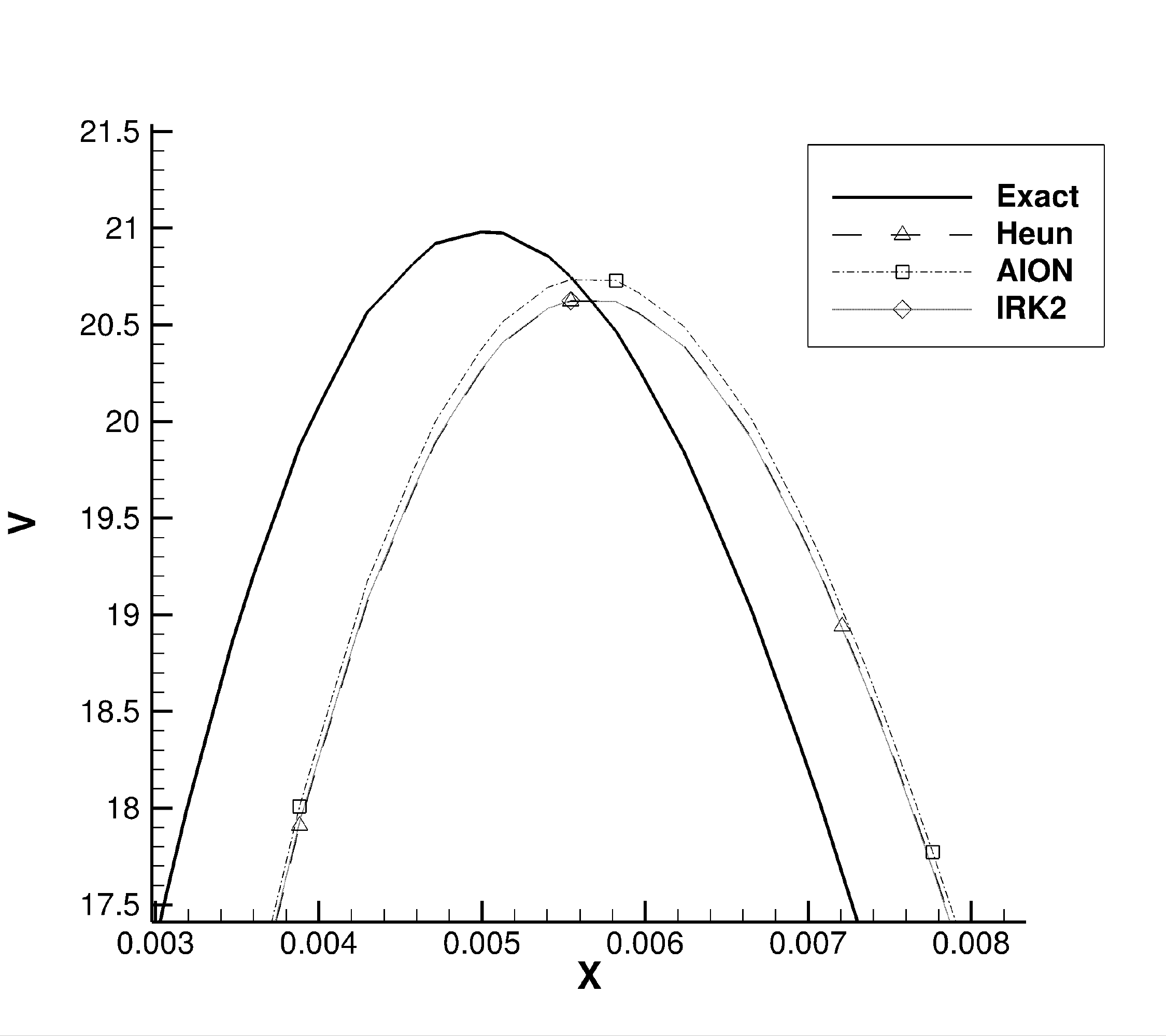}
\caption{Zoom on the velocity field (CFL=0.1)
\label{fig:compar_256_V_zoom}}
\end{center}
\end{figure}
\begin{figure}[!ht]
\begin{center}
\includegraphics [width=10cm]{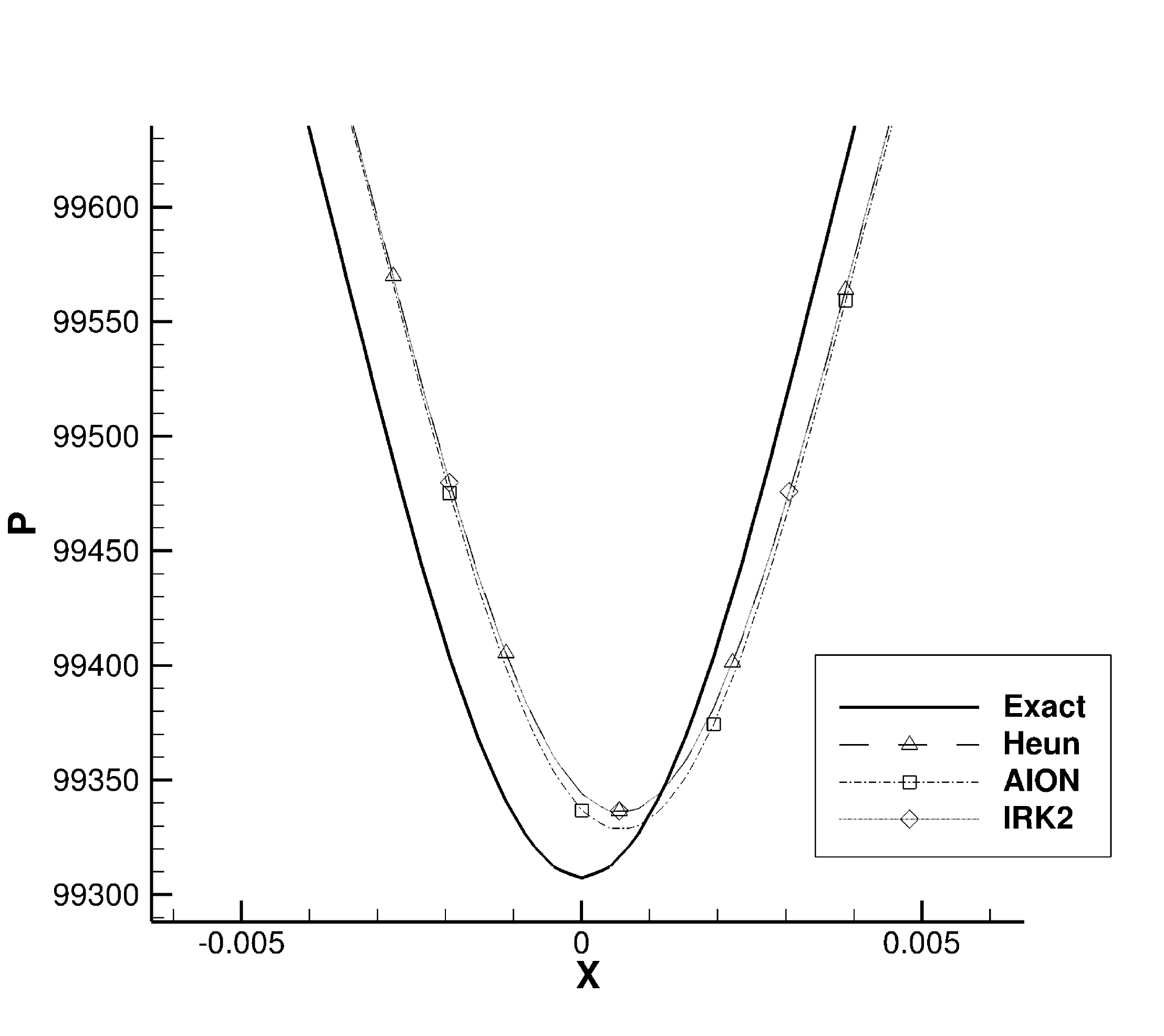}
\caption{Zoom on the pressure field (CFL=0.1)
\label{fig:compar_256_P_zoom}}
\end{center}
\end{figure}

According to the pressure and velocity field obtained with Heun, AION and IRK2 time integrator on grid $256^2$,
it appears that time integrators have slightly same properties of dissipation and dispersion.
The AION scheme tends to less dissipate than other second-order time integrator (according to Figs.~\ref{fig:compar_256_V_zoom} and ~\ref{fig:compar_256_P_zoom}).
\begin{figure}[!ht]
\begin{center}
\includegraphics [width=10cm]{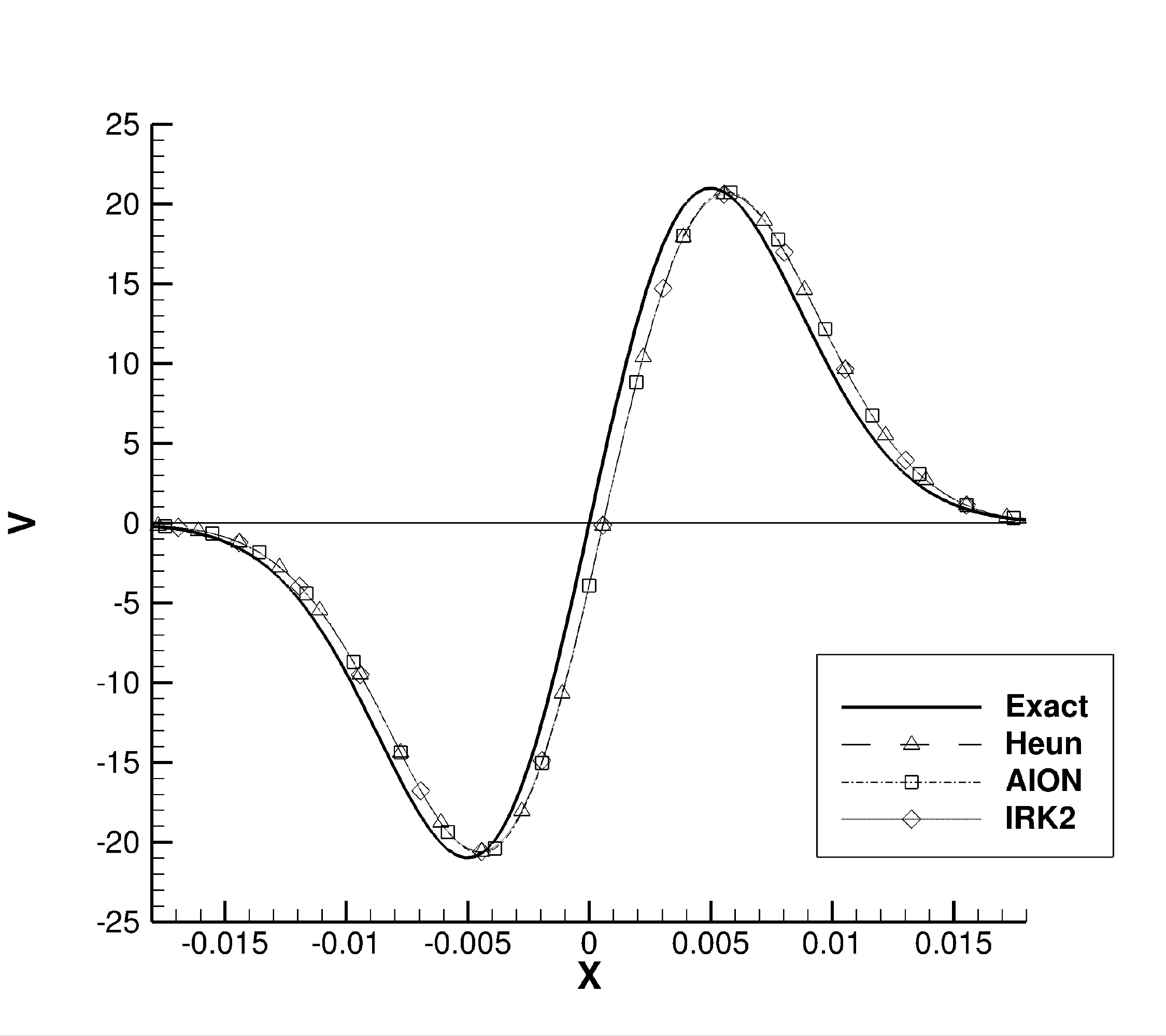}
\caption{Velocity field (CFL=0.1)
\label{fig:compar_256_V}}
\end{center}
\end{figure}
\begin{figure}[!ht]
\begin{center}
\includegraphics [width=10cm]{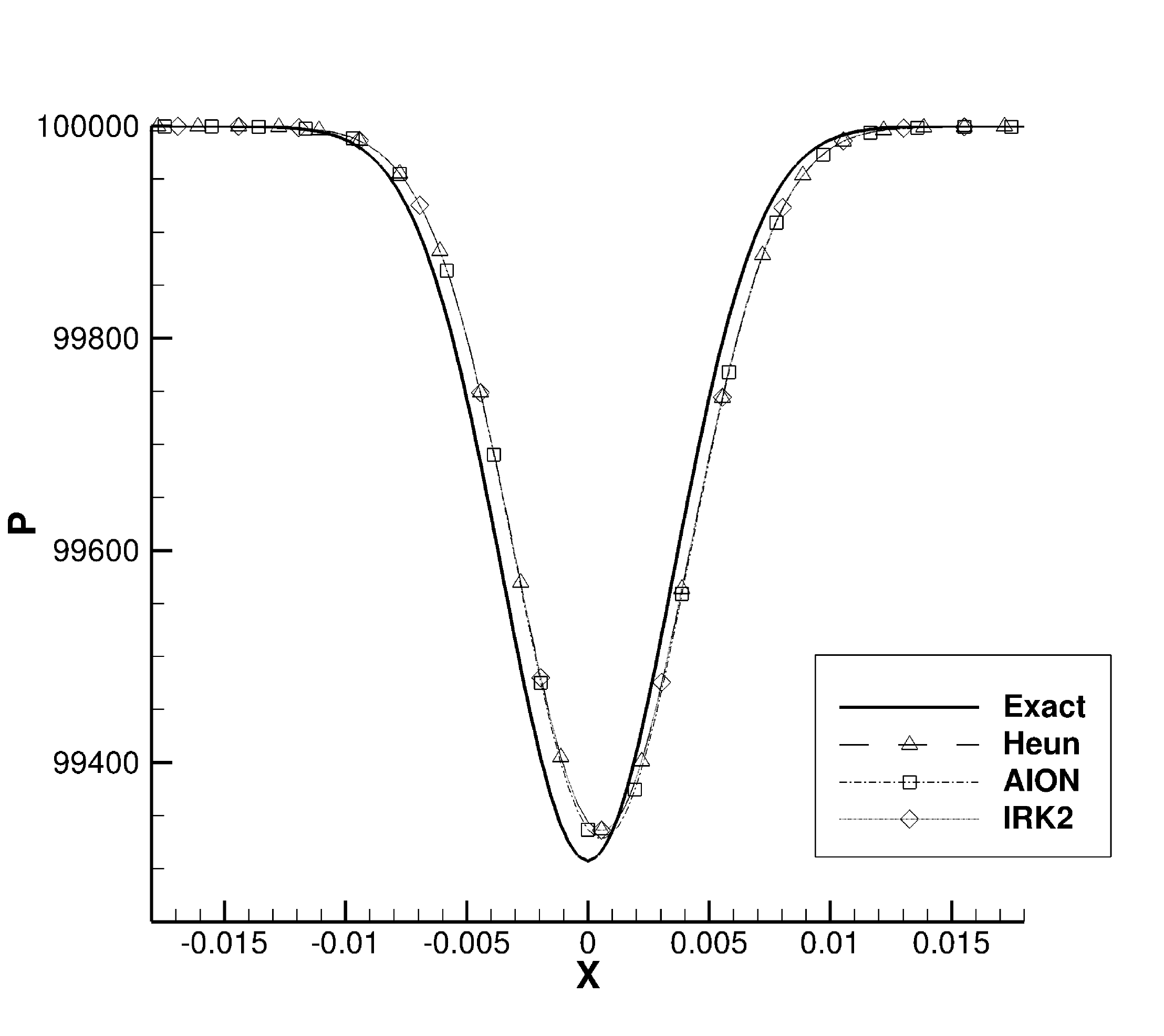}
\caption{Pressure field (CFL=0.1)
\label{fig:compar_256_P}}
\end{center}
\end{figure}
This feature is quite different regardless the ones found on one-dimensional test case with second-order space reconstruction,
where the AION was more dissipative than other time integrators. Here the third-order accurate space reconstruction tends to
improve the space-time spectral characteristics of the computation, time integrated with AION scheme.
According to the computation of $L^2$-norm of the error using the two velocity-vector components (see Fig.~\ref{fig:compar_256_V_zoom}),
it appears, as expected, that the AION time integrator conserves second-order space-time accuracy (see Tab.~\ref{tab:order_tab_FLUSEPA}). 
Nevertheless, according to the slope of $log(E_{total})$, it seems that the time integrator has a significant impact on the error computation 
(see Tabs.~\ref{tab:order_tab_FLUSEPA_1}-\ref{tab:order_tab_FLUSEPA_2}). Indeed the value of the slope is quite different between $log(h)=[-2.4,-2.1]$ 
and $log(h)=[-2.1,-1.8]$, particularly in case of the Heun scheme.   
The computational cost of the AION and the IRK2 time integrator at same CFL is also compared in Tab.~\ref{tab:CPU_time}.
The ratio between cells with $\omega_j<1$ and cells with $\omega_j=1$ (I/E cells) is equal to $18\%$ for all grids.
It appears that the AION scheme reaches an overall cost that is about nearly $50\%$ lower than the pure implicit IRK2 scheme for grid
with $256^2$ DOFs at same CFL.
\begin{table}[!htbp]
\caption{Elapsed CPU time to reach 3 time periods with AION and IRK2 time integrators \label{tab:CPU_time}.}
\begin{center}
\begin{tabular}[b]{|l|c|c|c|c|}
\hline
 DOFs & $64^2$ & $128^2$ & $256^2$ \\
\hline
I/E Cells & $18\%$ & $18\%$ & $18\%$ \\
\hline
AION/IRK2 & $0.75$& $0.66$ & $0.52$ \\
\hline
\end{tabular}
\end{center}
\end{table}

\begin{figure}[!ht]
\begin{center}
\includegraphics [width=10cm]{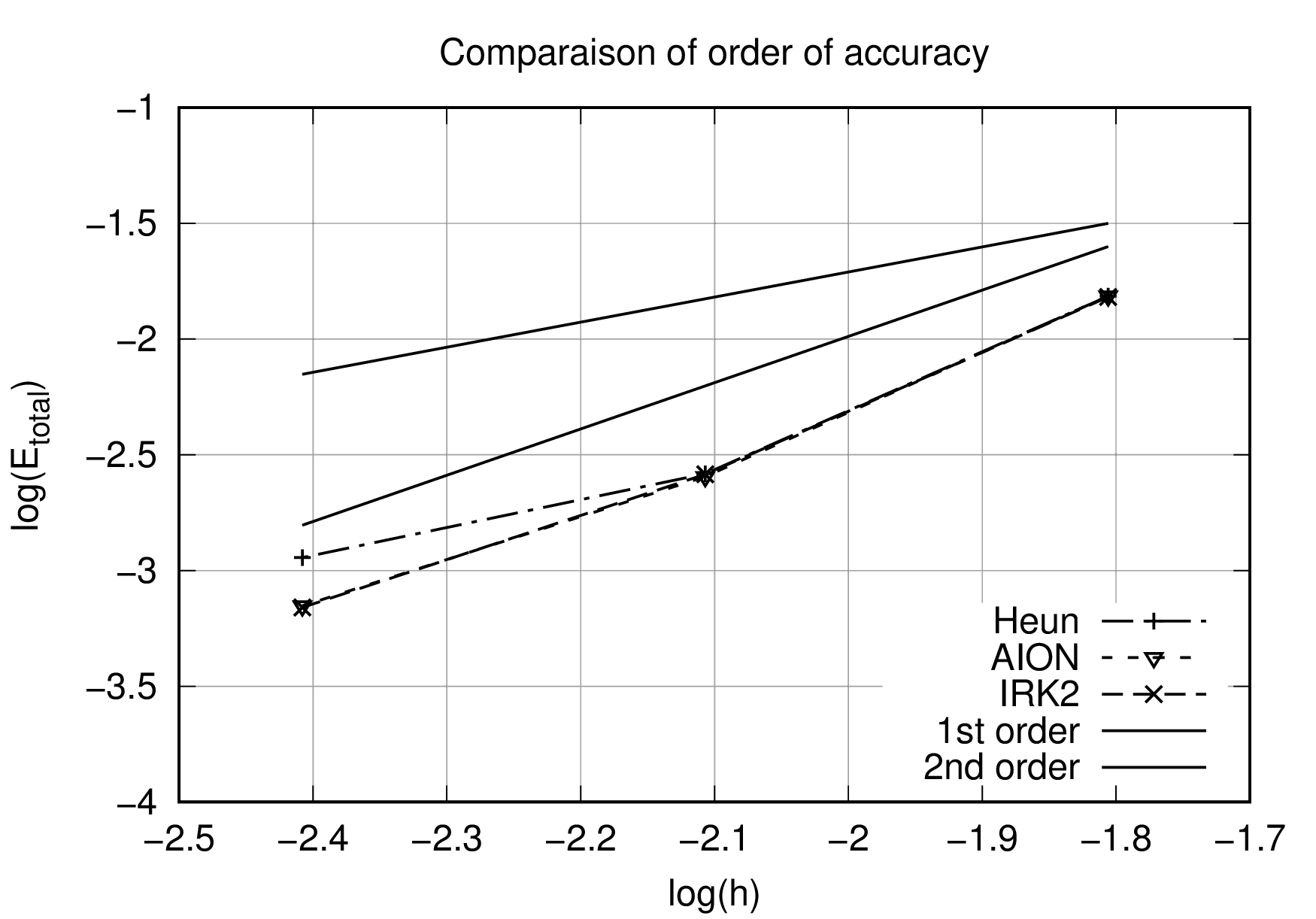}
\caption{Order of accuracy
\label{fig:accuracy_FLUSEPA}}
\end{center}
\end{figure}
\begin{table}[!ht]
\begin{center}
\begin{tabular}{|l|c|}
  \hline
  Time integrator & $log(E_{total})$ slope \\
  \hline
  Heun & 1.87  \\
  AION & 2.22  \\
  IRK2 & 2.22  \\
  \hline
\end{tabular}
\caption{Total error slopes \label{tab:order_tab_FLUSEPA} }
\end{center}
\end{table}

\begin{table}[!ht]
\begin{center}
\begin{tabular}{|l|c|}
  \hline
  Time integrator & $log(E_{total})$ slope \\
  \hline
  Heun & 1.2 \\
  AION & 1.84  \\
  IRK2 & 1.91  \\
  \hline
\end{tabular}
\caption{Total error slopes for $log(h)=[-2.4,-2.1]$ \label{tab:order_tab_FLUSEPA_1} }
\end{center}
\end{table}

\begin{figure}[!ht]
\begin{center}
\includegraphics [width=10cm]{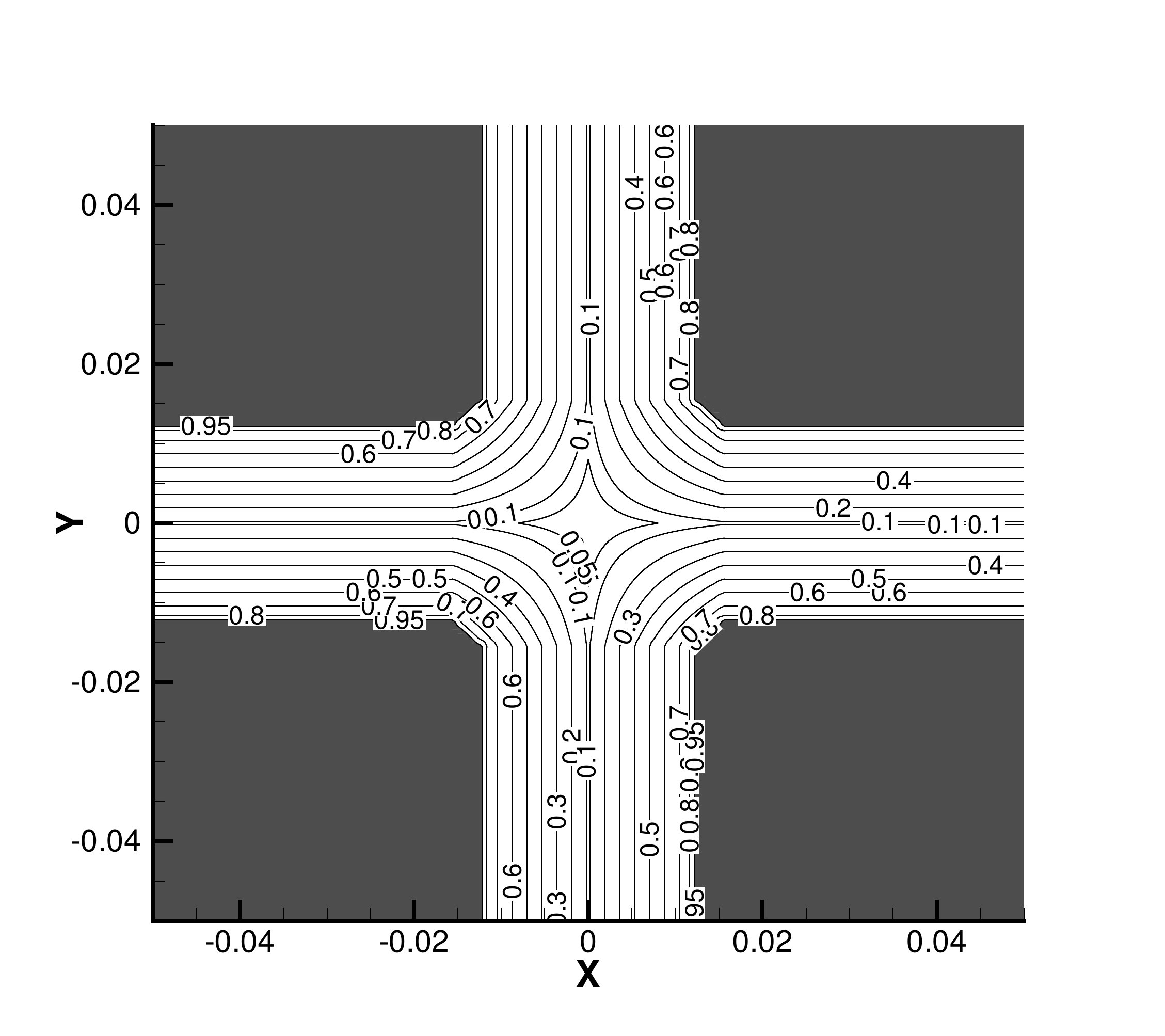}
\caption{Value of $\omega_j$
\label{fig:omega_j_260}}
\end{center}
\end{figure}

\begin{table}[!ht]
\begin{center}
\begin{tabular}{|l|c|}
  \hline
  Time integrator & $log(E_{total})$ slope \\
  \hline
  Heun & 2.54  \\
  AION & 2.59  \\
  IRK2 & 2.54  \\
  \hline
\end{tabular}
\caption{Total error slopes for $log(h)=[-2.1,-1.8]$ \label{tab:order_tab_FLUSEPA_2} }
\end{center}
\end{table}
The convection of a vortex on an irregular grid of $260^2$ DOFs is also performed.
The time integration is performed according to CFL condition of the biggest cells of the domain such as
\begin{equation}
\begin{aligned}
\Delta t =CFL \frac{\underset{j}{\text{max}}(h)}{\| \vec{v_j}\| },
\end{aligned}
\end{equation}
In this irregular domain, the size ratio between the largest and the smallest cell is equal to $11$.
Hence, the constant time step of computation will be determined according to the CFL number of the 
largest cell, which leads to the CFL number $11$ time higher for the smallest cells than for the larger ones.
The computation is performed at several CFL values, during one time period, in order to compare stability 
properties of the several time integrators. The parameter $\omega_j$ is configured such as
\begin{equation}
\begin{aligned}
\omega_j =\frac{|\Omega_j|}{\underset{j}{\text{max}}(|\Omega_j|)},
\end{aligned}
\end{equation}
The parameter $\omega_j$ is defined according to the size of cells in the domain (see Fig.~\ref{fig:omega_j_260} with grey zone corresponding to $\omega_j=1$).
In this case the proportion of cells with $\omega_j<1$ corresponding to hybrid and implicit cells is equal to $73\%$.
The final computational times for the several time integrator and CFL numbers is provided in Tab.~\ref{tab:CPU_time_260}.
The values are normalized with respect to computational cost of the Heun time integrator at CFL=0.1.
According to Tab.~\ref{tab:CPU_time_260}, as expected, the explicit time integrator is unstable for high CFL number due to small cells.
Then, the AION scheme, as expected, tends to have better stability properties than full Heun time integrator, due to implicit time treatment in the smallest cells.
The computational gain compared to full implicit computation is equal to $7\%$, due to high percentage of hybrid and implicit cells.
\begin{table}[!htbp]
\caption{Elapsed CPU time to reach 1 time \LM{period} with AION and IRK2 time integrators on a $260^2$ irregular mesh \label{tab:CPU_time_260}.}
\begin{center}
\begin{tabular}[b]{|l|c|c|c|}

  \hline
   &  CFL=0.1 & CFL=0.4 & CFL=0.9\\
  \hline
  Heun & 1 &  $\times$ & $\times$ \\
    \hline
  AION & 3.405  &  0.897 & 0.412\\
    \hline
  IRK2 & 3.64 &  0.948&0.434 \\
  \hline
\end{tabular}
\end{center}
\end{table}
In the following a focus is made on testing the hybrid treatment of the gradient for the diffusion scheme.

\subsection{Three-dimensional Taylor Green Vortex}

A three-dimensional laminar flow computation is now performed to assess the performance of 
the AION integration scheme for a viscous test case. The proposed test case is the Taylor Green vortex 
at Reynolds number ($Re=\frac{\rho_{\infty}U_{\infty}L}{\mu}$) equal to $1600$ and Mach number equal to 0.1.
The initial solution is defined inside the periodic cubic domain $[-\pi L,\pi L]^3$ ($L=1$):
\begin{equation}
\begin{aligned}
u&=U_{\infty}sin\big( \frac{x}{L}\big)cos\big( \frac{y}{L}\big)sin\big( \frac{z}{L}\big), \\
v&=-U_{\infty}cos\big( \frac{x}{L}\big)sin\big( \frac{y}{L}\big)cos\big( \frac{z}{L}\big),\\
w&= 0, \\
p&=P_{\infty}+\frac{\rho_{\infty}U^2_{\infty}}{16}\bigg( cos\big( \frac{2x}{L}\big)+cos\big( \frac{2y}{L}\big) \bigg) \bigg(cos\big( \frac{2z}{L}\big) +2\bigg)
\end{aligned}
\end{equation}
This test case is defined in order to observe a kind of vortex cascade, as encountered 
in turbulent flow computations. For the computation, the fluid is assumed to be a compressible perfect 
gas with $\gamma=1.4$ and the Prandtl number ($Pr=\frac{\mu.c_p}{\kappa}$) is equal to $0.71$,
with $c_p$ and $\kappa$, respectively, the heat capacities at constant pressure and the heat conductivity. 
The initial temperature is kept constant $T_{\infty}=\frac{P_{\infty}}{R_{gas}\rho_{\infty}}$ with 
$R_{gas}$ the perfect gas constant. From the local pressure and temperature is deduced 
the density $\rho=\frac{p}{R_{gas}T_{\infty}}$. 

\begin{figure}[!ht]
\begin{center}
\includegraphics [width=10cm]{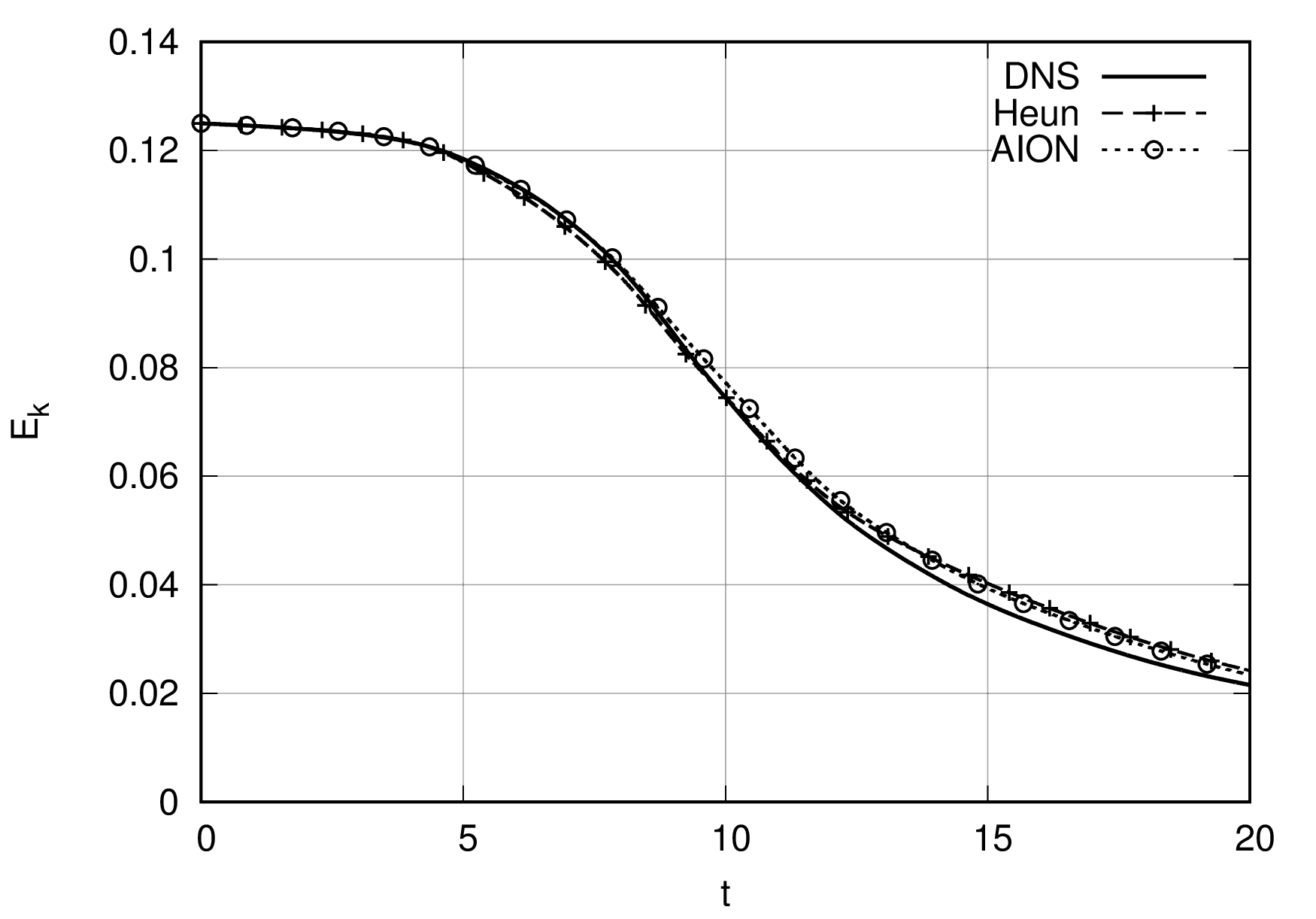}
\caption{The temporal evolution of the kinetic energy
\label{fig:Ec_256}}
\end{center}
\end{figure}

\begin{figure}[!ht]
\begin{center}
\includegraphics [width=10cm]{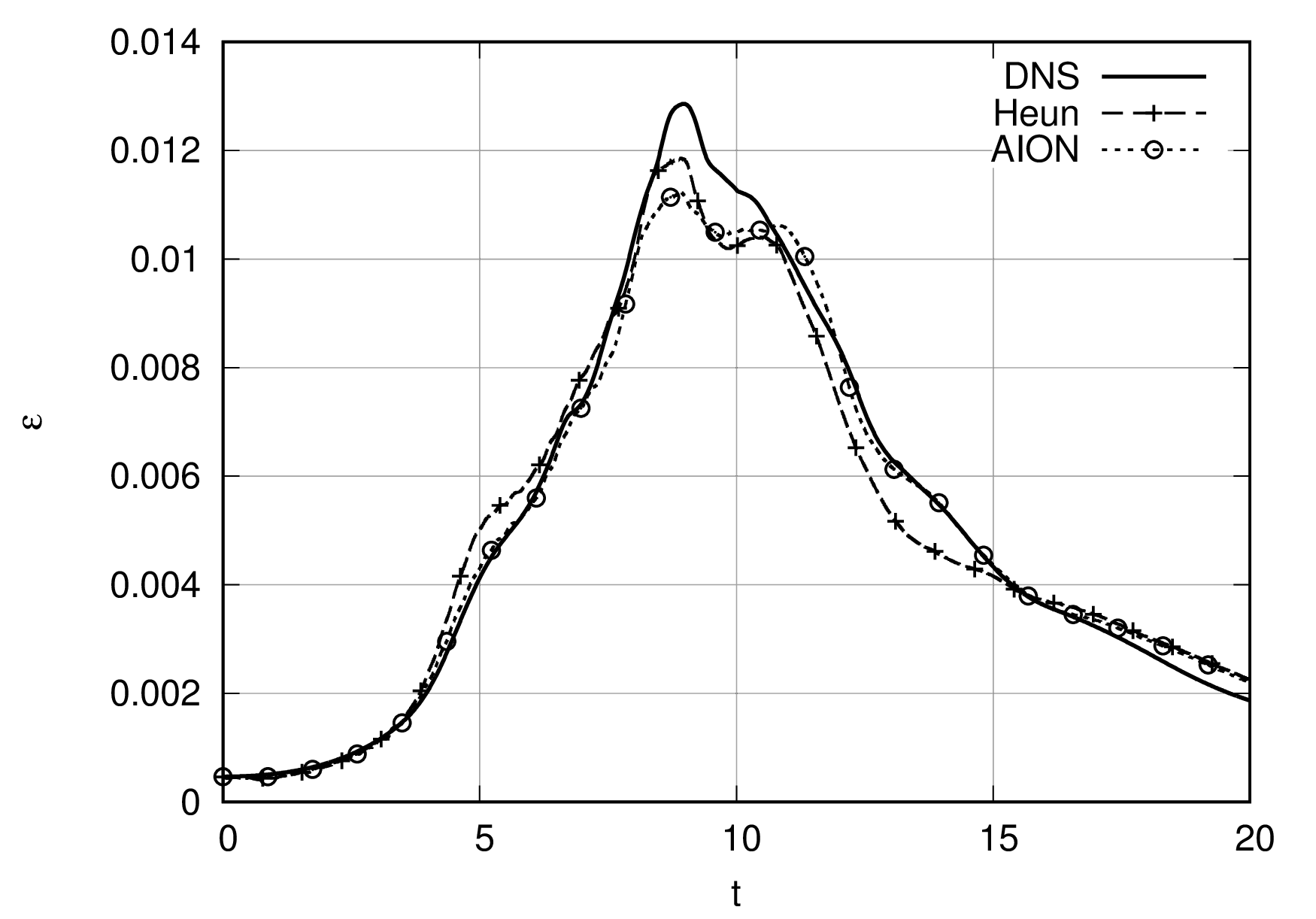}
\caption{The temporal evolution of the kinetic energy dissipation rate
\label{fig:dEc_256}}
\end{center}
\end{figure}

\begin{figure}[!ht]
\begin{center}
\includegraphics [width=10cm]{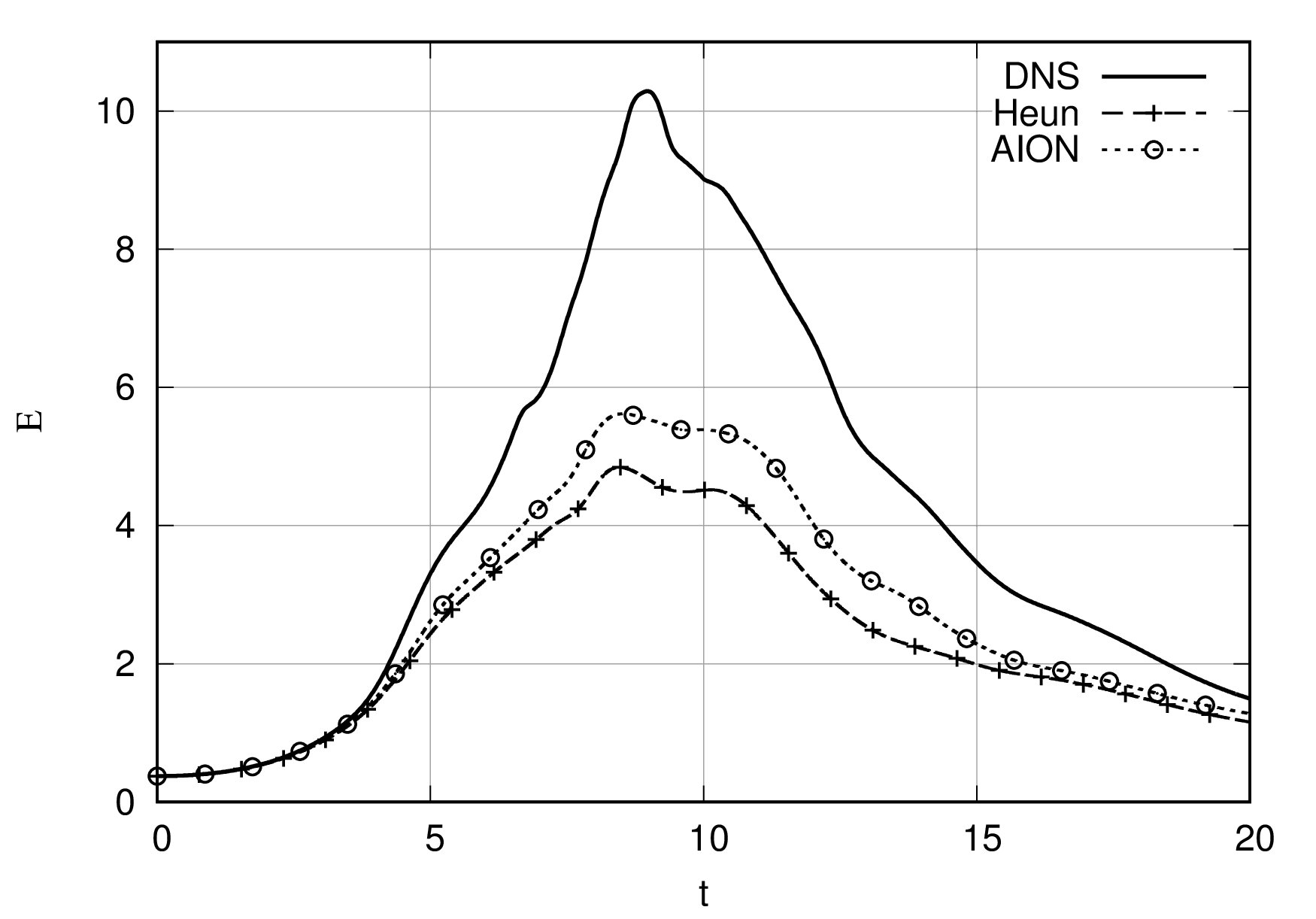}
\caption{The temporal evolution of the enstrophy
\label{fig:Ens_256}}
\end{center}
\end{figure}

The simulation is performed until the physical time of $t=20 t_{c}$ with $t_{c}=\frac{L}{U_{\infty}}$ the 
characteristic convective time is attained. The computation is performed with a 2-exact formulation for 
the spatial discretisation (three order accurate spatially) and time-integrated by Heun and AION schemes on 
a $256^3$ grid at $CFL=0.9$. The following outputs are performed:
\begin{itemize}
\item The temporal evolution of the kinetic energy integrated over the whole domain:
\begin{equation}
\begin{aligned}
E_k=\frac{1}{\rho_{\infty}\Omega}\int_{\Omega}\rho \frac{U \cdot U}{2}d\Omega.
\end{aligned}
\end{equation}

\item The temporal evolution of the kinetic energy dissipation rate:
\begin{equation}
\begin{aligned}
\varepsilon=-\frac{dE_k}{dt}.
\end{aligned}
\end{equation}

\item The temporal evolution of the kinetic energy dissipation rate on the whole domain:
\begin{equation}
\begin{aligned}
E=\frac{1}{\rho_{\infty}\Omega}\int_{\Omega}\rho \frac{w \cdot w}{2}d\Omega,
\end{aligned}
\end{equation}
with $w$ the vorticity.
\end{itemize}
These outputs are compared with the DNS results obtain on a $512^3$ grid with pseudo-spectral method, reference for the First International Workshop on High-Order CFD Methods held at the 50th AIAA Aerospace Meeting (https://www.grc.nasa.gov/hiocfd/).
It appears that the AION scheme seems to be quite performing for this kind of viscous case. Indeed the AION scheme fits better the DNS results than Heun scheme for time range $t=[0t_c,8t_c]$ and $t=[12t_c,20t_c]$ according to the $E_k$ and $\varepsilon$ (see Fig.~\ref{fig:Ec_256} -~\ref{fig:dEc_256}) when Heun scheme seems to be more dissipative in these time ranges. 
Nevertheless the Heun scheme seems to better reach the maximum of $\varepsilon$ than the AION scheme on the 
time range $t=[8t_c,12t_c]$. On the contrary, the AION scheme gives better result for the enstrophy $E$, 
on the whole time range than Heun scheme (see Fig.~\ref{fig:Ens_256}). This results confirm the 
particularity that the Heun time integrator has a significant importance on accuracy of the numerical 
results. Indeed this conclusion was already observed on the slope of $log(E_{total})$ obtained in the 
previous case of the convected vortex. This comparison allows us to validate our hybridation of the 
viscous gradients presented previously.

\newpage
\section{Conclusion}

RANS and LES equations have almost the same shape and their own pros and cons. In an industrial environment, 
it could be useful to perform LES far from the wall and RANS near the wall in order to account easily for the turbulence effects, 
while keeping an acceptable mesh size. Standard time integration schemes for LES and RANS equations do not follow the same 
constraints. For LES, an explicit time integration is chosen in order to control simply and efficiently spectral properties 
(dissipation and dispersion). For RANS equations, it is of paramount importance to reach the steady state as fast as possible, 
using an implicit time integration procedure. 

Indeed, the current paper specifically addresses how explicit and implicit time integrators can be coupled spatially. Here, 
two standard time integration schemes (Heun and second order implicit Runge-Kutta schemes -IRK2-) are hybridized / blended using 
a transition function $\omega$, while keeping the standard expected properties (spectral behaviour). In the first part of the paper, 
a way to couple the proposed schemes, adapted from the literature, was first introduced but lead to instability for some wavenumbers and 
CFL values. 
A new alternative approach, named AION scheme, was proposed and designed in order to correct the unexpected behaviour of the first 
coupling procedure. The spectral analysis performed on the coupled space / AION schemes enabled us to check the stability of 
the coupling procedure. In addition, in order to minimize the CPU cost, transition to explicit and implicit schemes was performed by 
reducing the transition area, playing with the values of $\omega$. 
After the spectral analysis, attention was paid on simulations of increasing complexity, from the \GP{1D Gaussian hump advection} 
to 3D Taylor Green Vortex. \GP{Three aspects are of paramount importnace. 
First, the results with the new scheme were shown to have same or better quality than the standard basic schemes. 
Moreover, starting from the reference full-implicit time integration, the hybrid formulation enables to reduce the CPU cost. 
Finally, the AION formulation seems less dependent on the Newton's algorithm 
convergence than the standard IRK2 scheme. Moreover, the convergence of the IRK2 
in a fully unsteady simulation needs many steps, something which was not necessary 
with the proposed AION scheme.}

We are working today on two new improvements. The first one deals with the use of an adaptive time-step procedure. 
In this case, explicit cells are separated into several classes depending of their local maximum stable time step. All the cells 
inside the same class are time-integrated using the same time step and for two adjacent classes, time step for the smallest cells is 
half the one for the largest cells. Attention must be paid on the computation of the flux for faces having the cell of one side in a 
given class and the cell on the other side in another one. This work is ongoing and will be summarized in a new paper. The last 
improvement concerns the application of the proposed technique to a coupled RANS/LES computation for demonstrating 
the capability in an industrial environment. We can assume that the corner stone will be the scaling of the hybrid 
parameter $\omega$ with the function that switches between RANS and LES models.



\bibliographystyle{abbrv}
\bibliography{biblio}

\begin{thebibliography}{10}
\expandafter\ifx\csname url\endcsname\relax
  \def\url#1{\texttt{#1}}\fi
\expandafter\ifx\csname urlprefix\endcsname\relax\def\urlprefix{URL }\fi
\expandafter\ifx\csname href\endcsname\relax
  \def\href#1#2{#2} \def\path#1{#1}\fi

\bibitem{Chapman_1979_AIAA}
D.~Chapman, {Computational Aerodynamics Development and Outlook}, AIAA Journal
  17~(12) (1979) 1293--1313.
\newblock \href {http://dx.doi.org/10.2514/3.61311}
  {\path{doi:10.2514/3.61311}}.

\bibitem{JKuntzmann_1961_ZAMM}
J.~Kuntzmann, {Neure Entwicklungen der Methoden von Runge und Kutta},
  Zeitschrift f\"ur angewandte Mathematik und Physik 41 (1961) 29--31.
\newblock \href {http://dx.doi.org/10.1002/zamm.19610411317}
  {\path{doi:10.1002/zamm.19610411317}}.

\bibitem{Butcher_1964_MC}
J.~Butcher, {Implicit Runge-Kutta processes}, Mathematics of Computation
  18~(85) (1964) 50--50.
\newblock \href {http://dx.doi.org/10.1090/s0025-5718-1964-0159424-9}
  {\path{doi:10.1090/s0025-5718-1964-0159424-9}}.

\bibitem{Gear_Book_1971}
G.~Gear, Numerical Initial Value Problems in Ordinary Differential Equations,
  1971.

\bibitem{Dahlquist_1956_MS}
G.~Dahlquist, Convergence and stability in the numerical integration of
  ordinary differential equations, MATHEMATICA SCANDINAVICA 4~(0) (1956)
  33--53.
\newblock \href {http://dx.doi.org/10.7146/math.scand.a-10454}
  {\path{doi:10.7146/math.scand.a-10454}}.

\bibitem{Sengupta_2003_JCP}
T.~K. Sengupta, G.~Ganeriwal, S.~De, {Analysis of central and upwind compact
  schemes}, Journal of Computational Physics 192 (2003) 677--694.
\newblock \href {http://dx.doi.org/10.1016/j.jcp.2003.07.015}
  {\path{doi:10.1016/j.jcp.2003.07.015}}.

\bibitem{Sengupta_2011_JCP}
T.~K. Sengupta, M.~J. Rajpoot, Y.~G. Bhumkar, {Space-time discretizing optimal
  DRP schemes for flow and wave propagation problems}, Journal of Computational
  Physics 47 (2011) 144--154.
\newblock \href {http://dx.doi.org/10.1016/j.compfluid.2011.03.003}
  {\path{doi:10.1016/j.compfluid.2011.03.003}}.

\bibitem{Runge_MathAnn_1895}
C.~Runge, \"uber die numerische aufl\"osung von differentialgleichungen.,
  Mathematische Annalen 46 (1895) 167--178.

\bibitem{Kutta_ZMathPhys_1901}
W.~Kutta, {Beitrag zur n\"aherungsweisen Integration totaler
  Differentialgleichungen}, Zeitschrift f\"ur angewandte Mathematik und Physik
  46 (1901) 435–453.

\bibitem{Bogey_2004_JCP}
C.~Bogey, C.~Bailly, {A family of low dispersive and low dissipative explicit
  schemes for flow and noise computations}, Journal of Computational Physics
  194~(1) (2004) 194--214.
\newblock \href {http://dx.doi.org/10.1016/j.jcp.2003.09.003}
  {\path{doi:10.1016/j.jcp.2003.09.003}}.

\bibitem{Cockburn_MC_52_1989}
B.~Cockburn, C.~Shu, {TVB} {R}unge-{K}utta local projection {D}iscontinuous
  {G}alerkin finite method for conservation laws {II} : General framework,
  Mathematics of Computation 52~(186) (1989) 411--435.
\newblock \href {http://dx.doi.org/10.1016/0021-9991(89)90183-6}
  {\path{doi:10.1016/0021-9991(89)90183-6}}.

\bibitem{Cockburn_JCP_84_1989}
B.~Cockburn, S.~Lin, C.~Shu, {TVB} {R}unge-{K}utta local projection
  discontinuous {G}alerkin finite element method for conservation laws {III}:
  one-dimensional systems, Journal of Computational Physics 84 (1989) 90--113.
\newblock \href {http://dx.doi.org/10.1016/0021-9991(89)90183-6}
  {\path{doi:10.1016/0021-9991(89)90183-6}}.

\bibitem{Cockburn_MC_54_1990}
B.~Cockburn, S.~Hou, C.~Shu, The {R}unge-{K}utta local projection discontinuous
  {G}alerkin finite element method for conservation laws {IV}: the
  multidimensional case, Mathematics of Computation 84~(190) (1990) 545--581.
\newblock \href {http://dx.doi.org/10.2307/2008501}
  {\path{doi:10.2307/2008501}}.

\bibitem{Cockburn_JCP_141_1998}
B.~Cockburn, C.~Shu, The {R}unge-{K}utta discontinuous {G}alerkin method for
  conservation laws {V}: Multidimensional systems, Journal of Computational
  Physics 141~(2) (1998) 199--224.
\newblock \href {http://dx.doi.org/10.1.1.158.1556}
  {\path{doi:10.1.1.158.1556}}.

\bibitem{Gottlieb_MoC_67_1998}
S.~Gottlieb, C.~Shu, Total variation diminishing {R}unge-{K}utta schemes,
  Mathematics of Computation 67~(221) (1998) 73--85.
\newblock \href {http://dx.doi.org/10.1.1.105.4521}
  {\path{doi:10.1.1.105.4521}}.

\bibitem{Williamson_1980_JCP}
J.~Williamson, {Low-storage Runge-Kutta schemes}, Journal of Computational
  Physics 35~(1) (1980) 48--56.
\newblock \href {http://dx.doi.org/10.1016/0021-9991(80)90033-9}
  {\path{doi:10.1016/0021-9991(80)90033-9}}.

\bibitem{Norsett_1969_Springer}
S.~Norsett, {An A-stable modification of the Adams-Bashforth methods}, in:
  {Lecture Notes in Mathematics}, Springer Berlin Heidelberg, 1969, pp.
  214--219.
\newblock \href {http://dx.doi.org/10.1007/bfb0060031}
  {\path{doi:10.1007/bfb0060031}}.

\bibitem{Higham_1993_BIT}
D.~J. Higham, L.~N. Trefethen, {{Stiffness of {ODEs}}}, BIT Numerical
  Mathematics 33~(2) (1993) 285--303.
\newblock \href {http://dx.doi.org/10.1007/bf01989751}
  {\path{doi:10.1007/bf01989751}}.

\bibitem{Catchirayer2018}
M.~Catchirayer, {{{J-F}.} Boussuge}, P.~Sagaut, M.~Montagnac, D.~Papadogiannis,
  X.~Garnaud, Extended integral wall-model for large-eddy simulations of
  compressible wall-bounded turbulent flows, Physics of Fluids 30~(6) (2018)
  065106.
\newblock \href {http://dx.doi.org/10.1063/1.5030859}
  {\path{doi:10.1063/1.5030859}}.

\bibitem{Spalart_2009}
P.~R. Spalart, Detached-eddy simulation, Annual Review of Fluid Mechanics
  41~(1) (2009) 181--202.
\newblock \href {http://dx.doi.org/10.1146/annurev.fluid.010908.165130}
  {\path{doi:10.1146/annurev.fluid.010908.165130}}.

\bibitem{TerracolSagautDeck}
P.~Sagaut, S.~Deck, M.~Terracol, Multiscale and Multiresolution Approaches in
  Turbulence, 2006.

\bibitem{Limare_2016_AIAA}
A.~Limare, P.~Brenner, H.~Borouchaki, An adaptive remeshing strategy for
  unsteady aerodynamics applications, in: 46th AIAA Fluid Dynamics Conference
  Washington, D.C., 13-17 June, AIAA Paper 2016-3180, 2016.
\newblock \href {http://dx.doi.org/10.2514/6.2016-3180}
  {\path{doi:10.2514/6.2016-3180}}.

\bibitem{Pont_JCP_2017}
G.~Pont, P.~Brenner, P.~Cinnella, B.~Maugars, J.~Robinet, {Multiple-correction
  hybrid k-exact schemes for high-order compressible RANS-LES simulations on
  fully unstructured grids}, Journal of Computational Physics 350 (2017)
  45--83.
\newblock \href {http://dx.doi.org/10.1016/j.jcp.2017.08.036}
  {\path{doi:10.1016/j.jcp.2017.08.036}}.

\bibitem{Charrier_2018_1}
L.~Charrier, G.~Pont, S.~Mari{\'e}, P.~Brenner, F.~Grasso, Hybrid {RANS/LES}
  simulation of a supersonic coaxial {He}/air jet experiment at various
  turbulent lewis numbers, in: Progress in Hybrid RANS-LES Modelling, Springer
  International Publishing, 2018, pp. 337--346.
\newblock \href {http://dx.doi.org/10.1007/978-3-319-70031-1_28}
  {\path{doi:10.1007/978-3-319-70031-1_28}}.

\bibitem{COUTEYENCARPAYE_2018_JCS}
J.~C. Carpaye, J.~Roman, P.~Brenner, Design and analysis of a task-based
  parallelization over a runtime system of an explicit finite-volume cfd code
  with adaptive time stepping, Journal of Computational Science 28 (2018) 439
  -- 454.
\newblock \href {http://dx.doi.org/10.1016/j.jocs.2017.03.008}
  {\path{doi:10.1016/j.jocs.2017.03.008}}.

\bibitem{Amandine_2018_AIAA}
A.~Menasria, P.~Brenner, P.~Cinnella, G.~Pont, Toward an improved wall
  treatment for multiple-correction k-exact schemes, in: 2018 Fluid Dynamics
  Conference, AIAA Aviation Forum, Atlanta, Georgia, June 25-29, AIAA Paper
  2018-4164, 2018.
\newblock \href {http://dx.doi.org/10.2514/6.2018-4164}
  {\path{doi:10.2514/6.2018-4164}}.

\bibitem{Amandine_CS_XX_2019}
A.~Menasria, P.~Brenner, P.~Cinnella, Improving the treatment of near-wall
  regions for multiple-correction k-exact schemes, Computers \& Fluids\href
  {http://dx.doi.org/10.1016/j.compfluid.2019.01.009}
  {\path{doi:10.1016/j.compfluid.2019.01.009}}.

\bibitem{Heun_ZMathPhys_1900}
K.~Heun, {{Neue Methoden zur approximativen Integration der
  Differentialgleichungen einer unabh\"angigen ver\"anderlichen}}, Zeitschrift
  f\"ur angewandte Mathematik und Physik 45 (1900) 23--38.

\bibitem{Krivodonova_JCP_229_2010}
L.~Krivodonova, An efficient local time-stepping scheme for solution of
  nonlinear conservation laws, Journal of Computational Physics 229~(22) (2010)
  8537--8551.
\newblock \href {http://dx.doi.org/10.1016/j.jcp.2010.07.037}
  {\path{doi:10.1016/j.jcp.2010.07.037}}.

\bibitem{Strang_1968}
G.~Strang, On the construction and comparison of difference schemes, {SIAM}
  Journal on Numerical Analysis 5~(3) (1968) 506--517.
\newblock \href {http://dx.doi.org/10.1137/0705041}
  {\path{doi:10.1137/0705041}}.

\bibitem{Kim_1985_JCP}
J.~Kim, P.~Moin, Application of a fractional-step method to incompressible
  {Navier-Stokes} equations, Journal of Computational Physics 59~(2) (1985)
  308--323.
\newblock \href {http://dx.doi.org/10.1016/0021-9991(85)90148-2}
  {\path{doi:10.1016/0021-9991(85)90148-2}}.

\bibitem{Ascher_1995_SIAM}
U.~Ascher, S.~Ruuth, B.~Wetton, Implicit-explicit methods for time-dependent
  partial differential equations, SIAM Journal on Numerical Analysis 32~(3)
  (1995) 797--823.
\newblock \href {http://dx.doi.org/10.1137/0732037}
  {\path{doi:10.1137/0732037}}.

\bibitem{Fryxell_1986_JCP}
B.~A. Fryxell, P.~R. Woodward, P.~Colella, K.-H. Winkler, An implicit-explicit
  hybrid method for lagrangian hydrodynamics, Journal of Computational Physics
  63~(2) (1986) 283 -- 310.
\newblock \href {http://dx.doi.org/10.1016/0021-9991(86)90195-6}
  {\path{doi:10.1016/0021-9991(86)90195-6}}.

\bibitem{Dai_1996_JCP}
W.~Dai, P.~R. Woodward, {A second-order iterative implicit-explicit hybrid
  scheme for hyperbolic systems of conservation laws}, Journal of Computational
  Physics 129~(0202) (1996) 181--196.
\newblock \href {http://dx.doi.org/10.1006/jcph.1996.0202}
  {\path{doi:10.1006/jcph.1996.0202}}.

\bibitem{Collins_1995_JCP}
J.~Collins, P.~Colella, H.~Glaz, {An Implicit-Explicit Eulerian Godunov Scheme
  for Compressible Flow}, Journal of Computational Physics 116~(2) (1995)
  195--211.
\newblock \href {http://dx.doi.org/10.1006/jcph.1995.1021}
  {\path{doi:10.1006/jcph.1995.1021}}.

\bibitem{Men_Shov_2004_AIAA}
I.~Men'shov, Y.~Nakamura, {Hybrid Explicit-Implicit, Unconditionally Stable
  Scheme for Unsteady Compressible Flows}, {AIAA} Journal 42~(3) (2004)
  551--559.
\newblock \href {http://dx.doi.org/10.2514/1.9109} {\path{doi:10.2514/1.9109}}.

\bibitem{Timofeev_2016_inProcessing}
E.~Timofeev, F.~Norouzi, Hybrid, explicit-implicit, finite-volume schemes on
  unstructured grids for unsteady compressible flows, in: AIP Conference
  Proceedings, Vol. 1738, 2016, p. 030002.
\newblock \href {http://dx.doi.org/10.1063/1.4951758}
  {\path{doi:10.1063/1.4951758}}.

\bibitem{Toth_2006_JCP}
G.~T\'{o}th, D.~{De Zeeuw}, T.~Gombosi, K.~Powell, A parallel explicit/implicit
  time stepping scheme on block-adaptive grids, Journal of Computational
  Physics 217~(2) (2006) 722--758.
\newblock \href {http://dx.doi.org/10.1016/j.jcp.2006.01.029}
  {\path{doi:10.1016/j.jcp.2006.01.029}}.

\bibitem{May_2016_JSC}
S.~May, M.~Berger, {An Explicit Implicit Scheme for Cut Cells in Embedded
  Boundary Meshes}, Journal of Scientific Computing 71~(3) (2017) 919--943.
\newblock \href {http://dx.doi.org/10.1007/s10915-016-0326-2}
  {\path{doi:10.1007/s10915-016-0326-2}}.

\bibitem{Persson_AIAA_2011}
P.-O. Persson, High-Order LES Simulations using Implicit-Explicit Runge-Kutta
  Schemes, Aerospace Sciences Meetings, 2011, Ch. 49th AIAA Aerospace Sciences
  Meeting including the New Horizons Forum and Aerospace Exposition.
\newblock \href {http://dx.doi.org/10.2514/6.2011-684}
  {\path{doi:10.2514/6.2011-684}}.

\bibitem{Haider_2014_SpringerBook}
F.~Haider, P.~Brenner, B.~Courbet, J.~Croisille, {Parallel Implementation of
  k-Exact Finite Volume Reconstruction on Unstructured Grids}, in: {Lecture
  Notes in Computational Science and Engineering}, Springer International
  Publishing, 2014, pp. 59--75.
\newblock \href {http://dx.doi.org/10.1007/978-3-319-05455-1_4}
  {\path{doi:10.1007/978-3-319-05455-1_4}}.

\bibitem{Gooch_2002_JCP}
C.~Ollivier-Gooch, {M. Van Altena}, {A High-Order-Accurate Unstructured Mesh
  Finite-Volume Scheme for the Advection{\textendash}Diffusion Equation},
  Journal of Computational Physics 181~(2) (2002) 729--752.
\newblock \href {http://dx.doi.org/10.1006/jcph.2002.7159}
  {\path{doi:10.1006/jcph.2002.7159}}.

\bibitem{Crank_1947_PCP}
J.~Crank, P.~Nicolson, A practical method for numerical evaluation of solutions
  of partial differential equations of the heat conduction type, Proceedings of
  the Cambridge Philosophical Society 43 (1947) 50--67.
\newblock \href {http://dx.doi.org/10.1017/S0305004100023197}
  {\path{doi:10.1017/S0305004100023197}}.

\bibitem{Norouzi_AIAAP_3046_2011}
F.~Norouzi, E.~Timofeev, A hybrid, explicit-implicit, second order in space and
  time {TVD}scheme for one-dimensional scalar conservation laws, in: 20th AIAA
  CFD Conference, Honolulu, Hawaii, June 27-30, AIAA Paper 2011-3046, 2011.
\newblock \href {http://dx.doi.org/10.2514/6.2011-3046}
  {\path{doi:10.2514/6.2011-3046}}.

\bibitem{Timofeev_ISSW_2011}
E.~Timofeev, F.~Norouzi, Application of a new hybrid explicit-implicit flow
  solver to 1d unsteady flows with shock waves, in: K.~Kontis (Ed.), Proc. of
  the 28th ISSW, Manchester, UK, 17-22 July, Vol.~2, 2011, pp. 245--250.
\newblock \href {http://dx.doi.org/10.1007/978-3-642-25685-1}
  {\path{doi:10.1007/978-3-642-25685-1}}.

\bibitem{Norouzi_WCCM_XI-ECCM_2014}
F.Norouzi, E.Timofeev, A hybrid,explicit-implicit,second order in space and
  time {TVD}~scheme for two-dimensional compressible flows, in: E.~O. {\it et
  al.} (Ed.), Proc. of the WCCM XI-ECCM V-ECCM VI, Barcelona, Spain, July
  20-25, Vol.~5, 2014, pp. 4820--4831.

\bibitem{Sengupta_2012_AMC}
T.~K. Sengupta, Y.~G. Bhumkar, M.~K. Rajpoot, V.~Suman, S.~Saurabh, {Spurious
  waves in discrete computation of wave phenomena and flow problems}, Applied
  Mathematics and Computation 218~(18) (2012) 9035--9065.
\newblock \href {http://dx.doi.org/10.1016/j.amc.2012.03.030}
  {\path{doi:10.1016/j.amc.2012.03.030}}.

\bibitem{Vanharen_JCP_2017}
J.~Vanharen, G.~Puigt, X.~Vasseur, {{{J-F}.} Boussuge}, P.~Sagaut, Revisiting
  the spectral analysis for high-order spectral discontinuous methods, Journal
  of Computational Physics 337 (2017) 379--402.
\newblock \href {http://dx.doi.org/10.1016/j.jcp.2017.02.043}
  {\path{doi:10.1016/j.jcp.2017.02.043}}.

\bibitem{Vichnevetsky_MCS_23_1981}
R.~Vichnevetsky, Energy and group velocity in semin discretizations of
  hyperbolic equations, Mathematics and Computers in Simulation 23 (1981)
  333--343.
\newblock \href {http://dx.doi.org/10.1016/0378-4754(81)90020-3}
  {\path{doi:10.1016/0378-4754(81)90020-3}}.

\bibitem{Vichnevetsky_1982_book}
R.~Vichnevetsky, J.~B. Bowles, {Fourier Analysis of Numerical Approximations of
  Hyperbolic Equations}, Society for Industrial and Applied Mathematics, 1982.
\newblock \href {http://dx.doi.org/10.1137/1.9781611970876}
  {\path{doi:10.1137/1.9781611970876}}.

\bibitem{Poinsot_2005_book}
T.~Poinsot, D.~Veynante, {Theoretical and Numerical Combustion}, second ed
  Edition, R.T. Edwards Inc., 2005.

\bibitem{Trefethen_1992_SIAM}
{{L. N}. Trefethen}, {Group velocity in finite difference schemes}, SIAM Review
  24~(2) (1992) 113--136.
\newblock \href {http://dx.doi.org/10.1137/1024038}
  {\path{doi:10.1137/1024038}}.

\bibitem{Cinnella_2016_JCP}
P.~Cinnella, C.~Content, {High-order implicit residual smoothing time scheme
  for direct and large eddy simulations of compressible flows}, Journal of
  Computational Physics 326 (2016) 1--29.
\newblock \href {http://dx.doi.org/10.1016/j.jcp.2016.08.023}
  {\path{doi:10.1016/j.jcp.2016.08.023}}.

\bibitem{MARTIN_2006_JCP}
M.~P. Martín, G.~V. Candler, A parallel implicit method for the direct
  numerical simulation of wall-bounded compressible turbulence, Journal of
  Computational Physics 215~(1) (2006) 153 -- 171.
\newblock \href {http://dx.doi.org/10.1016/j.jcp.2005.10.017}
  {\path{doi:10.1016/j.jcp.2005.10.017}}.

\bibitem{Roe_1986_ARFM}
P.~L. Roe, Characteristic-based schemes for the euler equations, Annual Review
  of Fluid Mechanics 18~(1) (1986) 337--365.
\newblock \href {http://dx.doi.org/10.1146/annurev.fl.18.010186.002005}
  {\path{doi:10.1146/annurev.fl.18.010186.002005}}.

\end{thebibliography}

\end{document}